\documentclass[review]{elsarticle}
\usepackage[latin1]{inputenc} 
\usepackage{listings, tcolorbox}

\usepackage{bbm}
\usepackage{}

\usepackage{lineno,hyperref}
\usepackage{amssymb}
\usepackage{color}
\usepackage{amsmath}
\usepackage{tikz}
\usepackage{amsfonts}
\usepackage{mathrsfs}
\usepackage{amsthm}

\usepackage{bbding}
\usepackage{mathrsfs}
\usepackage{amsmath, amsfonts, amssymb, mathrsfs, txfonts}
\usepackage{graphicx, subfigure}
\usepackage{wasysym}
\usepackage{color}
\usepackage{bm}
\usepackage{enumerate}

\usepackage{pifont}
\usepackage{txfonts}
\usepackage{amsmath}
\usepackage{graphicx}
\usepackage{amsfonts}
\usepackage{amssymb}
\usepackage{mathrsfs,psfrag,eepic,epsfig}
\usepackage{epstopdf}

\biboptions{numbers,sort&compress}
\newtheorem{thm}{Theorem}[section]

\theoremstyle{definition}

\modulolinenumbers[5]

\journal{Journal of \LaTeX\ Templates}

\makeatletter \@addtoreset{equation}{section}

\begin{document}

\begin{frontmatter}
\title{Riemann-Hilbert approach and $N$-soliton formula for the $N$-component  Fokas-Lenells
equations}
\tnotetext[mytitlenote]{Project supported by the Fundamental Research Fund for the Central Universities under the grant No. 2019ZDPY07.\\
\hspace*{3ex}$^{*}$Corresponding author.\\
\hspace*{3ex}\emph{E-mail addresses}:  sftian@cumt.edu.cn and
shoufu2006@126.com (S. F. Tian)}

\author{Wei-Kang Xun and Shou-Fu Tian$^{*}$}
\address{
School of Mathematics and Institute of Mathematical Physics, China University of Mining and Technology, Xuzhou 221116, People's Republic of China
 }

\begin{abstract}
  In this work,    the generalized $N$-component  Fokas-Lenells(FL) equations, which have been studied by Guo and Ling  (2012 J. Math. Phys.  53 (7)  073506) for $N=2$,  are first investigated via  Riemann-Hilbert(RH) approach. The main purpose of this is to study the soliton solutions of the coupled Fokas-Lenells(FL) equations for any positive integer  $N$,  which have more complex linear relationship  than the analogues reported before.  We first analyze  the spectral analysis of the Lax pair associated with a $(N+1)\times (N+1)$  matrix spectral problem for the $N$-component  FL  equations. Then,   a kind of RH problem is successfully formulated. By introducing the special conditions of irregularity and reflectionless case,    the $N$-soliton solution formula  of  the equations are derived through solving the corresponding RH problem. Furthermore, take $N=2,3$ and $4$ for examples, the localized structures and dynamic propagation behavior of their soliton solutions and their interactions are  discussed by some graphical analysis.
\end{abstract}


\end{frontmatter}


\section{Introduction}
   As we all known,  the nonlinear Schr\"{o}dinger equation  (NLS) is an important integrable  system, which plays an improtant role in  nonlinear optics, water waves and plasma physics. Futher more,  the Fokas-Lenells(FL) equation  is colsly related to  the NLS equation in the same way as the Camassa-Holm equation associated with the KdV equation. Since the FL system is  one of the important models from both mathematical and physical considerations,  obtaining a series solutions of FL system has been always a focusing subject for many  scholars, and  much of the research has been carried out on the coupled FL system. The single-component FL equation was first constructured  by  Fokas\cite{fokas1995class}.  After that,   the bi-hamiltonian structure, the Lax pair and conservation  laws were constructured  by Fokas and Lenells\cite{lenells2008novel}.  Besides, many other scholars have obtained  a series of solutions of single-component FL equation, such as dark soliton\cite{matsuno2012direct},   algebraic geometry  solution\cite{zhao2013algebro}  and long-time asymptotic behavior of solutions\cite{xu2015long}.  As for multi-component FL equations,  Yang has constructured  the generalized Darboux transformation(DT)  method for the generalized two-component FL equations  to  get the high-order rogue wave solutions\cite{yang2018higher}.  Zhang  et al have obtained the soliton, breather and  rogue waves solutions for a special two-component FL equations  via DT method\cite{zhang2017solitons}.  With the aid of Riemann-Hilbert(RH)  approach,  Kang et al have solve the two-component FL equations to get multi-soliton solutions formula\cite{kang2018multi}.  Hu et al have considered  the initial  boundary value problem for the two-component FL(FL) equations on the half-line via RH approach\cite{hu2017coupled}.  In addition, the multi-soliton solutions of  a $m$-component  FL   equations  with vanishing boundary conditions and nonvanishing  boundary conditions  are also given by bilinear transformation method\cite{matsuno2018multi}.

    Of particular concern in the field of nonlinear science is to find multi-soliton solutions for nonlinear partial differential equations, and a number of effective methods have been produced  to solve this problem,  such as   Hirota bilinear method\cite{hirota1980direct},  Darboux and B\"{a}cklund transformation\cite{matveev1979darboux},  inverse scattering transformation\cite{ablowitz1981solitons,beals1984scattering,beals1988direct} and RH approach.  Moreover,  many scholars show an increasing interest in using the RH approach as a powerful tool to solve certain important problem.  For instance, applying  the RH approach  can  slove the soliton solutions of a series of nonlinear evolution equations\cite{fokas2012unified,lenells2012initial,de2013riemann,geng2016riemann,yan2017initial,ma2018riemann,ma2019inverse,wang2010integrable,zhang2017riemann,tian2017initial,tian2018initial,tian2016mixed,
     xia2018initial,peng2019riemann,yang2019n},  study  integrable systems with non-zero boundaries\cite{ablowitz2018inverse,vekslerchik1992discrete,biondini2014inverse,prinari2015inverse,yang2019riemann},  disscuss the asymptoticity of integrable system solutions\cite{deift1992steepest,tian2018long,wang2019long,liu2019long} and so on. The main purpose of our work is to use   RH approach which is a powerful tool to solve the multi-soliton solutions of a new class of multi-component FL equations.

   In this work, we mainly consider a generalized  $N$-component  FL equations
   \begin{equation}\label{fangcheng}
            u_{x,t}-3u+i (u_x u^{\dagger} A u + u u^{\dagger} A u_x) =0,
   \end{equation}
    where $u(x,t)=\left( u_1(x,t),u_2(x,t),\dots,u_N(x,t) \right)^{T}$ is a $N$-component vector function and matrix $A$ is Hermitian.  In \cite{guo2012riemann}, the  $N$-component  FL equations  are given as a by-product of the coupled derivative Schr\"{o}dinger equations(cDNLS). In addition,  the  $N$-component  FL equations\cite{guo2012riemann} is equivalent to the coupled FL system in \cite{zhang2017solitons} by a gauge transformation.  Since the authors only considered the multi-soliton solutions under  the simplest non-trivial case, i.e.,
    \begin{equation}
   u_{x,t}-3u+i (u_x u^{\dagger} A u + u u^{\dagger} A u_x) =0, \quad  u(x,t)=(u_1, u_2)^{T}, \quad   A = \begin{pmatrix}
        1  &  0  \\
        0  & \sigma
      \end{pmatrix},             \sigma= \pm 1,
  \end{equation}
    we decided to generalize the author's result to obtain  the $N$-soliton solutions in more complex case. In our work, we consider  the more complex case that the FL system is a $N$-component equations and the relationship matrix $A$ is promoted to the general Hermitian matrix, i.e.,
  \begin{equation}
   u_{x,t}-3u+i (u_x u^{\dagger} A u + u u^{\dagger} A u_x) =0, \quad  u(x,t)=(u_1, u_2, \dots,u_{N-1},u_{N})^{T},
  \end{equation}
  and
  \begin{equation}
   A = \begin{pmatrix}
    a_{1,1} & a_{2,1}^{*} & a_{3,1}^{*} & \dots & a_{N,1}^{*}  \\
    a_{2,1} & a_{2,2} & a_{3,2}^{*} & \dots & a_{N,2}^{*} \\
     a_{3,1} & a_{3,2}  & a_{3,3} & \dots & a_{N,3}^{*} \\
      \vdots & \vdots & \vdots  & \ddots & \vdots  \\
        a_{N,1} &a_{N,2} & a_{N,3} & \dots & a_{N,N} \\
      \end{pmatrix},
\end{equation}
with  $a_{i,i}(1 \leq i \leq N+1)$ are real constants  and $a_{i,j}(i \neq j)$   are complex  constants.  Via carrying out the  RH approach, we  successfully obtain the multi-soliton solutions  of the new generalized FL equations and get some certain interesting phenomena about the soltions.

The outline of this work is as follows. In Section 2,  starting from  analyzing  the spectral problem of the Lax pair  and analyticity of scattering matrix,  a  RH  problem for the  $N$-component  FL equations  is formulated.  Next, via solving the RH problem, we obtain the explicit multi-soliton solutions of   the  $N$-component  FL equations. In Section 3,  we consider   the solutions under the special case that $N$ and the elements of $A$ are taken as fixed values.    Moreover, the localized structures and dynamic propagation behavior of these solutions are presented vividly by some graphics.   Finally,  the conclusions  are given in the last section.

\section{Riemann-Hilbert  problem}
We begin our discussion by considering the Lax pair representation of  the $N$-component  FL equations
\begin{equation} \label{lax1}
 \left\{
   \begin{aligned}
       \Phi_x & = U \Phi, &  U &= i \frac{1}{\zeta^2} \sigma_0 + i \frac{1}{\zeta} U_{1,x}, \\
       \Phi_t & = V \Phi, &  V &= - \frac{1}{3} \zeta^2  i \sigma_0 + \zeta \sigma_1 U_1 - i \sigma_1 U_1^2,
   \end{aligned}
   \right.
\end{equation}
where
\begin{equation}
   \sigma_0=\begin{pmatrix} -2 & 0_{1\times N} \\ 0_{N\times 1} & I_{N\times N}    \end{pmatrix}, \quad \sigma_1=\begin{pmatrix} 1 & 0_{1\times N} \\ 0_{N\times 1} & - I_{N\times N}    \end{pmatrix},  \quad  U_1=\begin{pmatrix} 0 & v^T \\ u  & 0_{N\times N}  \end{pmatrix},
\end{equation}
with  $u=(u_1,u_2,\dots,u_N)^T$  and  $v=(v_1,v_2,\dots,v_N)^T$. If taking 
$v=A^T u^{\ast}$, Eq. \eqref{fangcheng} can be derived by  the compatibility condition of  Eq. \eqref{lax1}.
The Lax pair \eqref{lax1} can be rewritten  as  the equivalent form
\begin{equation}\label{lax2}
\left\{
   \begin{aligned}
    \Phi_x &= i \frac{1}{\zeta^2} \sigma_0 \Phi+ P \Phi, \\
    \Phi_t &= -\frac{1}{3} i \zeta^2 \sigma_0 \Phi +  Q   \Phi,
   \end{aligned}
\right.
\end{equation}
where $P=i \frac{1}{\zeta} U_{1,x},   Q= \zeta \sigma_1 U_1 - i \sigma_1 U_1^2$. From Eq. \eqref{lax2}, when $|x| \to \infty$, we have
\begin{equation}
 \Phi \varpropto \exp(i \frac{1}{\zeta^2} \sigma_0 x -\frac{1}{3}\zeta^2 i\sigma_0 t).
\end{equation}
Let $J = \Phi \exp(-i \frac{1}{\zeta^2} \sigma_0 x + \frac{1}{3} \zeta^2 i\sigma_0 t)$, then we obtain
\begin{equation}\label{lax3}
\left\{
     \begin{aligned}
        J_x & = i \frac{1}{\zeta^2} [\sigma_0, J] +P J,   \\
        J_t &= - \frac{1}{3} i \zeta^2 [\sigma_0, J] +Q J,
     \end{aligned}
\right.
\end{equation}
where $[\sigma_0, J]=\sigma_0 J - J \sigma_0$  is the commutator. Moreover, we can get the following formula
\begin{equation}\label{eq1}
  d(e^{-i(\frac{1}{\zeta^2} - \frac{1}{3} \zeta^2) \hat{\sigma}_0   } J)= e^{-i(\frac{1}{\zeta^2} - \frac{1}{3}\zeta^2 ) \hat{\sigma}_0} (P dx +Q dt) J.
\end{equation}
Now, let us construct two matrix solutions of Eq. \eqref{lax3}
\begin{equation}
\begin{aligned}
   J_{-} &= ([J_{-}]_{1},[J_{-}]_{2},\dots,[J_{-}]_{N+1}),  \\
    J_{+} &= ([J_{+}]_{1},[J_{+}]_{2},\dots,[J_{+}]_{N+1}),
\end{aligned}
\end{equation}
where each $[J_{\pm}]_{l}$ denotes  the $l$-th  column of  $J_{\pm}$.   In addition,  $J_{\pm}$  are determined by
\begin{equation}
 \begin{aligned}
     J_{-}&=\mathbb{I}+ \int_{-\infty}^{x}e^{i \frac{1}{\zeta^2} \sigma_0 (x- \xi)} P(\xi) J_{-}(\xi, \zeta)e^{-i \frac{1}{\zeta^2} \sigma_0 (x- \xi)} d\xi,  \\
     J_{+}&=\mathbb{I}- \int_{x}^{+\infty}e^{i \frac{1}{\zeta^2} \sigma_0 (x- \xi)} P(\xi) J_{+}(\xi, \zeta)e^{-i \frac{1}{\zeta^2} \sigma_0 (x- \xi)} d\xi,
 \end{aligned}
\end{equation}
which satisfy the asymptotic conditions
\begin{equation}
\begin{aligned}
   J_{-} \to \mathbb{I}, & \quad \zeta \to -\infty,   \\
    J_{+} \to \mathbb{I}, & \quad \zeta \to +\infty.
\end{aligned}
\end{equation}
It is easy to find that $[J_{-}]_{1},[J_{+}]_{2},[J_{+}]_{3},\dots,[J_{+}]_{N+1}$ are analytic  for  $\zeta \in \mathbb{D}^{+}$,   and $[J_{+}]_{1},[J_{-}]_{2},[J_{-}]_{3},\dots,[J_{-}]_{N+1}$ are analytic  for  $\zeta \in \mathbb{D}^{-}$, where
\begin{equation}
  \begin{aligned}
      \mathbb{D}^{+} & = \left\{  \zeta   |   \arg \zeta  \in (0,\frac{\pi}{2}) \cup (\pi,\frac{3\pi}{2})    \right\}, \\
      \mathbb{D}^{-} & = \left\{  \zeta   |   \arg \zeta  \in (\frac{\pi}{2}, \pi ) \cup (\frac{3\pi}{2}, 2 \pi )    \right\}.
  \end{aligned}
\end{equation}

Next, we pay attention to the properties of $J_{\pm}$. According to the Abel's identity and $tr(U)=0$, we obtain that $\det(J_{\pm})$  are  independent of $x$ and $\det(J_{\pm})=1$.  Moreover, since $J_{-}E$ and $J_{+}E$  are the matrix solutions of the spectral problem,  where $E=e^{i \frac{1}{\zeta^2} \sigma_0 x}$, they are linearly related by a $(N+1)\times (N+1)$ scattering matrix $S(\zeta)=(s_{kj})_{(N+1)\times (N+1)}$, namely
\begin{equation}\label{linear}
   J_{-}E=J_{+}ES(\zeta), \quad \zeta \in \mathbb{R}\cup i  \mathbb{R},
\end{equation}
where
\begin{equation}
S(\zeta)=\begin{pmatrix}
s_{1,1} & \dots & s_{1,N}  & s_{1,N+1}  \\
\vdots  &  \ddots  & \vdots & \vdots    \\
s_{N,1}  & \dots  & s_{N,N} & s_{N,N+1} \\
s_{N+1,1}  & \dots  & s_{N+1,N} & s_{N+1,N+1}
\end{pmatrix}
\end{equation}
 represents the inverse scattering matrix, $\zeta \in \mathbb{R}\cup i  \mathbb{R}$.
 Futhermore, it is easy to know $\det(S(\zeta))=1$.

A Riemann-Hilbert problem  to be formulated for the $N$-component  FL equations need two matrix functions: one is analytic in $\mathbb{D}^{+}$ and the other is analytic in $\mathbb{D}^{-}$.  Let $\Gamma_1=\Gamma_1(x,\zeta)$ be an  alalytic function of $\zeta$
\begin{equation}
   \Gamma_1(x,\zeta)=([J_{-}]_{1},[J_{+}]_{2},[J_{+}]_{3},\dots,[J_{+}]_{N+1})(x,\zeta),
\end{equation}
defining in $\mathbb{D}^{+}$, with the asymptotic behavior $\Gamma_1 \to  \mathbb{I}$,   $\zeta \in \mathbb{D}^{+} \to \infty$.

To formulate a Riemann-Hilbert probelm for the $N$-component  FL equations, we also need to consider the inverse matrices of $J_{\pm}$.
We write the matrices $J_{\pm}^{-1}$ as a collection of rows
\begin{equation}
  J_{\pm}^{-1}=\begin{pmatrix}   [J_{\pm}^{-1}]_{1} \\ [J_{\pm}^{-1}]_{2}  \\ \vdots    \\ [J_{\pm}^{-1}]_{N} \\ [J_{\pm}^{-1}]_{N+1}                  \end{pmatrix}.
\end{equation}
It can be seen that $J_{\pm}^{-1}$ satisfy
 \begin{equation}
 K_x =i \frac{1}{\zeta^2} [\sigma_0, K] - K P,
 \end{equation}
and meet the boundary condition $J_{\pm}^{-1} \to \mathbb{I}$, $x \to \pm \infty$. The matrix function $\Gamma_2(x,\zeta)$ which is analytic in $\mathbb{D}^{-}$, can be defined as follows:
\begin{equation}
\Gamma_2(x,\zeta)=\begin{pmatrix} [J_{-}^{-1}]_{1} \\ [J_{+}^{-1}]_{2} \\ [J_{+}^{-1}]_{3} \\ \vdots  \\ [J_{+}^{-1}]_{N+1}  \end{pmatrix} (x,\zeta).
\end{equation}
Similar to $\Gamma_1(x,\zeta)$, we obtain  that $\Gamma_2(x,\zeta) \to  \mathbb{I}$ as $\zeta \in \mathbb{D}^{-} \to \infty$. Besides, we can get the linear relationship
\begin{equation}
   e^{-i \frac{1}{\zeta^2} \sigma_0 x} J_{-}^{-1} = R(\zeta)  e^{-i \frac{1}{\zeta^2} \sigma_0 x}  J_{+}^{-1}, \quad \zeta \in \mathbb{R} \cup i \mathbb{R},
\end{equation}
where
\begin{equation}
R(\zeta)=\begin{pmatrix}
r_{1,1} & \dots & r_{1,N}  & r_{1,N+1}  \\
\vdots  &  \ddots  & \vdots & \vdots    \\
r_{N,1}  & \dots  & r_{N,N} & r_{N,N+1} \\
r_{N+1,1}  & \dots  & r_{N+1,N} & r_{N+1,N+1}
\end{pmatrix}
\end{equation}
 represents the inverse scattering matrix, $\zeta \in \mathbb{R}\cup i \mathbb{R}$. As for the analyticity  of the scattering matrix and the inverse scattering matrix, we have the following theorem,
\begin{thm}
    The spectral function $\Gamma_1(x,\zeta)$, the element $s_{1,1}$ and  $r_{k,j}(2 \leq k,j \leq N+1)$ are analytic in $\mathbb{D}^{+}$;  The spectral function $\Gamma_2(x,\zeta)$, the element $r_{1,1}$ and  $s_{k,j}(2 \leq k,j \leq N+1)$ are analytic in $\mathbb{D}^{-}$; In addition,  $s_{1,k}(2 \leq k \leq N+1)$ and  $s_{j,1}(2 \leq j \leq N+1)$ are  not analytic in   $\mathbb{D}^{+}$ or $\mathbb{D}^{-}$  but continuous to the real axis and image axis, and $r_{1,k}(2 \leq k \leq N+1)$ and  $r_{j,1}(2 \leq j \leq N+1)$ are  not analytic in   $\mathbb{D}^{+}$ or $\mathbb{D}^{-}$  and not continuous to the real axis and image axis.
\end{thm}
Based on the above analysis, the Riemann-Hilbert  problem of the $N$-component  FL equations can be formulated.
\begin{thm}
     Let's make the convention that  the limit of $\Gamma_1(x,\zeta)$  when $\zeta \in \mathbb{D}^{+}$  approaches  $\mathbb{R} \cup  i \mathbb{R}$ is $\Gamma^{+}$,  and the limit of $\Gamma_2(x,\zeta)$  when $\zeta \in \mathbb{D}^{-}$  approaches  $\mathbb{R} \cup  i \mathbb{R}$ is $\Gamma^{-}$, then the Riemann-Hilbert problem can be set up as follows:
\begin{equation}
\begin{aligned}
                 &  \Gamma^{\pm}    \mbox{ are  analytic  in }    \mathbb{D}^{\pm},  \\
                  &    \Gamma^{-} \Gamma^{+}=G(x,\zeta),& \quad  \zeta \in \mathbb{R},  \\
                 &    \Gamma^{\pm} \to  \mathbb{I},   & \quad   \zeta \to  \infty,
\end{aligned}
\end{equation}
where $G= \begin{pmatrix}
 1  &  r_{1,2}e^{-\frac{3i}{\zeta^2}x} & r_{1,3}e^{-\frac{3i}{\zeta^2}x}  & \dots & r_{1,N}e^{-\frac{3i}{\zeta^2}x} & r_{1,N+1}e^{-\frac{3i}{\zeta^2}x}  \\
 s_{2,1}e^{\frac{3i}{\zeta^2}x} & 1 & 0  & \dots & 0 & 0 \\
 s_{3,1}e^{\frac{3i}{\zeta^2}x} & 0 & 1  & \dots & 0 & 0  \\
 \vdots & \vdots  & \vdots & \ddots & \vdots & \vdots  \\
 s_{N,1}e^{\frac{3i}{\zeta^2}x} & 0 & 0  & \dots & 1 & 0  \\
 s_{N+1,1}e^{\frac{3i}{\zeta^2}x} & 0 & 0  & \dots & 0 & 1
\end{pmatrix}$,
and the canonical  normalization condition of the RH problem is
\begin{equation}
     \begin{aligned}
            \Gamma_1(x,\zeta) \to \mathbb{I},   & \quad \zeta \in \mathbb{D}^{+} \to \infty,  \\
            \Gamma_2(x,\zeta) \to \mathbb{I},   & \quad \zeta \in \mathbb{D}^{-} \to \infty.
     \end{aligned}
\end{equation}
\end{thm}
To solve the RH problem, we suppose that it is irregular, which means that $\det(\Gamma_1)$ and  $\det(\Gamma_2)$ have certain zeros in the analytic domains, respectively. According to  the definition of them, we can obtain
\begin{equation}
\begin{aligned}
  \det(\Gamma_1(x,\zeta))= s_{1,1}(\zeta), \quad  & \zeta \in \mathbb{D}^{+}, \\
  \det(\Gamma_2(x,\zeta))= r_{1,1}(\zeta), \quad   & \zeta \in \mathbb{D}^{-}. \\
\end{aligned}
\end{equation}
Because of the above important results, it is necessary to consider the characteristics of zeros by the symmetry of $U_(x,\zeta)$, which is helpful for classfying the soliton solutions of  the  $N$-component  FL equations. At frist, we can see  that the matrix $U_1$ satisfy
\begin{equation}\label{symme1}
  U_1^{\dagger}= - B U_1 B^{-1},
\end{equation}
where $B=\begin{pmatrix}
-1 &  0  \\
0  &  A
\end{pmatrix}$, and the symbol  $\dagger$  represents the  Hermitian transporse of one matrix. According to Eqs. \eqref{lax3} and \eqref{symme1}, we obtain
\begin{equation}
 J_{\pm}^{\dagger}(\zeta^{*})= B  J_{\pm}^{-1}(\zeta) B^{-1},
\end{equation}
futhermore, the scattering matrix $S(\zeta)$  meet the condition
\begin{equation}
  B^{-1} S^{\dagger}(\zeta^{*}) B =S^{-1} (\zeta) = R(\zeta),
\end{equation}
moreover,
\begin{equation}\label{sym1}
\Gamma_1^{\dagger}(\zeta^{*})= B  \Gamma_2(\zeta) B^{-1}.
\end{equation}

Besides, the  matrix $U_1$ meets the relation $U_1=-\sigma_1 U_1 \sigma_1 $, based on which we can conclude  that
\begin{equation}
J_{\pm}(-\zeta) = \sigma_1 J_{\pm}(\zeta)   \sigma_1,
\end{equation}
and
\begin{equation}\label{sym2}
\Gamma_1(\zeta) = \sigma_1 \Gamma_1(-\zeta)  \sigma_1.
\end{equation}

At this point, we suppose that $\det(\Gamma_1)$ has 2$N_1$ simple zeros $\{\zeta_{j}  \}(1\leq j \leq 2N_1)$  in $\mathbb{D}^{+}$ whcih  satisfy  $\zeta_{N+l}=-\zeta_{l}$, $1\leq l \leq N_1$. At the same time, $\det(\Gamma_{2})$ has  2$N_1$ simple zeros $\{\hat{\zeta_{j}}  \}(1\leq j \leq 2N_1)$ in $\mathbb{D}^{-}$, where $\hat{\zeta_{j}}=\zeta_{j}^{*}$, $1 \leq j \leq 2 N_1$.

In fact, the scattering data needed to  solve the RH problem include the continuous scattering data $\{s_{2,1},s_{3,1},\dots, s_{N+1,1}\}$ as well as the discrete data  $\{ \zeta_{j}, \hat{\zeta_{j}}, \vartheta_{j}, \hat{\vartheta_{j}}    \}(1\leq j \leq 2N_1)$, where $\vartheta_{j}$ and $\hat{\vartheta_{j}}$ are nonzero column vectors and row vectors, respectively, satisfying
\begin{equation}\label{scatter}
\begin{aligned}
  \Gamma_{1}(\zeta_{j}) \vartheta_{j} & = 0,  \\
  \hat{\vartheta_{j}} \Gamma_2(\hat{\zeta_{j}}) & =0.
\end{aligned}
\end{equation}
According to Eqs. \eqref{sym1} and \eqref{scatter}, we can  reveal the relation
\begin{equation}
  \hat{\vartheta_{j}} = \vartheta_{j}^{\dagger} B, \quad 1 \leq j \leq 2 N_1.
\end{equation}
Similarly, from  Eqs. \eqref{sym2} and \eqref{scatter}, we can obtain
\begin{equation}
  \vartheta_{N+j} =  \sigma_1  \vartheta_{j},  \quad  1 \leq j \leq  N_1.
\end{equation}
To obtain the explicit form of $\vartheta_{j}(1\leq j \leq 2N_1)$,   we take the derivatives of the first expression of   Eq. \eqref{scatter}
with respect to $x$ and  $ t$ and get
\begin{equation}
\left\{
    \begin{aligned}
        \frac{\partial{\vartheta_{j}}}{\partial{x}} & = i \frac{1}{\zeta_{j}^{2}} \sigma_0  \vartheta_{j}, \\
        \frac{\partial{\vartheta_{j}}}{\partial{t}} & = - i \frac{1}{3} \zeta_{j}^{2} \sigma_0  \vartheta_{j},
    \end{aligned}
    \right.
\end{equation}
then the $\vartheta_{j}$  and $\hat{\vartheta_{j}}$ are determined by
\begin{equation}
  \vartheta_{j}= \left\{
                        \begin{array}{ll}
                        e^{i(\frac{1}{\zeta_{j}^2} x -\frac{1}{3}\zeta_{j}^{2} t)\sigma_0} \vartheta_{j,0},  & \quad   1 \leq j \leq N_1, \\
                       \sigma_{3} e^{i(\frac{1}{\zeta_{j-N_1}^2} x -\frac{1}{3}\zeta_{j-N}^{2} t )\sigma_0} \vartheta_{j-N,0},  & \quad   N_1+1 \leq j \leq 2 N_1,
                        \end{array}
       \right.
\end{equation}
and
\begin{equation}\label{111}
  \hat{\vartheta_{j}}= \left\{
                        \begin{array}{ll}
                       \vartheta_{j,0}^{\dagger} e^{[i(\frac{1}{\zeta_{j}^2} x -\frac{1}{3}\zeta_{j}^{2} t ) ]^{*}\sigma_0} B,  & \quad   1 \leq j \leq N_1, \\
                    \vartheta_{j-N,0}^{\dagger}   e^{[i(\frac{1}{\zeta_{j-N_1}^2}  x -\frac{1}{3}\zeta_{j-N}^{2} t)]^{*} \sigma_0}  \sigma_{3}  B,  & \quad   N_1+1 \leq j \leq 2 N_1,
                        \end{array}
       \right.
\end{equation}
where $\vartheta_{j,0}$ are the complex constant vectors.

It is pointed out that the Riemann-Hilbert problem  examined corresponds to the reflectionless case, namely, $s_{2,1}=s_{3,1}=\dots =s_{N+1,1}=0$. We introduce a $2N_1 \times 2N_1$ matrix $M$  with elements
\begin{equation}
m_{k,j}=\frac{\hat{\vartheta_{k}}\vartheta_{j}}{\zeta_{j}-\hat{\zeta_{k}}},  \quad  1 \leq  k,j  \leq 2N_1,
\end{equation}
and suppose that the inverse matrix $M^{-1}$ exists, then the solutions of the RH problem can be given by
\begin{equation}\label{srh}
   \begin{aligned}
       \Gamma_1(x,\zeta) & = \mathbb{I} - \sum_{k=1}^{2N_1}\sum_{j=1}^{2N_1} \frac{\vartheta_{k}\hat{\vartheta_{j}}(M^{-1})_{k,j}}{\zeta-\hat{\zeta_{j}}}, \\
              \Gamma_2(x,\zeta) & = \mathbb{I} + \sum_{k=1}^{2N_1}\sum_{j=1}^{2N_1} \frac{\vartheta_{k}\hat{\vartheta_{j}}(M^{-1})_{k,j}}{\zeta-\zeta_{k}}.
   \end{aligned}
\end{equation}
Futhermore, we take the expansion for $\Gamma_1(x,\zeta)$
\begin{equation}\label{expan}
   \Gamma_1(x,\zeta)= \mathbb{I} + \frac{\Gamma_{1}^{(1)}}{\zeta}+\frac{\Gamma_{1}^{(2)}}{\zeta^2}+\frac{\Gamma_{1}^{(3)}}{\zeta^3}+ O\left(\frac{1}{\zeta^4} \right).
\end{equation}
Comparing Eq. \eqref{srh} and Eq. \eqref{expan}, we obtain
\begin{equation}\label{solu1}
   \Gamma_{1}^{(1)}= -\sum_{k=1}^{2N_1} \sum_{j=1}^{2N_1} \vartheta_{k}\hat{\vartheta_{j}}(M^{-1})_{k,j},
\end{equation}
substituting the above expression into Eq. \eqref{lax3}, the following relationship can be obtained
\begin{equation}
    U_1 = \frac{i}{3} \sigma_1 [\sigma_0, \Gamma_{1}^{(1)}],
\end{equation}
more explicitly,
\begin{equation}
  \left\{
 \begin{aligned}
   u_1(x,t) & =-i (\Gamma_{1}^{(1)})_{2,1},  \\
     u_2(x,t) & =-i (\Gamma_{1}^{(1)})_{3,1},  \\
     \dots  &  \dots,  \\
       u_N(x,t) & =-i (\Gamma_{1}^{(1)})_{N+1,1},  \\
 \end{aligned}
 \right.
\end{equation}
where $(\Gamma_{1}^{(1)})_{k,j}$  denotes the $(k,j)$-entry of matrix $\Gamma_{1}^{(1)}$. Besides, from Eq. \eqref{solu1}, the potential functions $u_j(x,t)(1 \leq j \leq N)$ can be recovered  as follows
\begin{equation}\label{exact}
  \left\{
 \begin{aligned}
   u_1(x,t) & =i \left(\sum_{k=1}^{2N_1} \sum_{j=1}^{2N_1} \vartheta_{k}\hat{\vartheta_{j}}(M^{-1})_{k,j}\right)_{2,1},  \\
     u_2(x,t) & =i \left(\sum_{k=1}^{2N_1} \sum_{j=1}^{2N_1} \vartheta_{k}\hat{\vartheta_{j}}(M^{-1})_{k,j}\right)_{3,1},  \\
     \dots  &  \dots,  \\
       u_N(x,t) & =i \left(\sum_{k=1}^{2N_1} \sum_{j=1}^{2N_1} \vartheta_{k}\hat{\vartheta_{j}}(M^{-1})_{k,j}\right)_{N+1,1},  \\
 \end{aligned}
 \right.
\end{equation}
where $(M^{-1})_{k,j}$ denotes the $(k,j)$-entry of inverse matrix of $M$.

\section{Multi-soliton solutions}
  To obtain the explicit  expression of the multi-soliton solutions, we should make more efforts. At first, we  set $\vartheta_{j,0}=(\mu_{j,1},\mu_{j,2},\dots,\mu_{j,N+1})^{T}$ , $\theta_{j}=i(\frac{1}{\zeta_{j}^2} x  -\frac{1}{3}\zeta_{j}^{2}  t)$ and $\zeta_j=a_j+i b_j$. When $1 \leq j \leq N_1$,
\begin{equation}\label{11}
  \vartheta_{j}= e^{\theta_{j} \sigma_0}  \vartheta_{j,0} =\begin{pmatrix}
  e^{-2 \theta_{j}}  &  0  & 0 &  \dots  & 0    \\
   0  &  e^{\theta_{j}} & 0   &  \dots  & 0    \\
   0  &  0 & e^{\theta_{j}}    &  \dots  & 0    \\
     \vdots  &  \vdots &  \vdots    &  \ddots  & \vdots    \\
     0  &  0 & 0   &  \dots  & e^{\theta_{j}}    \\
 \end{pmatrix}
 \begin{pmatrix}
   \mu_{j,1}  \\
   \mu_{j,2}  \\
   \vdots      \\
   \mu_{j,N}  \\
   \mu_{j,N+1}
 \end{pmatrix}=
 \begin{pmatrix}
   \mu_{j,1} e^{-2 \theta_{j}}  \\
   \mu_{j,2}  e^{ \theta_{j}} \\
   \vdots      \\
   \mu_{j,N}  e^{\theta_{j}} \\
   \mu_{j,N+1} e^{\theta_{j}}
 \end{pmatrix},
\end{equation}
when $N_1+1 \leq j \leq 2N_1$,
\begin{equation}\label{21}
  \vartheta_{j}= \sigma_1 e^{\theta_{j-N_1} \sigma_0}  \vartheta_{j-N_1,0} =
  \begin{pmatrix}
    1  &  0  & 0 &  \dots  & 0    \\
   0  &  -1 & 0   &  \dots  & 0    \\
   0  &  0 & -1  &  \dots  & 0    \\
     \vdots  &  \vdots &  \vdots    &  \ddots  & \vdots    \\
     0  &  0 & 0   &  \dots  & -1    \\
  \end{pmatrix}
  \begin{pmatrix}
   \mu_{j-N_1,1} e^{-2 \theta_{j-N_1}}  \\
   \mu_{j-N_1,2}  e^{ \theta_{j-N_1}} \\
   \vdots      \\
   \mu_{j-N_1,N}  e^{\theta_{j-N_1}} \\
   \mu_{j-N_1,N+1} e^{\theta_{j-N_1}}
 \end{pmatrix}=
 \begin{pmatrix}
   \mu_{j-N_1,1} e^{-2 \theta_{j-N_1}}  \\
   -\mu_{j-N_1,2}  e^{ \theta_{j-N_1}} \\
   \vdots      \\
   -\mu_{j-N_1,N}  e^{\theta_{j-N_1}} \\
   -\mu_{j-N_1,N +1} e^{\theta_{j-N_1}}
 \end{pmatrix}.
\end{equation}
According to Eqs .\eqref{11} , \eqref{21} and \eqref{111}, we can obtain the explicit expressions of $\vartheta_j$ and $\hat{\vartheta_j}$  $(1\leq j \leq 2 N_1)$  needed to solve the RH  problem. Inserting  these  data into  Eq. \eqref{exact}, we have the explicit expressions of multi-soliton solutions.

\subsection{Case 1: multi-soliton solutions of  two-component FL equations}
To observe the propagation behavior of the solutions, we take $ N=2$,  $A=\begin{pmatrix} 1   &  0 \\ 0 &  -1  \end{pmatrix}$, i.e. $v_1= u_1^*$ and $v_2=-u_2^*$,   Eq. \eqref{fangcheng}  is reduced into the following form
\begin{equation}
  \begin{aligned}
     u_{1,xt} & = 3 u_1 - i (2 |u_1|^2 u_{1,x} - u_2^* u_1 u_{2,x} - |u_2|^2 u_{1,x}), \\
     u_{2,xt} & = 3 u_2 - i (-2 |u_2|^2 u_{2,x} - u_1^* u_2 u_{1,x} - |u_1|^2 u_{2,x}),
  \end{aligned}
\end{equation}
when $N_1=1$,  we can express the solution to Eq. \eqref{exact}  explicitly
\begin{equation}\label{so1}
  \left\{
   \begin{aligned}
        u_1(x,t) =& i  ( -\mu_{1,2} \mu_{1,1}^{*} e^{\theta_1 -2\theta_1^*} (M^{-1})_{1,1} -\mu_{1,2} \mu_{1,1}^{*} e^{\theta_1 -2\theta_1^*} (M^{-1})_{1,2} \\
         & + \mu_{1,2} \mu_{1,1}^{*} e^{\theta_1 -2\theta_1^*} (M^{-1})_{2,1} + \mu_{1,2} \mu_{1,1}^{*} e^{\theta_1 -2\theta_1^*} (M^{-1})_{2,2} ),  \\
                 u_2(x,t) =& i (-\mu_{1,3} \mu_{1,1}^{*} e^{\theta_1 -2\theta_1^*} (M^{-1})_{1,1} -\mu_{1,3} \mu_{1,1}^{*} e^{\theta_1 -2\theta_1^*} (M^{-1})_{1,2} \\
         & + \mu_{1,3} \mu_{1,1}^{*} e^{\theta_1 -2\theta_1^*} (M^{-1})_{2,1} + \mu_{1,3} \mu_{1,1}^{*} e^{\theta_1 -2\theta_1^*} (M^{-1})_{2,2}            ),  \\
   \end{aligned}
\right.
\end{equation}
where
\begin{equation}
  \left\{
   \begin{aligned}
       m_{1,1}  & = \frac{-|\mu_{1,1}|^2 e^{-2\theta_1-2\theta_1^*} + |\mu_{1,2}|^2 e^{  \theta_1 + \theta_1^*} -|\mu_{1,3}|^2 e^{ \theta_1 +\theta_1^*} }{\zeta_1 -\zeta_1^*},  \\
              m_{1,2}  & = \frac{-|\mu_{1,1}|^2 e^{-2\theta_1-2\theta_1^*} - |\mu_{1,2}|^2 e^{  \theta_1 + \theta_1^*} + |\mu_{1,3}|^2 e^{ \theta_1 +\theta_1^*} }{\zeta_2 -\zeta_1^*},  \\
                m_{2,1}  & = \frac{-|\mu_{1,1}|^2 e^{-2\theta_1-2\theta_1^*} - |\mu_{1,2}|^2 e^{  \theta_1 + \theta_1^*} + |\mu_{1,3}|^2 e^{ \theta_1 +\theta_1^*} }{\zeta_1 + \zeta_1^*},  \\
               m_{2,2}  & = \frac{-|\mu_{1,1}|^2 e^{-2\theta_1-2\theta_1^*} + |\mu_{1,2}|^2 e^{  \theta_1 + \theta_1^*} - |\mu_{1,3}|^2 e^{ \theta_1 +\theta_1^*} }{\zeta_2 + \zeta_1^*},  \\
   \end{aligned}
\right.
\end{equation}

\noindent
{\rotatebox{0}{\includegraphics[width=3.6cm,height=3.0cm,angle=0]{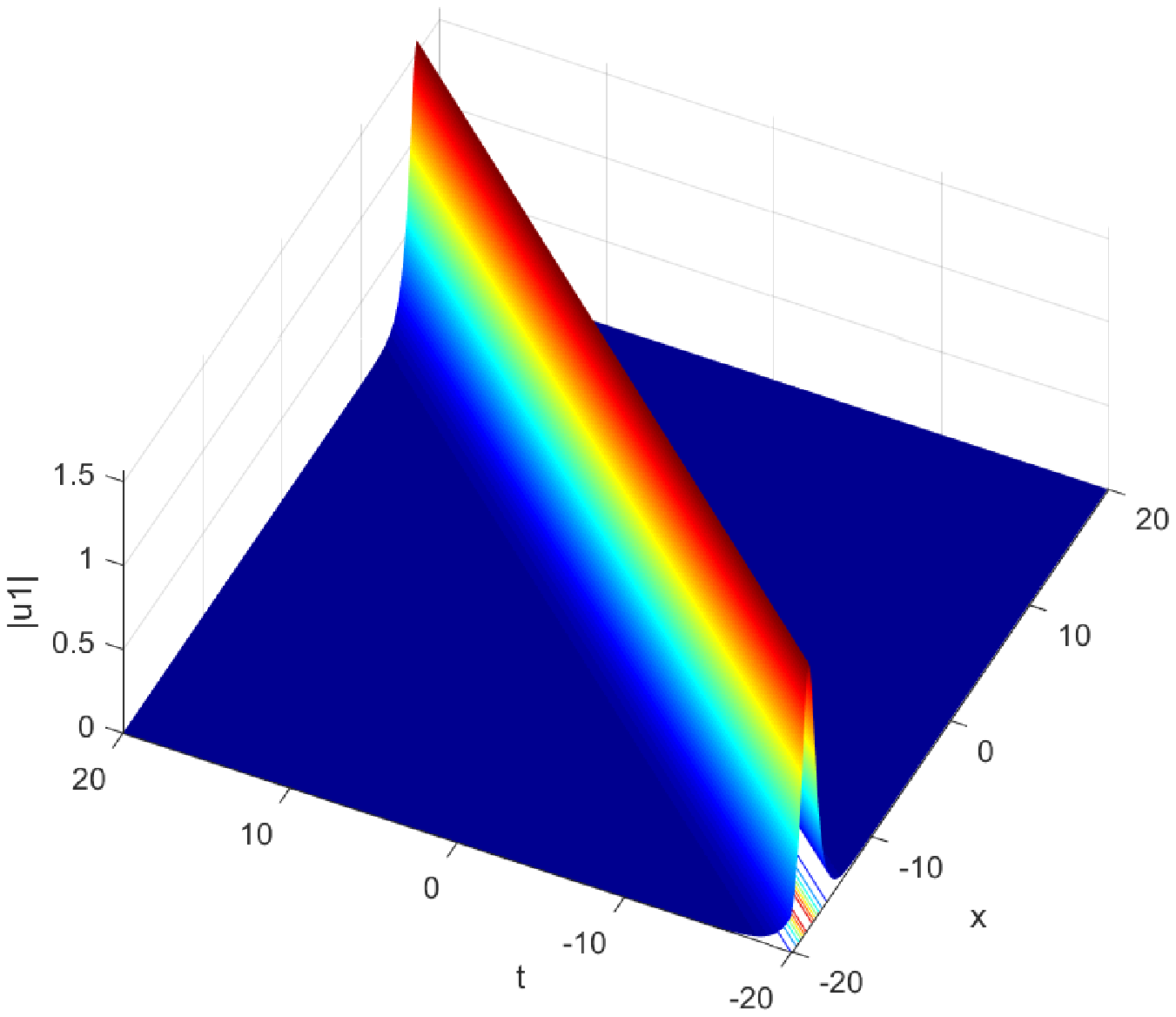}}}
~~~~
{\rotatebox{0}{\includegraphics[width=3.6cm,height=3.0cm,angle=0]{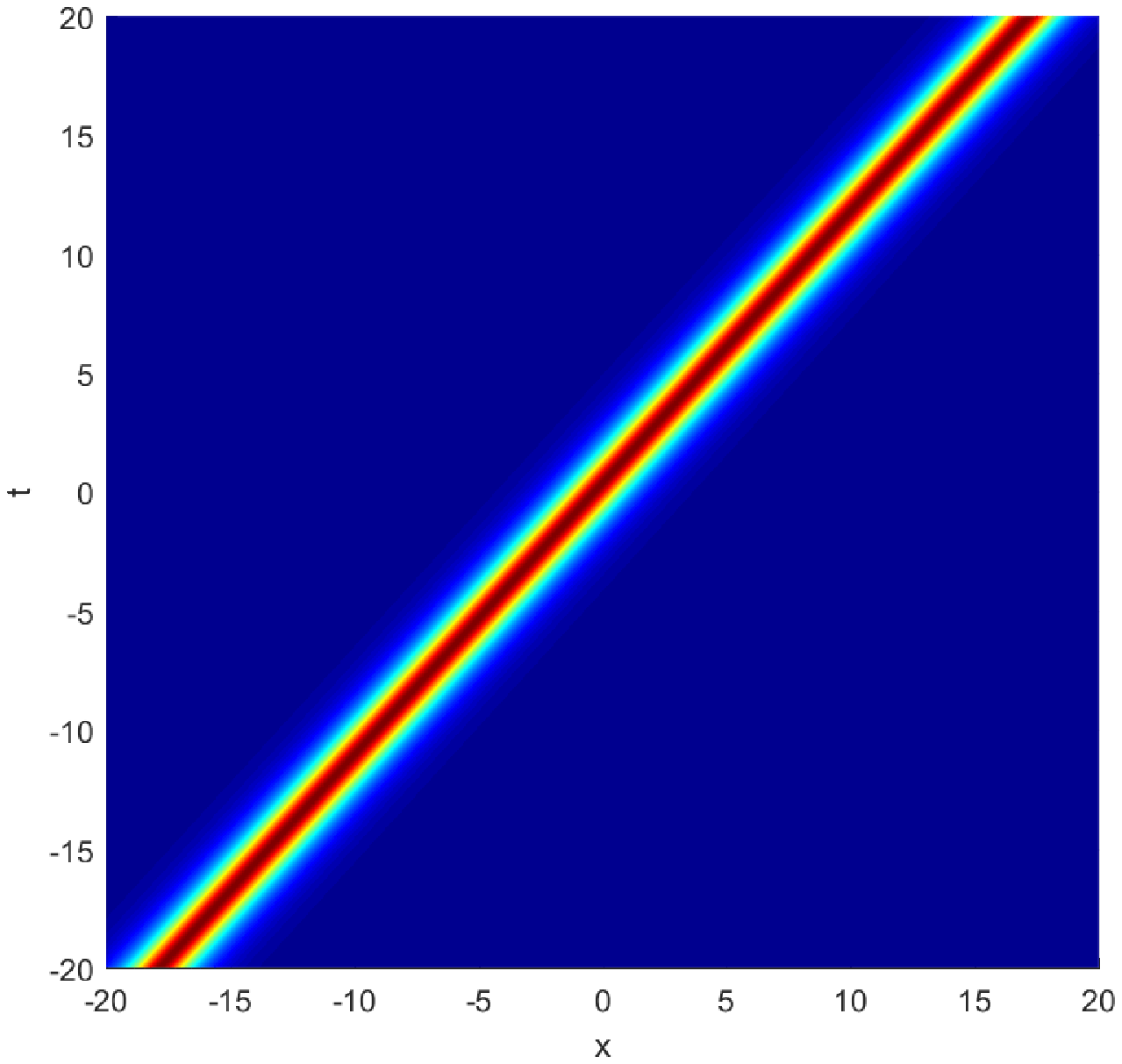}}}
~~~~
{\rotatebox{0}{\includegraphics[width=3.6cm,height=3.0cm,angle=0]{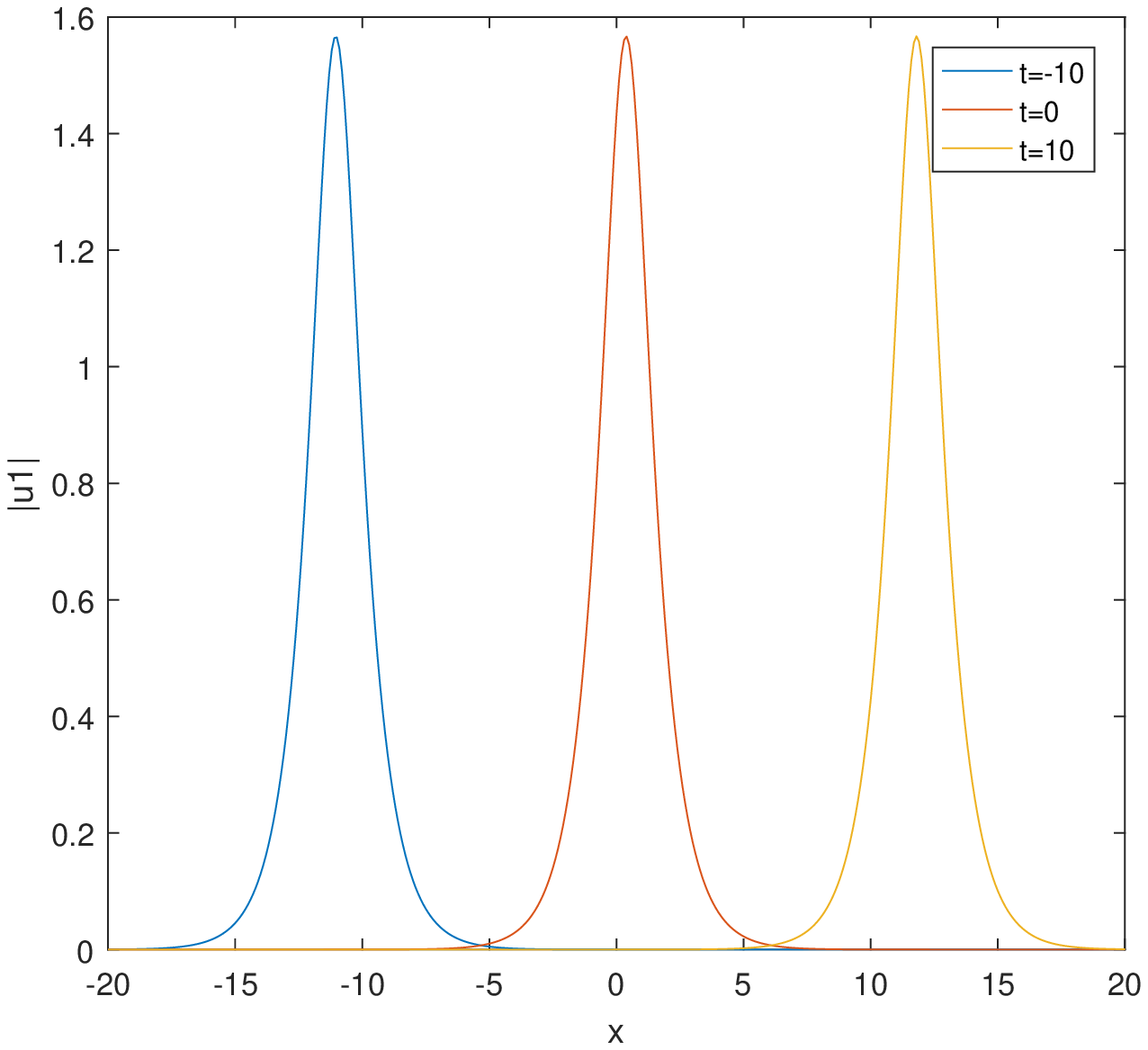}}}

$\ \qquad~~~~~~(\textbf{a})\qquad \ \qquad\qquad\qquad\qquad~(\textbf{b})
\ \qquad\qquad\qquad\qquad\qquad~(\textbf{c})$\\
\noindent
{\rotatebox{0}{\includegraphics[width=3.6cm,height=3.0cm,angle=0]{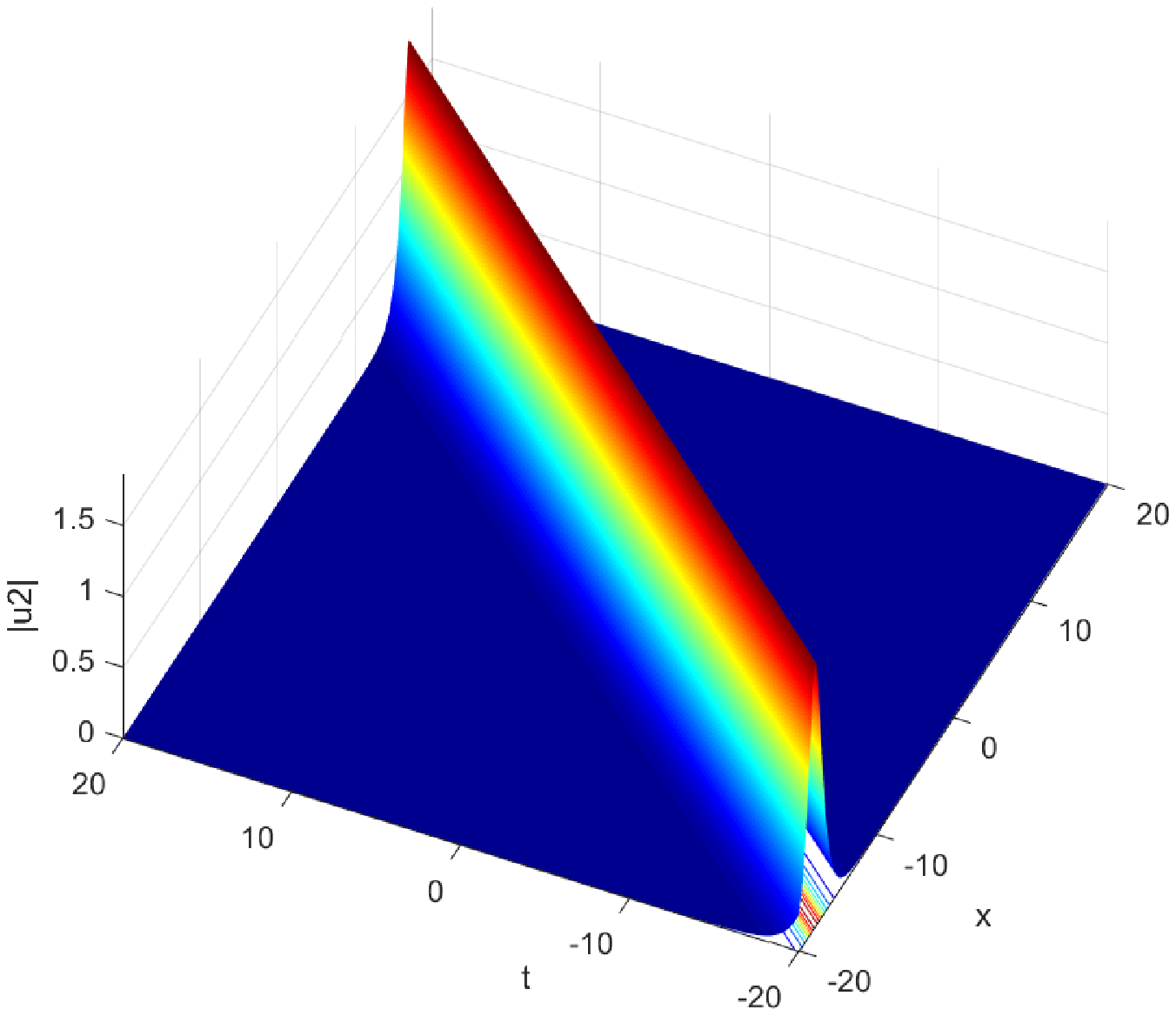}}}
~~~~
{\rotatebox{0}{\includegraphics[width=3.6cm,height=3.0cm,angle=0]{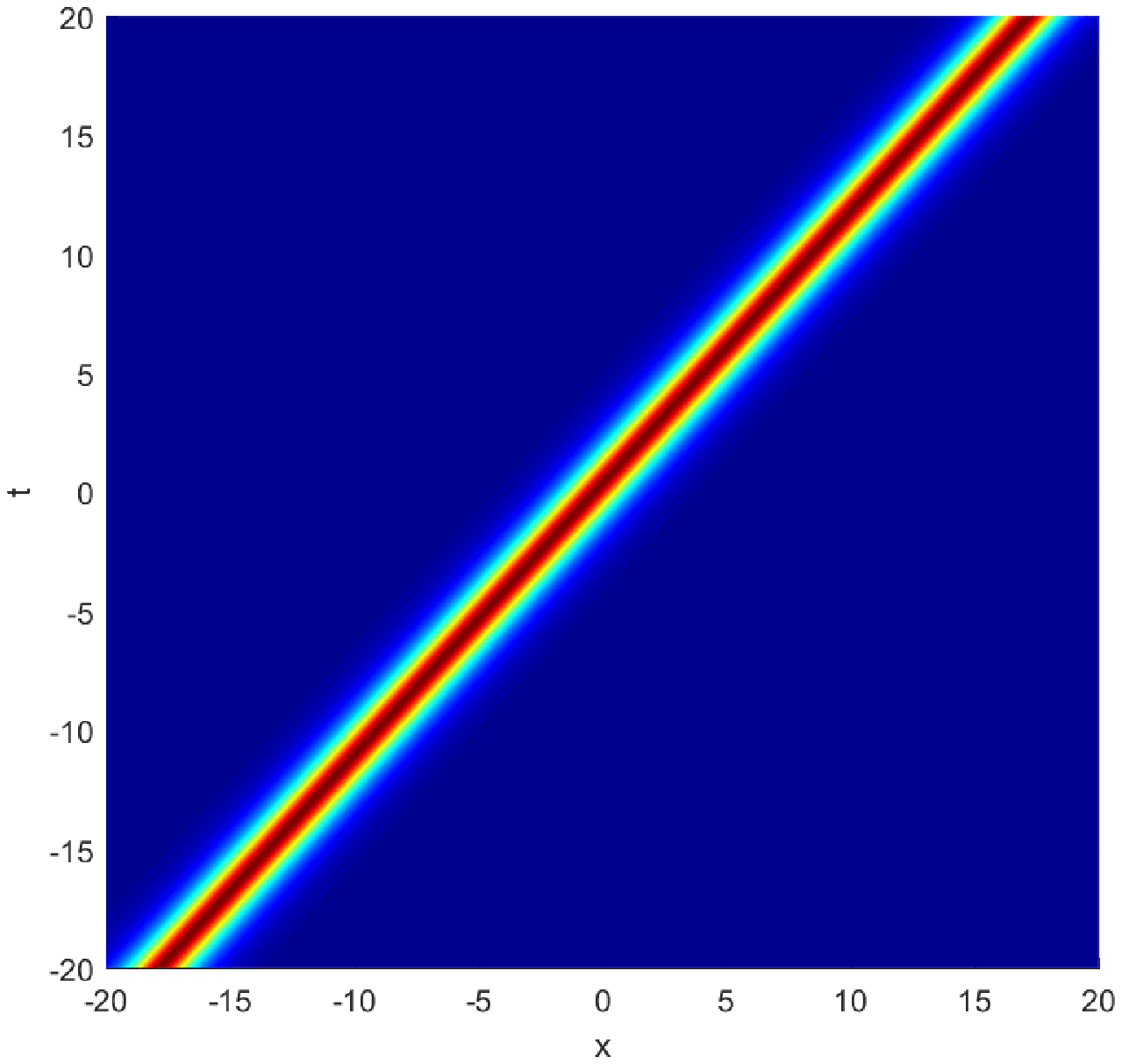}}}
~~~~
{\rotatebox{0}{\includegraphics[width=3.6cm,height=3.0cm,angle=0]{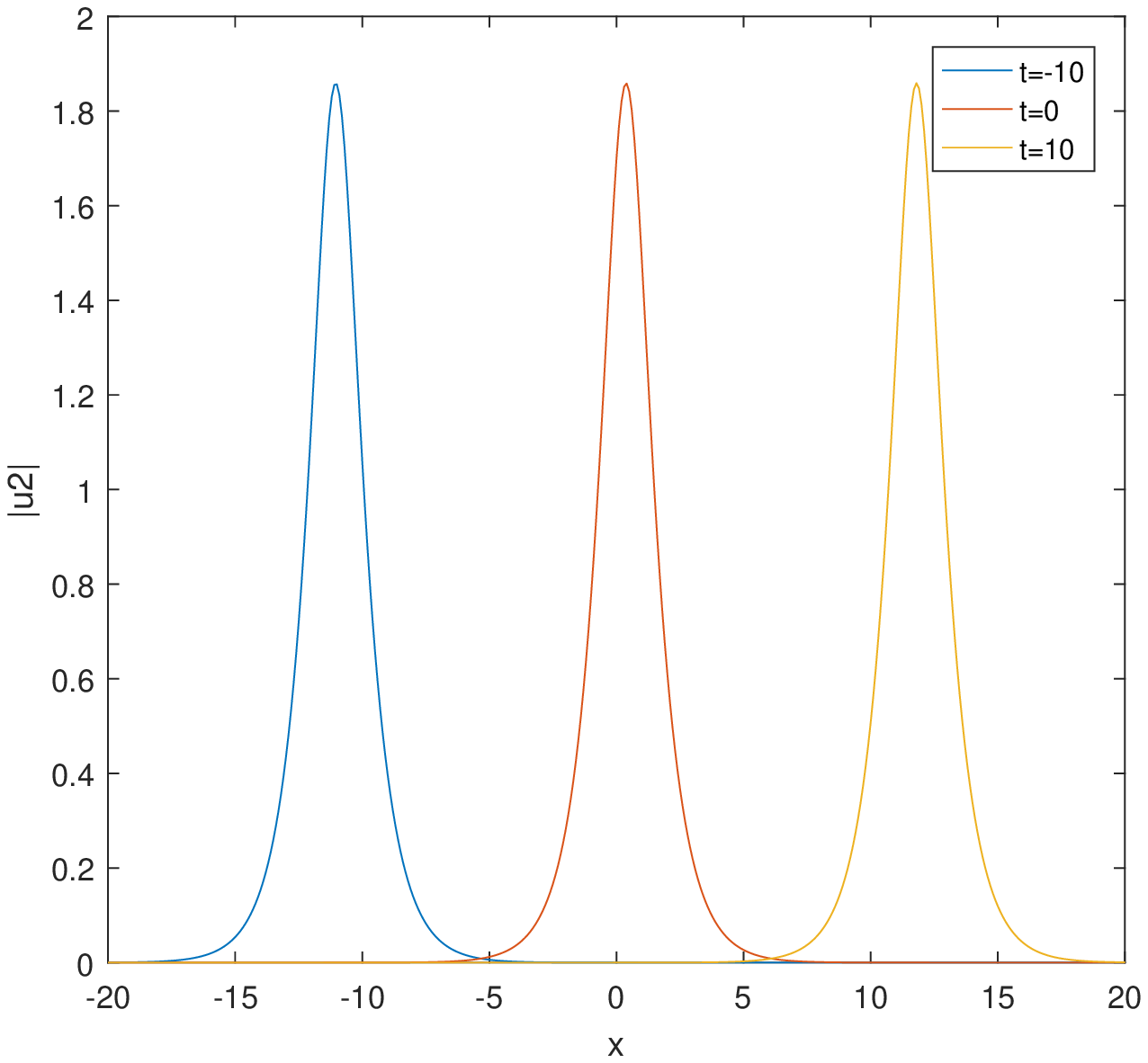}}}

$\ \qquad~~~~~~(\textbf{d})\qquad \ \qquad\qquad\qquad\qquad~(\textbf{e})
\ \qquad\qquad\qquad\qquad\qquad~(\textbf{f})$\\
\noindent { \small \textbf{Figure 1.} One-hump  solutions to Eq. \eqref{so1} with parameters $\zeta_1=1.2+0.5i$, $\mu_{1,1}=0.97$, $\mu_{1,2}=1.02$,  and $\mu_{1,3}=1.21$ .
$\textbf{(a)(b)(c)}$: the local structure, density and wave propagation of the one-hump  solution $|u_1(x,t)|$,
$\textbf{(d)(e)(f)}$: the local structure, density and wave propagation of the one-hump  solution $|u_2(x,t)|$.}  \\

\noindent
{\rotatebox{0}{\includegraphics[width=3.6cm,height=3.0cm,angle=0]{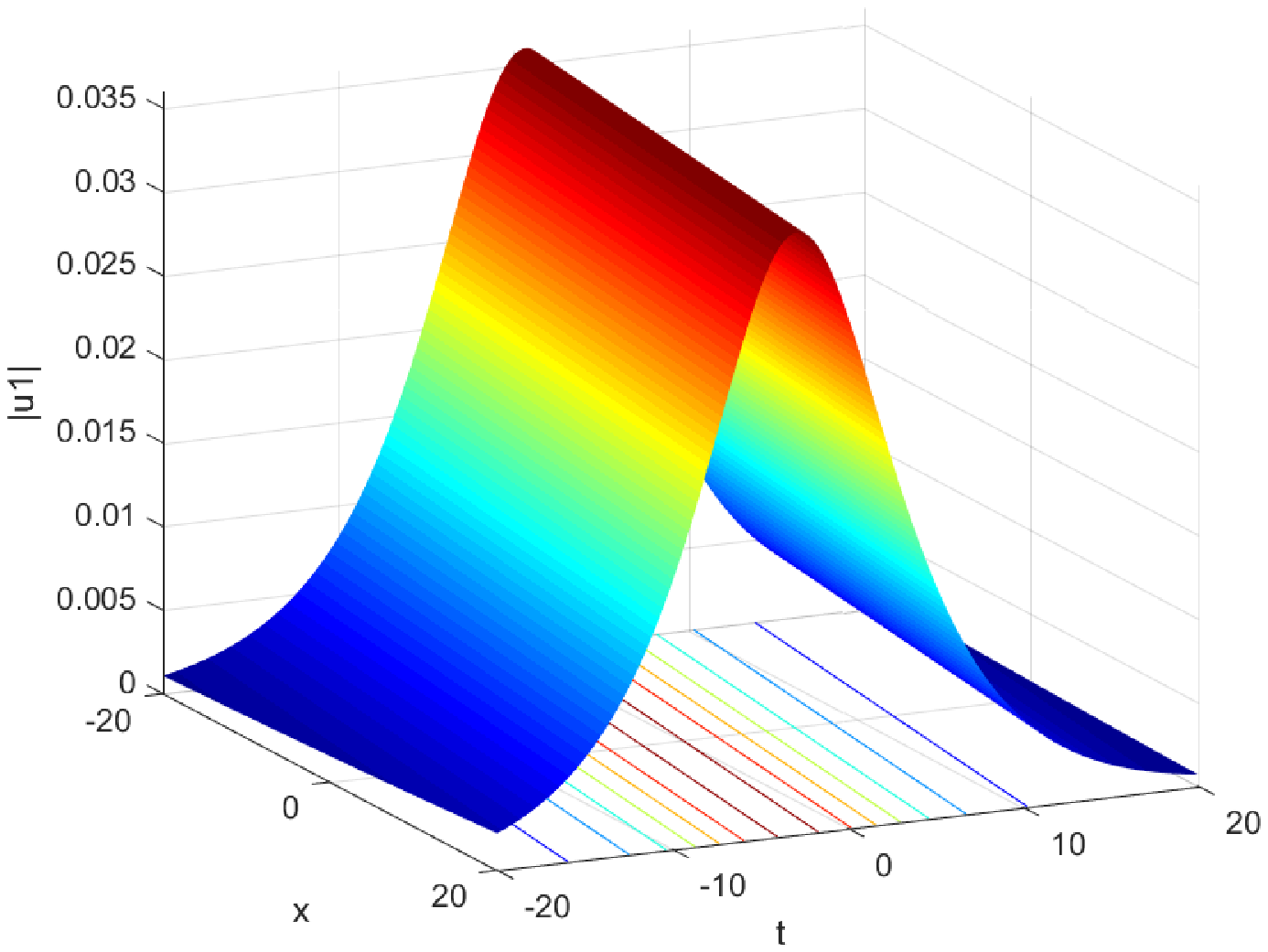}}}
~~~~
{\rotatebox{0}{\includegraphics[width=3.6cm,height=3.0cm,angle=0]{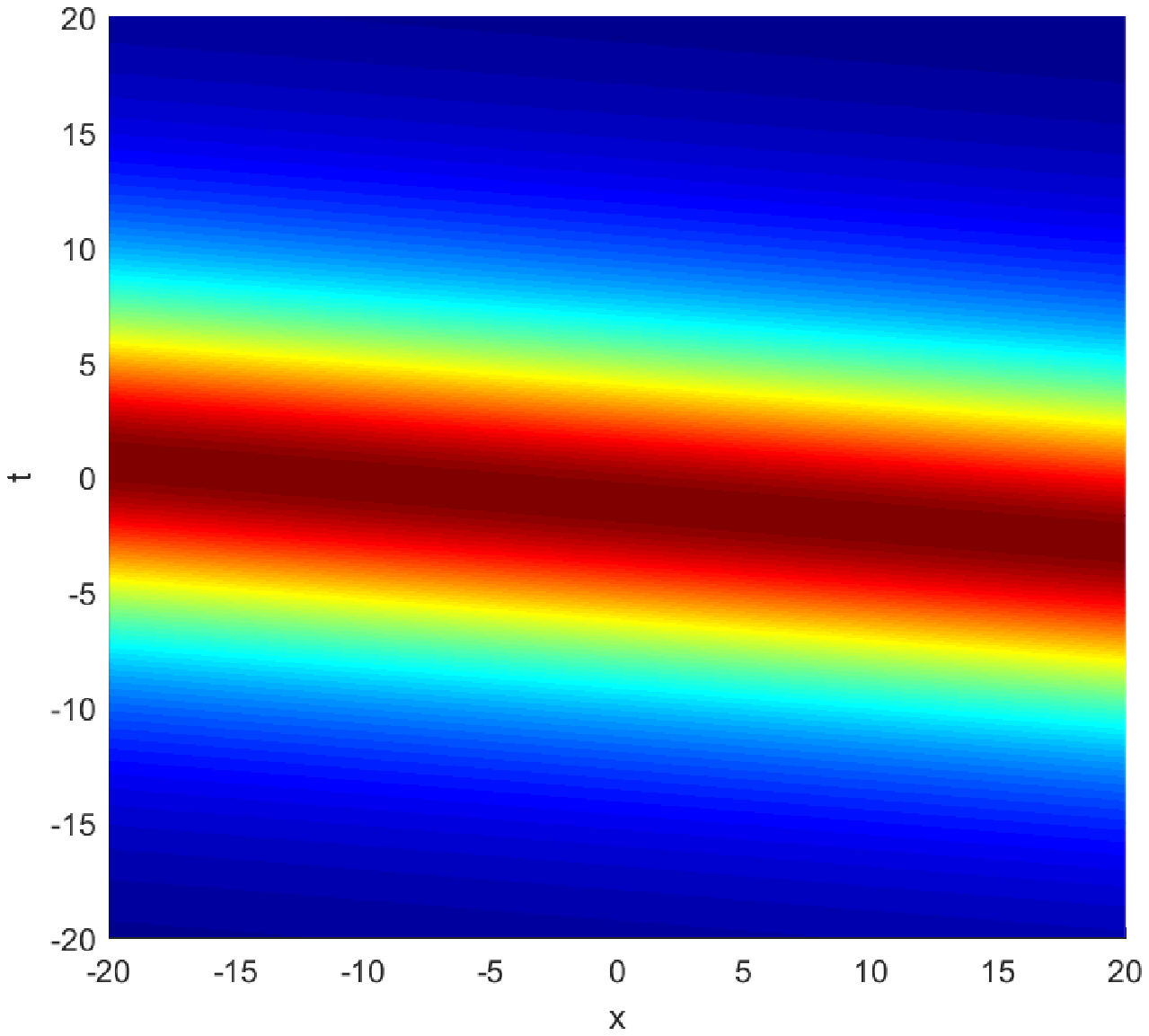}}}
~~~~
{\rotatebox{0}{\includegraphics[width=3.6cm,height=3.0cm,angle=0]{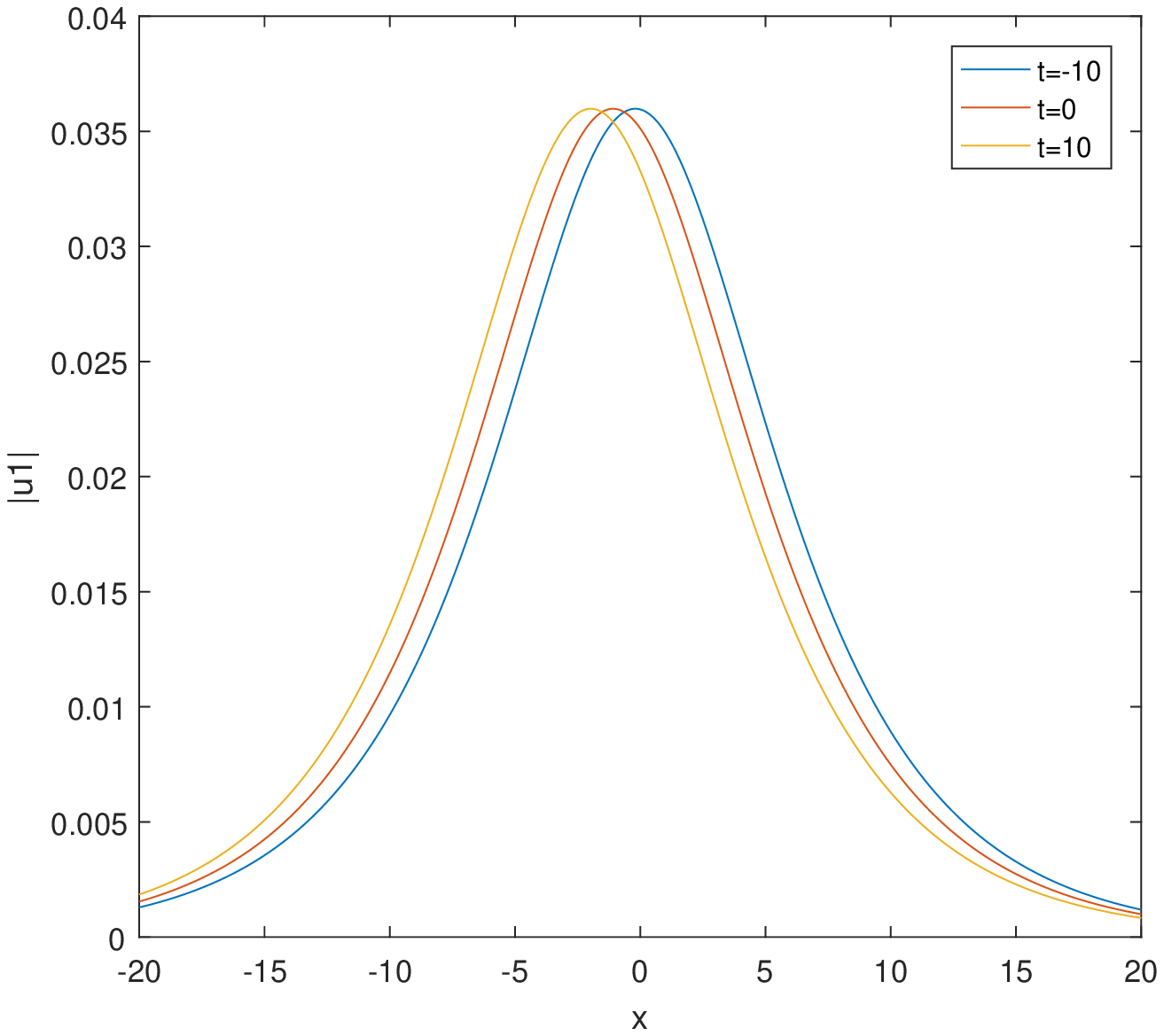}}}

$\ \qquad~~~~~~(\textbf{a})\qquad \ \qquad\qquad\qquad\qquad~(\textbf{b})
\ \qquad\qquad\qquad\qquad\qquad~(\textbf{c})$\\
\noindent
{\rotatebox{0}{\includegraphics[width=3.6cm,height=2.8cm,angle=0]{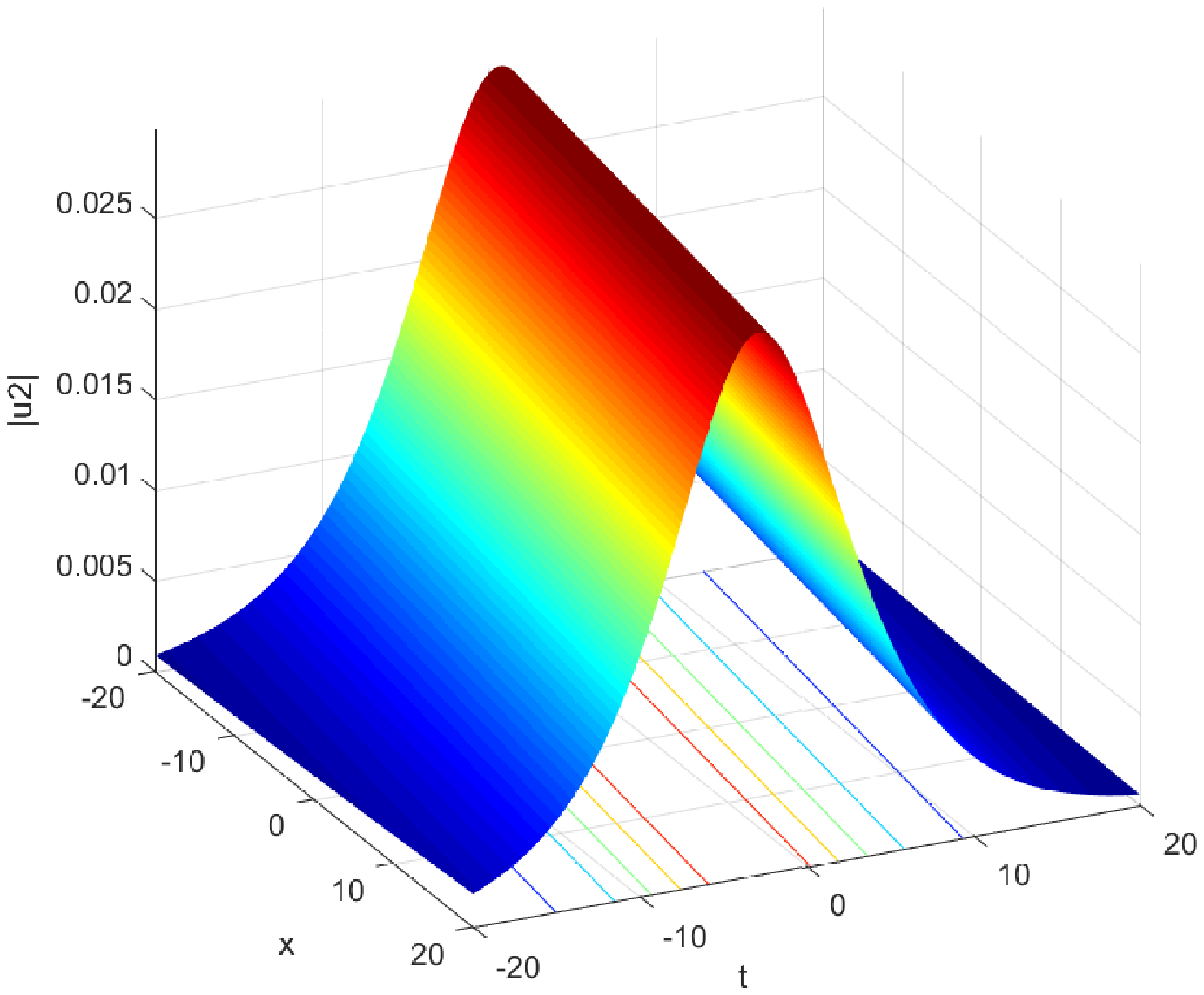}}}
~~~~
{\rotatebox{0}{\includegraphics[width=3.6cm,height=2.8cm,angle=0]{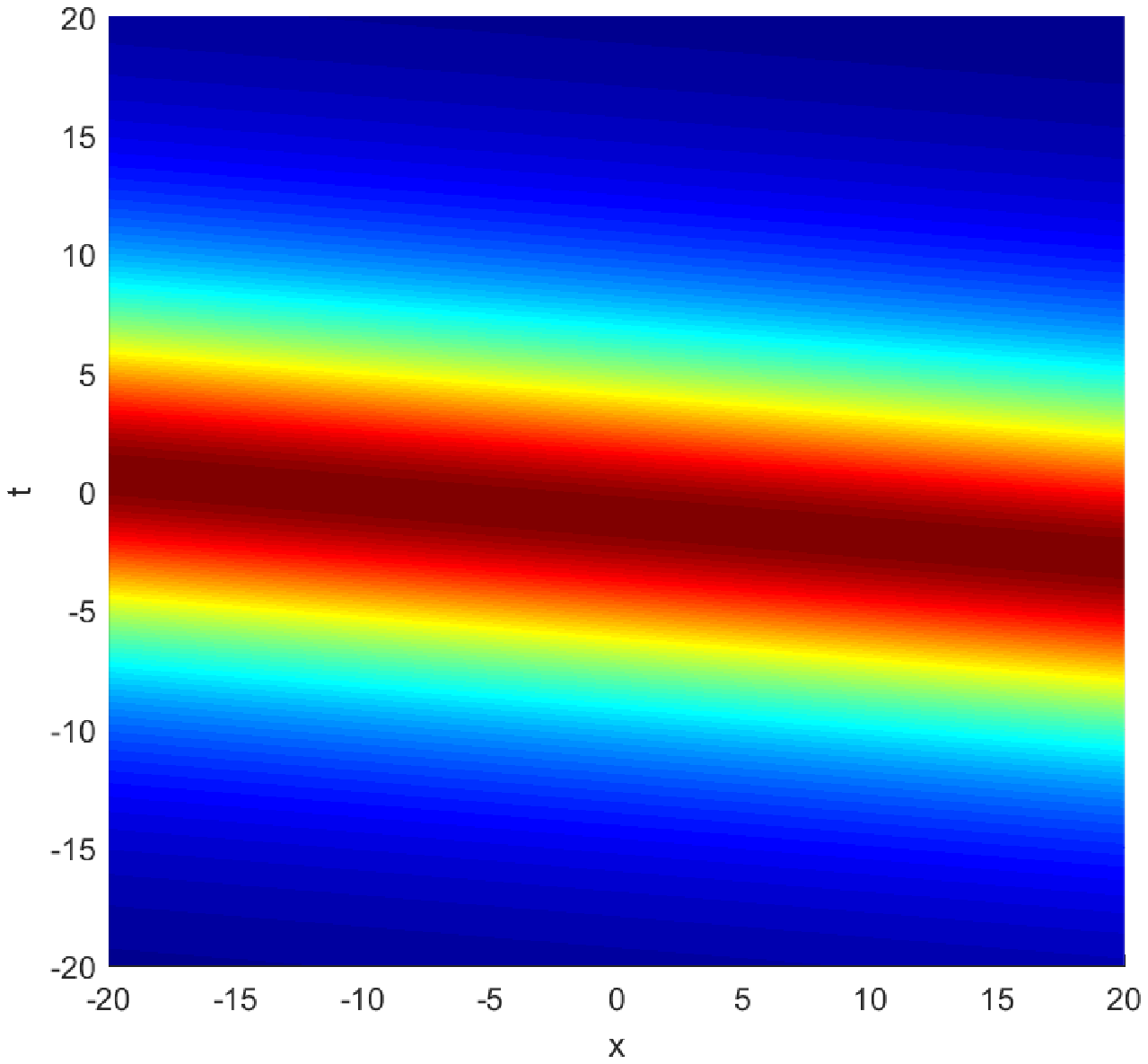}}}
~~~~
{\rotatebox{0}{\includegraphics[width=3.6cm,height=2.8cm,angle=0]{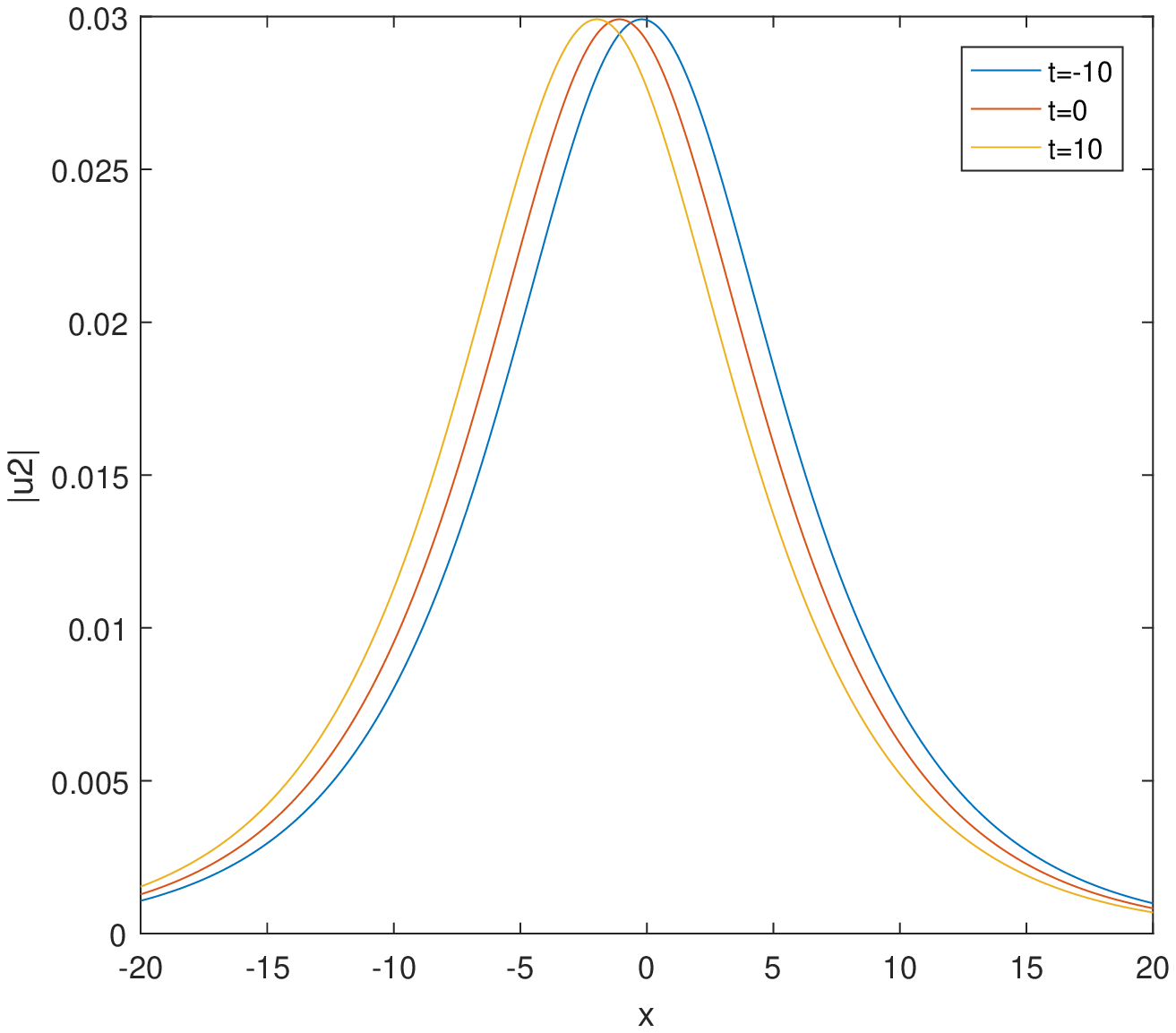}}}

$\ \qquad~~~~~~(\textbf{d})\qquad \ \qquad\qquad\qquad\qquad~(\textbf{e})
\ \qquad\qquad\qquad\qquad\qquad~(\textbf{f})$\\
\noindent { \small \textbf{Figure 2.} One-soliton  solutions to Eq. \eqref{so1} with parameters $\zeta_1=1.2+0.5i$, $\mu_{1,1}=0.97+0.5i$, $\mu_{1,2}=1.02+1.2i$,  and $\mu_{1,3}=1.21+0.5i$ .
$\textbf{(a)(b)(c)}$: the local structure, density and wave propagation of the one-soliton solution $|u_1(x,t)|$,
$\textbf{(d)(e)(f)}$: the local structure, density and wave propagation of the one-soliton  solution $|u_2(x,t)|$.} \\

The above results for the two-component FL equations is consistent with the results in \cite{guo2012riemann}.  Moreover,
the localized structures and dynamic  behavior of one-hump solutions and one-soliton solutions are shown in Fig. 1 and Fig. 2, respectively.  From the following  two figures,  we can see that  the single solitons  all travel in an exchanged direction,  but the amplitude of the single-hump solution  in Fig. 1  is larger than  the single soliton in Fig. 2,    and  the  single  soliton in Fig. 2 is  much wider than the single hump in Fig. 1.

Next, if we take $N_1=2$,    the solutions to Eq. \eqref{exact}  can be expressed  explicitly by
\begin{equation}\label{so2}
  \left\{
   \begin{aligned}
        u_1(x,t) =& i  (-\mu_{1,2} \mu_{1,1}^{*} e^{\theta_1 -2\theta_1^*} (M^{-1})_{1,1}
        -\mu_{1,2} \mu_{2,1}^{*} e^{\theta_1 -2\theta_2^*} (M^{-1})_{1,2} \\
         & - \mu_{1,2} \mu_{1,1}^{*} e^{\theta_1 -2\theta_1^*} (M^{-1})_{1,3}
         - \mu_{1,2} \mu_{2,1}^{*} e^{\theta_1 -2\theta_2^*} (M^{-1})_{1,4}           \\
         &  -\mu_{2,2} \mu_{1,1}^{*} e^{\theta_2 -2\theta_1^*} (M^{-1})_{2,1}
        -\mu_{2,2} \mu_{2,1}^{*} e^{\theta_2 -2\theta_2^*} (M^{-1})_{2,2} \\
         & - \mu_{2,2} \mu_{1,1}^{*} e^{\theta_2 -2\theta_1^*} (M^{-1})_{2,3}
         - \mu_{2,2} \mu_{2,1}^{*} e^{\theta_2 -2\theta_2^*} (M^{-1})_{2,4}      \\
         &  +\mu_{1,2} \mu_{1,1}^{*} e^{\theta_1 -2\theta_1^*} (M^{-1})_{3,1}
        +\mu_{1,2} \mu_{2,1}^{*} e^{\theta_1 -2\theta_2^*} (M^{-1})_{3,2} \\
         & +\mu_{1,2} \mu_{1,1}^{*} e^{\theta_1 -2\theta_1^*} (M^{-1})_{3,3}
         +\mu_{1,2} \mu_{2,1}^{*} e^{\theta_1 -2\theta_2^*} (M^{-1})_{3,4}      \\
                  &  +\mu_{2,2} \mu_{1,1}^{*} e^{\theta_2 -2\theta_1^*} (M^{-1})_{4,1}
        +\mu_{2,2} \mu_{2,1}^{*} e^{\theta_2 -2\theta_2^*} (M^{-1})_{4,2} \\
         & +\mu_{2,2} \mu_{1,1}^{*} e^{\theta_2 -2\theta_1^*} (M^{-1})_{4,3}
         +\mu_{2,2} \mu_{2,1}^{*} e^{\theta_2 -2\theta_2^*} (M^{-1})_{4,4} ),     \\
                 u_2(x,t) =& i (-\mu_{1,3} \mu_{1,1}^{*} e^{\theta_1 -2\theta_1^*} (M^{-1})_{1,1}
        -\mu_{1,3} \mu_{2,1}^{*} e^{\theta_1 -2\theta_2^*} (M^{-1})_{1,2} \\
         & - \mu_{1,3} \mu_{1,1}^{*} e^{\theta_1 -2\theta_1^*} (M^{-1})_{1,3}
         - \mu_{1,3} \mu_{2,1}^{*} e^{\theta_1 -2\theta_2^*} (M^{-1})_{1,4}           \\
         &  -\mu_{2,3} \mu_{1,1}^{*} e^{\theta_2 -2\theta_1^*} (M^{-1})_{2,1}
        -\mu_{2,3} \mu_{2,1}^{*} e^{\theta_2 -2\theta_2^*} (M^{-1})_{2,2} \\
         & - \mu_{2,3} \mu_{1,1}^{*} e^{\theta_2 -2\theta_1^*} (M^{-1})_{2,3}
         - \mu_{2,3} \mu_{2,1}^{*} e^{\theta_2 -2\theta_2^*} (M^{-1})_{2,4}      \\
         &  +\mu_{1,3} \mu_{1,1}^{*} e^{\theta_1 -2\theta_1^*} (M^{-1})_{3,1}
        +\mu_{1,3} \mu_{2,1}^{*} e^{\theta_1 -2\theta_2^*} (M^{-1})_{3,2} \\
         & +\mu_{1,3} \mu_{1,1}^{*} e^{\theta_1 -2\theta_1^*} (M^{-1})_{3,3}
         +\mu_{1,3} \mu_{2,1}^{*} e^{\theta_1 -2\theta_2^*} (M^{-1})_{3,4}      \\
                  &  +\mu_{2,3} \mu_{1,1}^{*} e^{\theta_2 -2\theta_1^*} (M^{-1})_{4,1}
        +\mu_{2,3} \mu_{2,1}^{*} e^{\theta_2 -2\theta_2^*} (M^{-1})_{4,2} \\
         & +\mu_{2,3} \mu_{1,1}^{*} e^{\theta_2 -2\theta_1^*} (M^{-1})_{4,3}
         +\mu_{2,3} \mu_{2,1}^{*} e^{\theta_2 -2\theta_2^*} (M^{-1})_{4,4} ),    \\
   \end{aligned}
\right.
\end{equation}
where
\begin{equation}
m_{k,j}=\frac{\hat{\vartheta}_{k}\vartheta_{j}}{\zeta_j-\hat{\zeta_{k}}}, \quad 1 \leq k,j \leq 4,
\end{equation}
with
\begin{equation}
  \zeta_3=-\zeta_1, \quad \zeta_4=-\zeta_2, \quad \hat{\zeta_j}=\zeta_j^{*}, \quad 1 \leq j \leq 4.
\end{equation}

\noindent
{\rotatebox{0}{\includegraphics[width=3.6cm,height=3.6cm,angle=0]{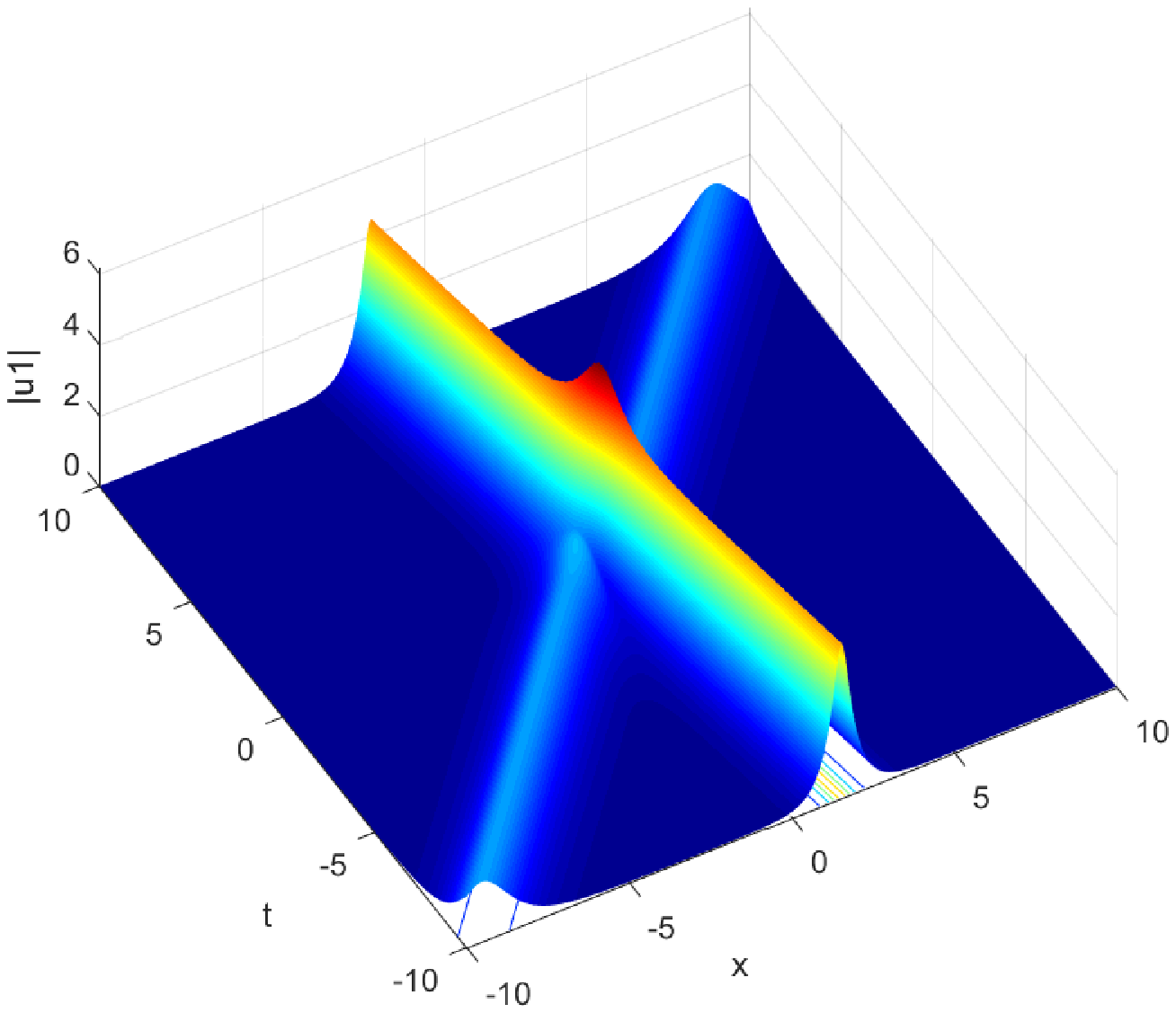}}}
~~~~
{\rotatebox{0}{\includegraphics[width=3.6cm,height=3.6cm,angle=0]{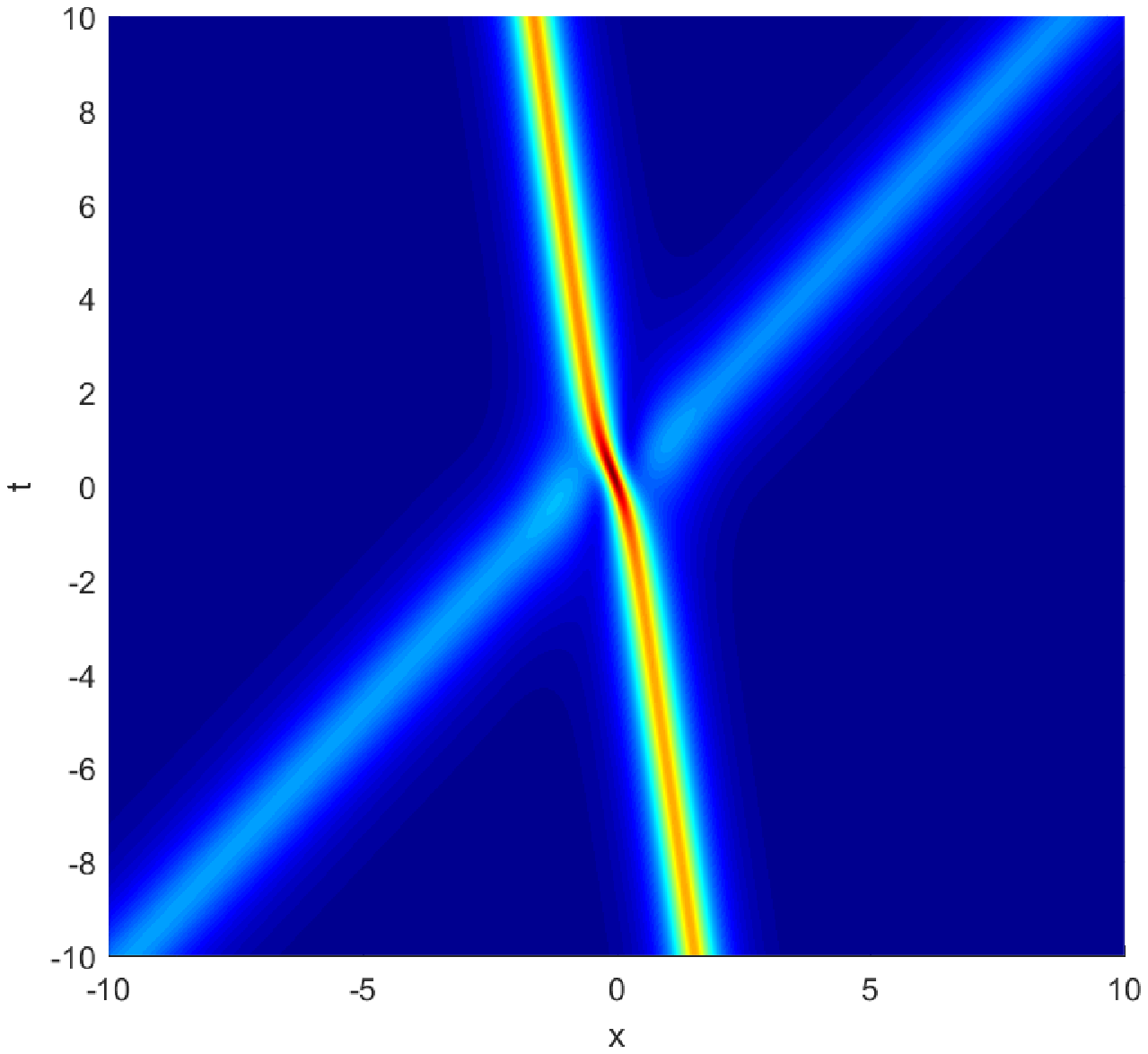}}}
~~~~
{\rotatebox{0}{\includegraphics[width=3.6cm,height=3.6cm,angle=0]{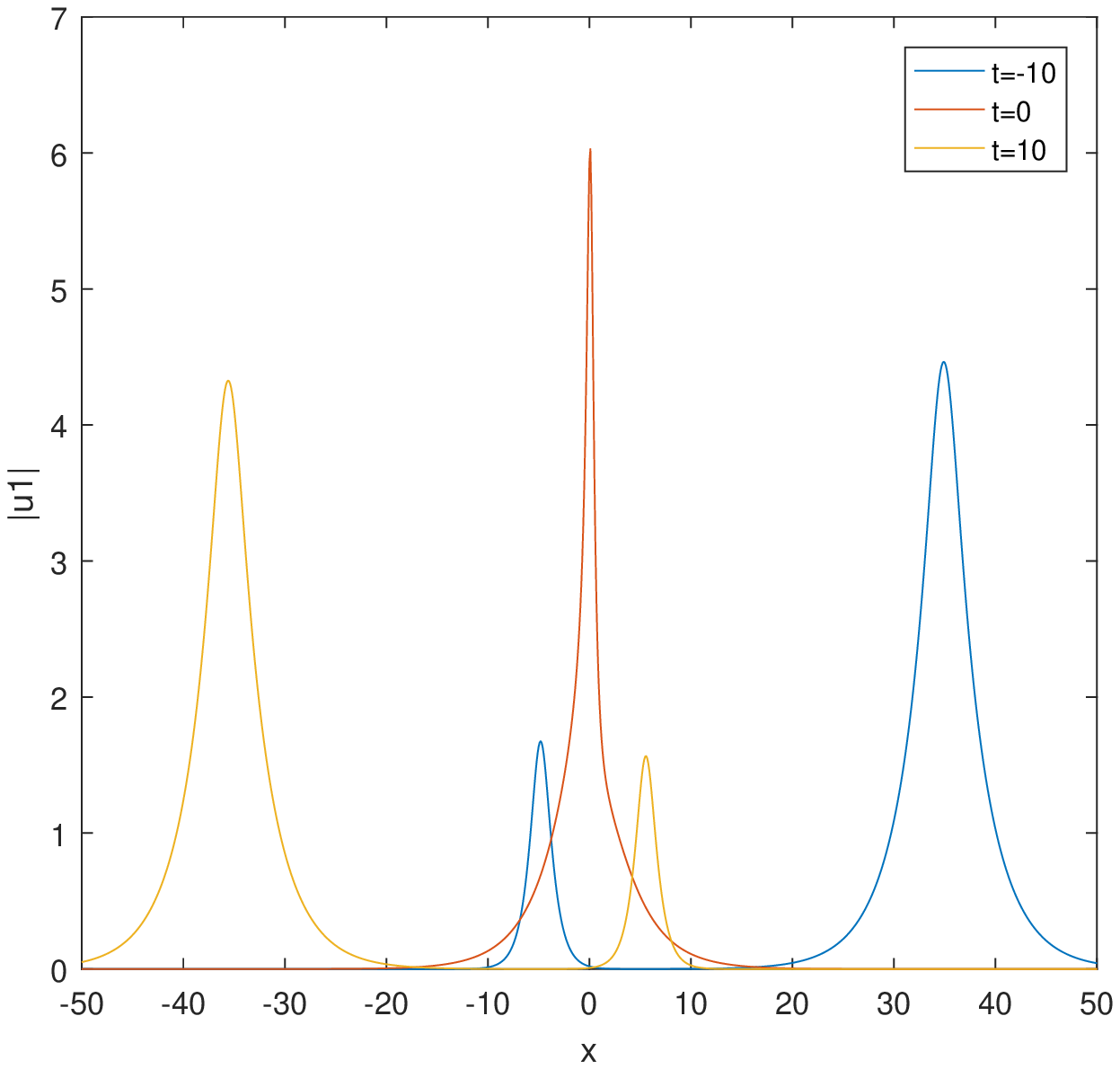}}}

$\ \qquad~~~~~~(\textbf{a})\qquad \ \qquad\qquad\qquad\qquad~(\textbf{b})
\ \qquad\qquad\qquad\qquad\qquad~(\textbf{c})$\\
\noindent
{\rotatebox{0}{\includegraphics[width=3.6cm,height=3.6cm,angle=0]{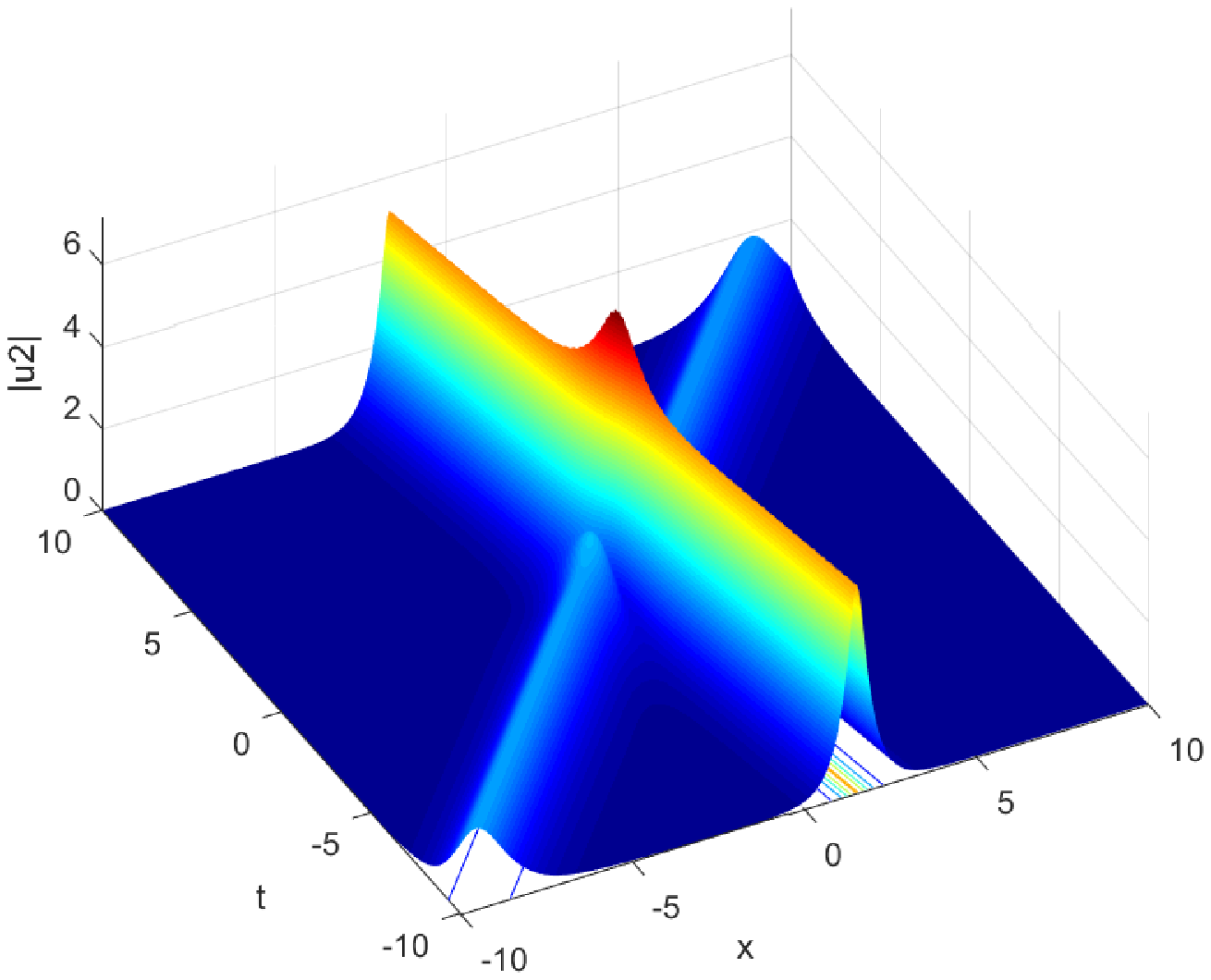}}}
~~~~
{\rotatebox{0}{\includegraphics[width=3.6cm,height=3.6cm,angle=0]{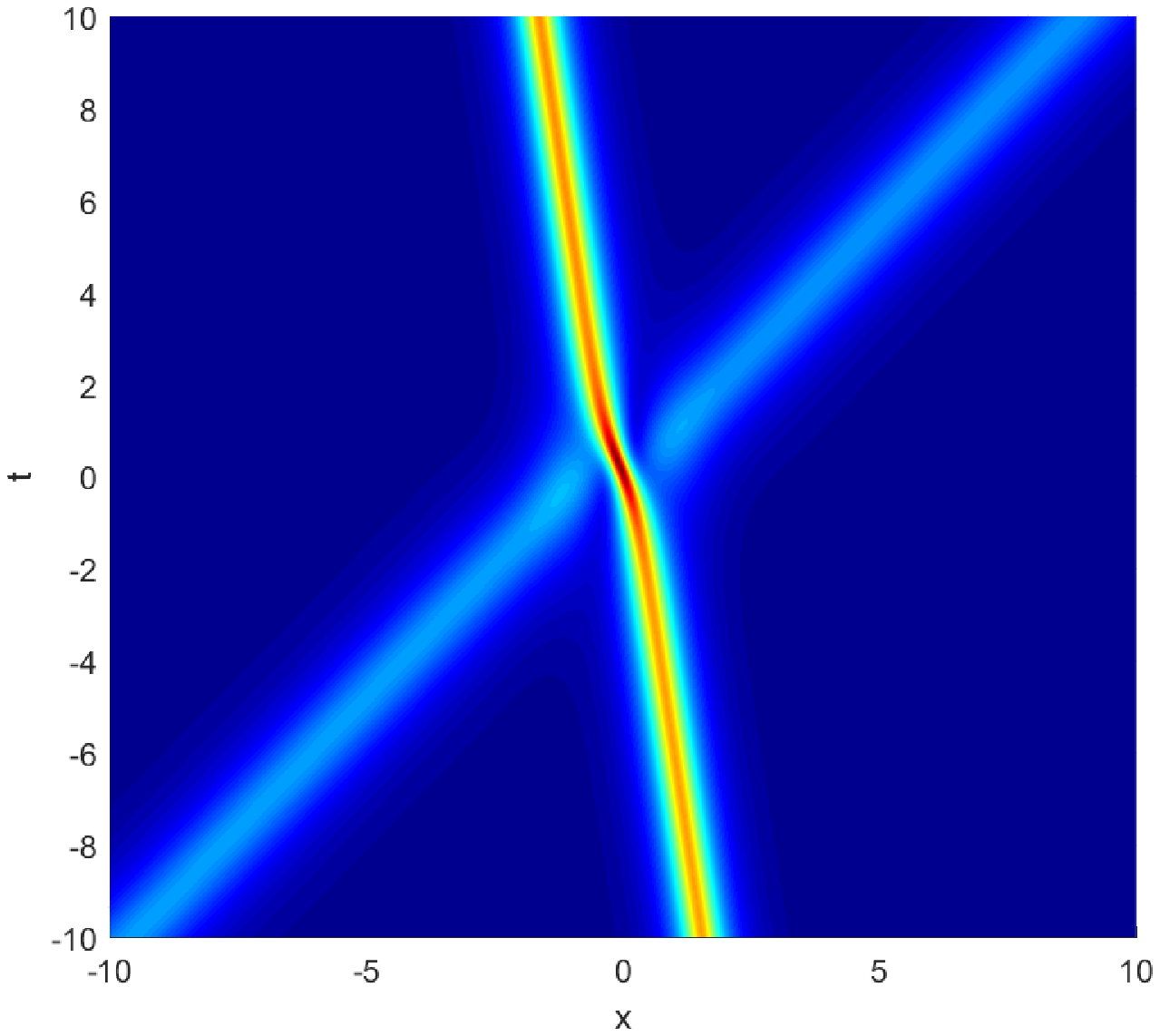}}}
~~~~
{\rotatebox{0}{\includegraphics[width=3.6cm,height=3.6cm,angle=0]{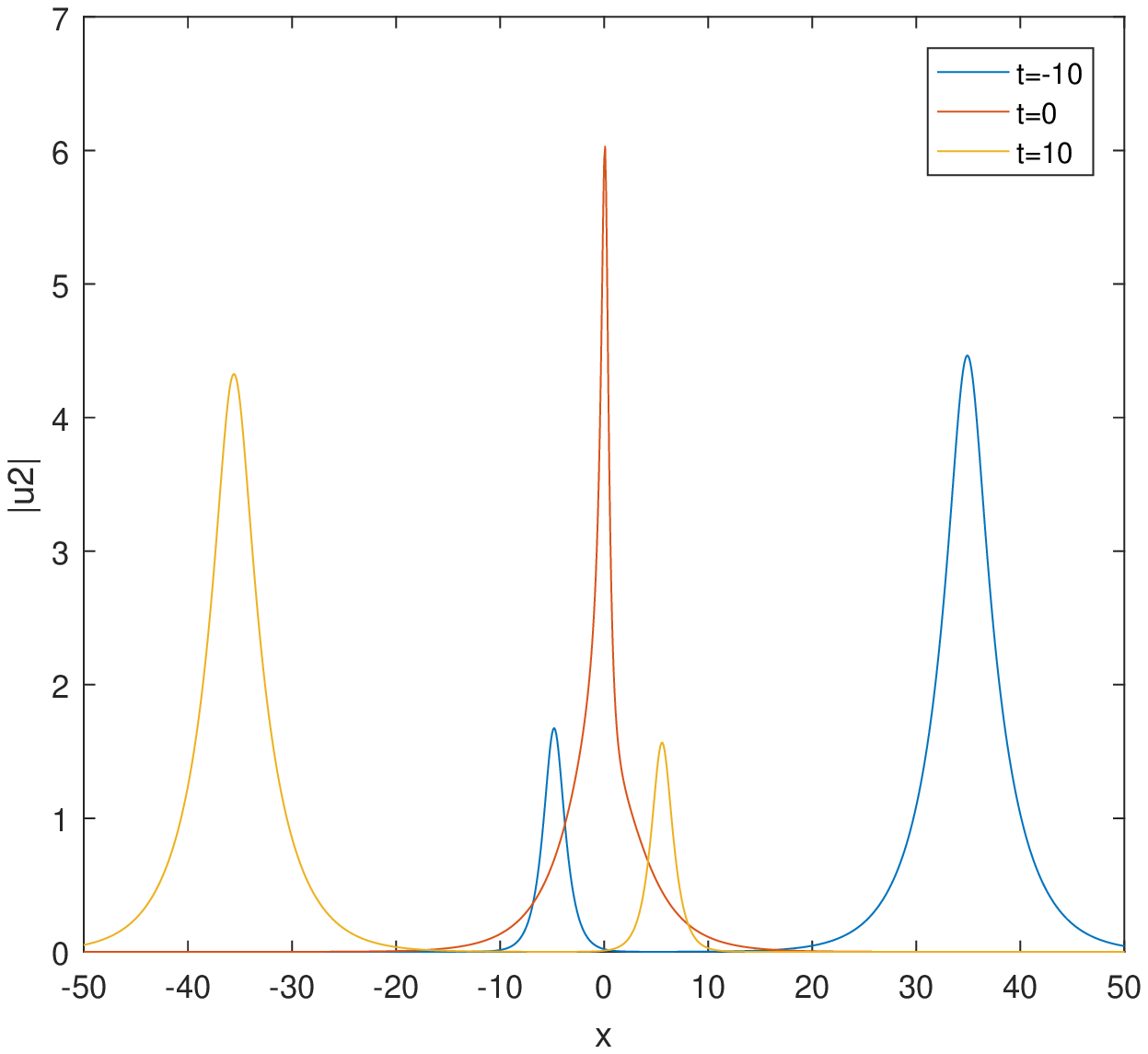}}}

$\ \qquad~~~~~~(\textbf{d})\qquad \ \qquad\qquad\qquad\qquad~(\textbf{e})
\ \qquad\qquad\qquad\qquad\qquad~(\textbf{f})$\\
\noindent { \small \textbf{Figure 3.} Two-soliton  solutions to Eq. \eqref{so2} with parameters $\zeta_1=1.2+0.5i$, $\zeta_2=0.9+1.3i$, $\mu_{1,1}=0.7$, $\mu_{1,2}=1.5$,  $\mu_{1,3}=0.9$ $\mu_{2,1}=0.8$, $\mu_{2,2}=1.2$, and $\mu_{2,3}=1.4$.
$\textbf{(a)(b)(c)}$: the local structure, density and wave propagation of the two-soliton  solution $|u_1(x,t)|$,
$\textbf{(d)(e)(f)}$: the local structure, density and wave propagation of the two-soliton  solution $|u_2(x,t)|$.} \\

The localized structures and dynamic propagation behavior of the two soliton  solutions are displayed in Fig. 3.   From Fig. 3,  it can be seen that before two solitons  colliside each other, they   spread forward in directions that cross each other. After they colliside each other,  the directions of two solitons are not exchanged, but the positions of them has been shifted and the enenry  of them has been swapped. \\

When  $N_1=3$,    we can rewrite  Eq. \eqref{exact} as the following form
\begin{equation}\label{three1}
\left\{
   \begin{aligned}
     u_1(x,t) = & i (  -\mu_{1,2} \mu_{1,1}^{*} e^{\theta_1 - 2 \theta_1^{*}} (M^{-1})_{1,1}
                    -\mu_{1,2} \mu_{2,1}^{*} e^{\theta_1 - 2 \theta_2^{*}} (M^{-1})_{1,2}
                    -\mu_{1,2} \mu_{3,1}^{*} e^{\theta_1 - 2 \theta_3^{*}} (M^{-1})_{1,3}  \\
                    & -\mu_{1,2} \mu_{1,1}^{*} e^{\theta_1 - 2 \theta_1^{*}} (M^{-1})_{1,4}
                    -\mu_{1,2} \mu_{2,1}^{*} e^{\theta_1 - 2 \theta_2^{*}} (M^{-1})_{1,5}
                     -\mu_{1,2} \mu_{3,1}^{*} e^{\theta_1 - 2 \theta_3^{*}} (M^{-1})_{1,6}    \\
                   &   -\mu_{2,2} \mu_{1,1}^{*} e^{\theta_2 - 2 \theta_1^{*}} (M^{-1})_{2,1}
                    -\mu_{2,2} \mu_{2,1}^{*} e^{\theta_2 - 2 \theta_2^{*}} (M^{-1})_{2,2}
                    -\mu_{2,2} \mu_{3,1}^{*} e^{\theta_2 - 2 \theta_3^{*}} (M^{-1})_{2,3}  \\
                  &  -\mu_{2,2} \mu_{1,1}^{*} e^{\theta_2 - 2 \theta_1^{*}} (M^{-1})_{2,4}
                    -\mu_{2,2} \mu_{2,1}^{*} e^{\theta_2 - 2 \theta_2^{*}} (M^{-1})_{2,5}
                     -\mu_{2,2} \mu_{3,1}^{*} e^{\theta_2 - 2 \theta_3^{*}} (M^{-1})_{2,6}    \\
                   &      -\mu_{3,2} \mu_{1,1}^{*} e^{\theta_3 - 2 \theta_1^{*}} (M^{-1})_{3,1}
                    -\mu_{3,2} \mu_{2,1}^{*} e^{\theta_3 - 2 \theta_2^{*}} (M^{-1})_{3,2}
                    -\mu_{3,2} \mu_{3,1}^{*} e^{\theta_3 - 2 \theta_3^{*}} (M^{-1})_{3,3}  \\
                   & -\mu_{3,2} \mu_{1,1}^{*} e^{\theta_3 - 2 \theta_1^{*}} (M^{-1})_{3,4}
                    -\mu_{3,2} \mu_{2,1}^{*} e^{\theta_3 - 2 \theta_2^{*}} (M^{-1})_{3,5}
                     -\mu_{3,2} \mu_{3,1}^{*} e^{\theta_3 - 2 \theta_3^{*}} (M^{-1})_{3,6}    \\
                 &    +\mu_{1,2} \mu_{1,1}^{*} e^{\theta_1 - 2 \theta_1^{*}} (M^{-1})_{4,1}
                    +\mu_{1,2} \mu_{2,1}^{*} e^{\theta_1 - 2 \theta_2^{*}} (M^{-1})_{4,2}
                    +\mu_{1,2} \mu_{3,1}^{*} e^{\theta_1 - 2 \theta_3^{*}} (M^{-1})_{4,3}\\
                  &  +\mu_{1,2} \mu_{1,1}^{*} e^{\theta_1 - 2 \theta_1^{*}} (M^{-1})_{4,4}
                    +\mu_{1,2} \mu_{2,1}^{*} e^{\theta_1 - 2 \theta_2^{*}} (M^{-1})_{4,5}
                     +\mu_{1,2} \mu_{3,1}^{*} e^{\theta_1 - 2 \theta_3^{*}} (M^{-1})_{4,6}   \\
                  &                        +\mu_{2,2} \mu_{1,1}^{*} e^{\theta_2 - 2 \theta_1^{*}} (M^{-1})_{5,1}
                    +\mu_{2,2} \mu_{2,1}^{*} e^{\theta_2 - 2 \theta_2^{*}} (M^{-1})_{5,2}
                    +\mu_{2,2} \mu_{3,1}^{*} e^{\theta_2 - 2 \theta_3^{*}} (M^{-1})_{5,3}  \\
                  &  +\mu_{2,2} \mu_{1,1}^{*} e^{\theta_2 - 2 \theta_1^{*}} (M^{-1})_{5,4}
                    +\mu_{2,2} \mu_{2,1}^{*} e^{\theta_2 - 2 \theta_2^{*}} (M^{-1})_{5,5}
                     +\mu_{2,2} \mu_{3,1}^{*} e^{\theta_2 - 2 \theta_3^{*}} (M^{-1})_{5,6}    \\
                   &                       +\mu_{3,2} \mu_{1,1}^{*} e^{\theta_3 - 2 \theta_1^{*}} (M^{-1})_{6,1}
                    +\mu_{3,2} \mu_{2,1}^{*} e^{\theta_3 - 2 \theta_2^{*}} (M^{-1})_{6,2}
                    +\mu_{3,2} \mu_{3,1}^{*} e^{\theta_3 - 2 \theta_3^{*}} (M^{-1})_{6,3}\\
                 &   +\mu_{3,2} \mu_{1,1}^{*} e^{\theta_3 - 2 \theta_1^{*}} (M^{-1})_{6,4}
                    +\mu_{3,2} \mu_{2,1}^{*} e^{\theta_3 - 2 \theta_2^{*}} (M^{-1})_{6,5}
                     +\mu_{3,2} \mu_{3,1}^{*} e^{\theta_3 - 2 \theta_3^{*}} (M^{-1})_{6,6}     ), \\
          u_2(x,t) = & i (  -\mu_{1,3} \mu_{1,1}^{*} e^{\theta_1 - 2 \theta_1^{*}} (M^{-1})_{1,1}
                    -\mu_{1,3} \mu_{2,1}^{*} e^{\theta_1 - 2 \theta_2^{*}} (M^{-1})_{1,2}
                    -\mu_{1,3} \mu_{3,1}^{*} e^{\theta_1 - 2 \theta_3^{*}} (M^{-1})_{1,3}  \\
                    & -\mu_{1,3} \mu_{1,1}^{*} e^{\theta_1 - 2 \theta_1^{*}} (M^{-1})_{1,4}
                    -\mu_{1,3} \mu_{2,1}^{*} e^{\theta_1 - 2 \theta_2^{*}} (M^{-1})_{1,5}
                     -\mu_{1,3} \mu_{3,1}^{*} e^{\theta_1 - 2 \theta_3^{*}} (M^{-1})_{1,6}    \\
                   &   -\mu_{2,3} \mu_{1,1}^{*} e^{\theta_2 - 2 \theta_1^{*}} (M^{-1})_{2,1}
                    -\mu_{2,3} \mu_{2,1}^{*} e^{\theta_2 - 2 \theta_2^{*}} (M^{-1})_{2,2}
                    -\mu_{2,3} \mu_{3,1}^{*} e^{\theta_2 - 2 \theta_3^{*}} (M^{-1})_{2,3}  \\
                  &  -\mu_{2,3} \mu_{1,1}^{*} e^{\theta_2 - 2 \theta_1^{*}} (M^{-1})_{2,4}
                    -\mu_{2,3} \mu_{2,1}^{*} e^{\theta_2 - 2 \theta_2^{*}} (M^{-1})_{2,5}
                     -\mu_{2,3} \mu_{3,1}^{*} e^{\theta_2 - 2 \theta_3^{*}} (M^{-1})_{2,6}    \\
                   &      -\mu_{3,3} \mu_{1,1}^{*} e^{\theta_3 - 2 \theta_1^{*}} (M^{-1})_{3,1}
                    -\mu_{3,3} \mu_{2,1}^{*} e^{\theta_3 - 2 \theta_2^{*}} (M^{-1})_{3,2}
                    -\mu_{3,3} \mu_{3,1}^{*} e^{\theta_3 - 2 \theta_3^{*}} (M^{-1})_{3,3}  \\
                   & -\mu_{3,3} \mu_{1,1}^{*} e^{\theta_3 - 2 \theta_1^{*}} (M^{-1})_{3,4}
                    -\mu_{3,3} \mu_{2,1}^{*} e^{\theta_3 - 2 \theta_2^{*}} (M^{-1})_{3,5}
                     -\mu_{3,3} \mu_{3,1}^{*} e^{\theta_3 - 2 \theta_3^{*}} (M^{-1})_{3,6}    \\
                 &    +\mu_{1,3} \mu_{1,1}^{*} e^{\theta_1 - 2 \theta_1^{*}} (M^{-1})_{4,1}
                    +\mu_{1,3} \mu_{2,1}^{*} e^{\theta_1 - 2 \theta_2^{*}} (M^{-1})_{4,2}
                    +\mu_{1,3} \mu_{3,1}^{*} e^{\theta_1 - 2 \theta_3^{*}} (M^{-1})_{4,3}\\
                  &  +\mu_{1,3} \mu_{1,1}^{*} e^{\theta_1 - 2 \theta_1^{*}} (M^{-1})_{4,4}
                    +\mu_{1,3} \mu_{2,1}^{*} e^{\theta_1 - 2 \theta_2^{*}} (M^{-1})_{4,5}
                     +\mu_{1,3} \mu_{3,1}^{*} e^{\theta_1 - 2 \theta_3^{*}} (M^{-1})_{4,6}   \\
                  &                        +\mu_{2,3} \mu_{1,1}^{*} e^{\theta_2 - 2 \theta_1^{*}} (M^{-1})_{5,1}
                    +\mu_{2,3} \mu_{2,1}^{*} e^{\theta_2 - 2 \theta_2^{*}} (M^{-1})_{5,2}
                    +\mu_{2,3} \mu_{3,1}^{*} e^{\theta_2 - 2 \theta_3^{*}} (M^{-1})_{5,3}  \\
                  &  +\mu_{2,3} \mu_{1,1}^{*} e^{\theta_2 - 2 \theta_1^{*}} (M^{-1})_{5,4}
                    +\mu_{2,3} \mu_{2,1}^{*} e^{\theta_2 - 2 \theta_2^{*}} (M^{-1})_{5,5}
                     +\mu_{2,3} \mu_{3,1}^{*} e^{\theta_2 - 2 \theta_3^{*}} (M^{-1})_{5,6}    \\
                   &                       +\mu_{3,3} \mu_{1,1}^{*} e^{\theta_3 - 2 \theta_1^{*}} (M^{-1})_{6,1}
                    +\mu_{3,3} \mu_{2,1}^{*} e^{\theta_3 - 2 \theta_2^{*}} (M^{-1})_{6,2}
                    +\mu_{3,3} \mu_{3,1}^{*} e^{\theta_3 - 2 \theta_3^{*}} (M^{-1})_{6,3}\\
                 &   +\mu_{3,3} \mu_{1,1}^{*} e^{\theta_3 - 2 \theta_1^{*}} (M^{-1})_{6,4}
                    +\mu_{3,3} \mu_{2,1}^{*} e^{\theta_3 - 2 \theta_2^{*}} (M^{-1})_{6,5}
                     +\mu_{3,3} \mu_{3,1}^{*} e^{\theta_3 - 2 \theta_3^{*}} (M^{-1})_{6,6}     ),
   \end{aligned}
\right.
\end{equation}
where
\begin{equation}
m_{k,j}=\frac{\hat{\vartheta}_{k}\vartheta_{j}}{\zeta_j-\hat{\zeta_{k}}}, \quad 1 \leq k,j \leq 6,
\end{equation}
with
\begin{equation}
  \zeta_4=-\zeta_1, \quad \zeta_5=-\zeta_2,\quad \zeta_6=-\zeta_3, \quad \hat{\zeta_j}=\zeta_j^{*}, \quad 1 \leq j \leq 6.
\end{equation}

\noindent
{\rotatebox{0}{\includegraphics[width=3.6cm,height=3.5cm,angle=0]{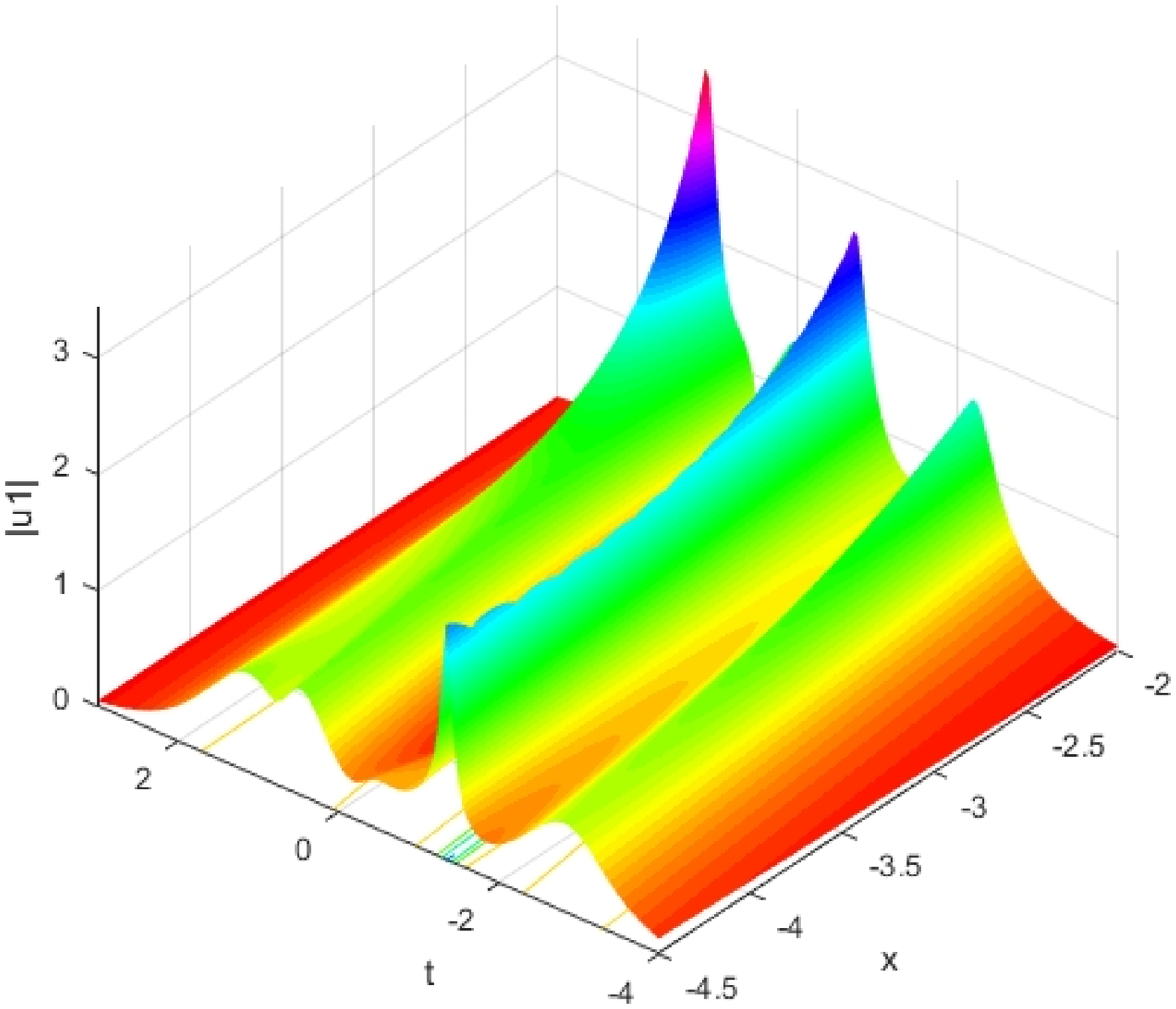}}}
~~~~
{\rotatebox{0}{\includegraphics[width=3.6cm,height=3.5cm,angle=0]{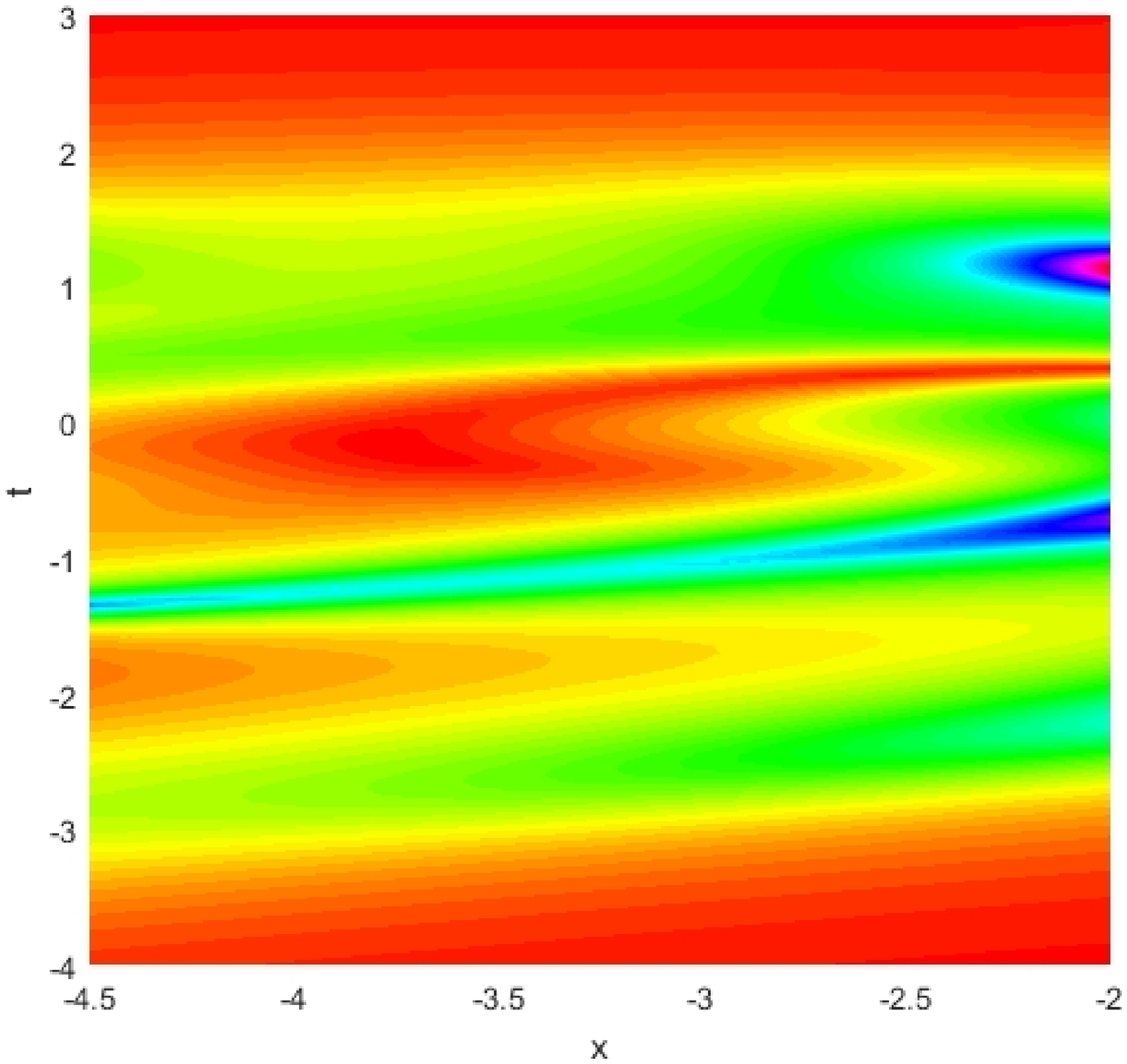}}}
~~~~
{\rotatebox{0}{\includegraphics[width=3.6cm,height=3.5cm,angle=0]{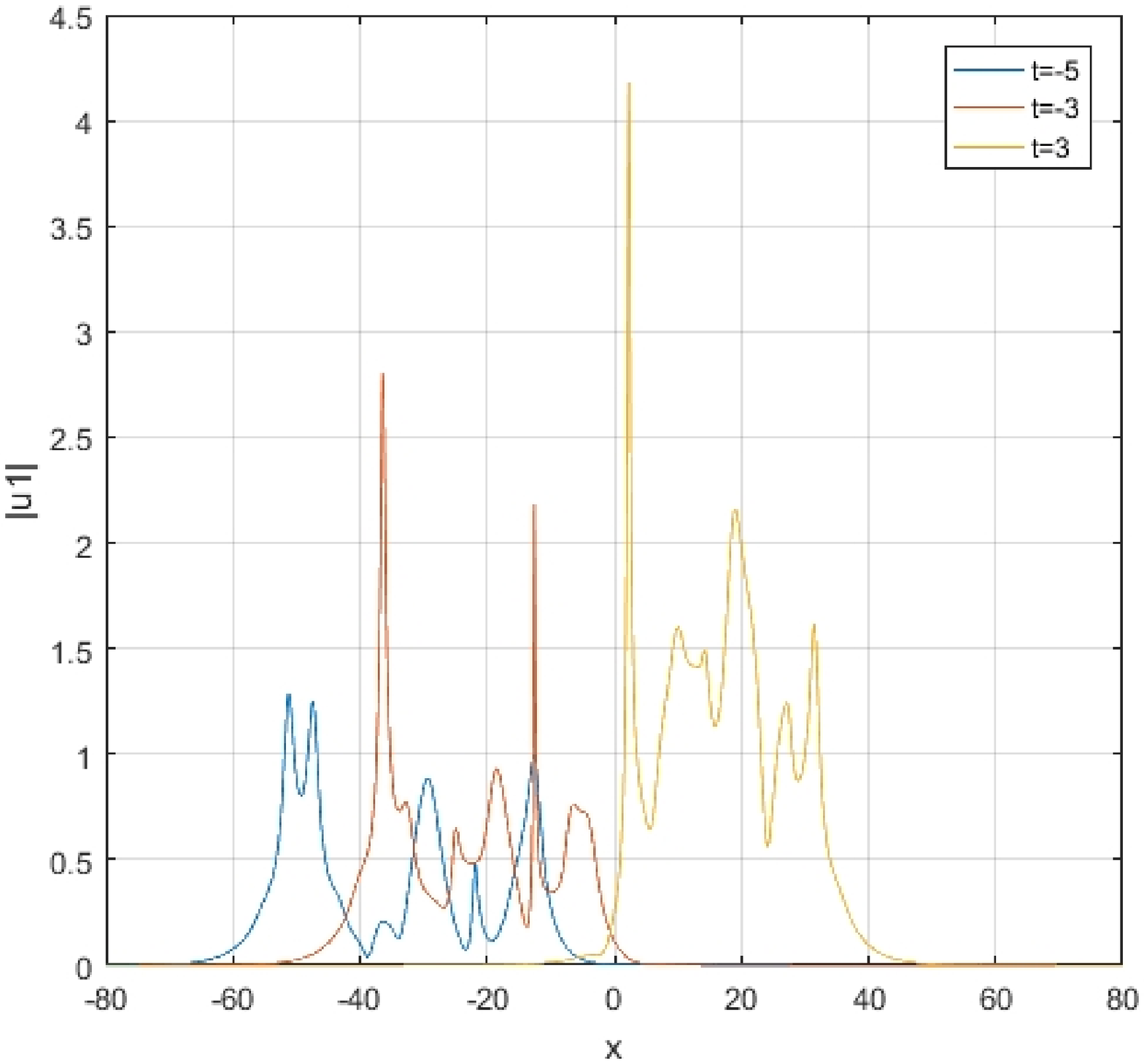}}}

$\ \qquad~~~~~~(\textbf{a})\qquad \ \qquad\qquad\qquad\qquad~(\textbf{b})
\ \qquad\qquad\qquad\qquad\qquad~(\textbf{c})$\\
\noindent { \small \textbf{Figure 4.} Three-soliton  solution to Eq. \eqref{three1} with parameters $\zeta_1=1.2+0.8i$,  $\zeta_2=1.5+0.7i$, $\zeta_3=1.1+0.9i$,  $\mu_{1,1}=0.1$, $\mu_{1,2}=\mu_{2,1}=0.2$,  $\mu_{1,3}=\mu_{3,1}=0.3$,  $\mu_{2,2}=0.4$,  $\mu_{2,3}=\mu_{3,2}=0.6$ and $\mu_{3,3}=0.9$.
$\textbf{(a)}$: the local structures of the three soliton solutions $u_1(x,t)$,
$\textbf{(b)}$: the density plot of $u_1(x,t)$,
$\textbf{(c)}$: the wave propagation of the  three soliton solutions $u_1(x,t)$.}  \\

\subsection{Case 2: multi-soliton  solutions of  three-component FL equations}

If we take $N=3$ and $v=A^{T}u^{*}$, where
   $ A=
       \begin{pmatrix}
      a_{11}  & a_{21}^{*}  & a_{31}^{*}    \\
      a_{21} & a_{22}  & a_{32}^{*}      \\
      a_{31}  & a_{32}  & a_{33}    \\
    \end{pmatrix}
   =\begin{pmatrix}
      1  & 2  & 3 \\
      2 & 1  & 0   \\
      3   & 0  & 1   \\
    \end{pmatrix}$,
when $N_1=1$,  we can express the solution to Eq. \eqref{exact}  explicitly
\begin{equation}\label{so44}
  \left\{
   \begin{aligned}
        u_1(x,t) =& i (-\mu_{1,2} \mu_{1,1}^{*} e^{\theta_1 -2\theta_1^*} (M^{-1})_{1,1} -\mu_{1,2} \mu_{1,1}^{*} e^{\theta_1 -2\theta_1^*} (M^{-1})_{1,2} \\
         & + \mu_{1,2} \mu_{1,1}^{*} e^{\theta_1 -2\theta_1^*} (M^{-1})_{2,1} + \mu_{1,2} \mu_{1,1}^{*} e^{\theta_1 -2\theta_1^*} (M^{-1})_{2,2}            ),  \\
                 u_2(x,t) =& i (-\mu_{1,3} \mu_{1,1}^{*} e^{\theta_1 -2\theta_1^*} (M^{-1})_{1,1} -\mu_{1,3} \mu_{1,1}^{*} e^{\theta_1 -2\theta_1^*} (M^{-1})_{1,2} \\
         & + \mu_{1,3} \mu_{1,1}^{*} e^{\theta_1 -2\theta_1^*} (M^{-1})_{2,1} + \mu_{1,3} \mu_{1,1}^{*} e^{\theta_1 -2\theta_1^*} (M^{-1})_{2,2}            ),  \\
                          u_3(x,t) =& i (-\mu_{1,4} \mu_{1,1}^{*} e^{\theta_1 -2\theta_1^*} (M^{-1})_{1,1} -\mu_{1,4} \mu_{1,1}^{*} e^{\theta_1 -2\theta_1^*} (M^{-1})_{1,2} \\
         & + \mu_{1,4} \mu_{1,1}^{*} e^{\theta_1 -2\theta_1^*} (M^{-1})_{2,1} + \mu_{1,4} \mu_{1,1}^{*} e^{\theta_1 -2\theta_1^*} (M^{-1})_{2,2}            ),
   \end{aligned}
\right.
\end{equation}
where
\begin{equation}
 \left\{
      \begin{aligned}
           m_{11}= \frac{-|\mu_{11}|^2 e^{-2\theta_1-2 \theta_1^*} + e^{ \theta_1 + \theta_1^*}(|\mu_{12}|^2+2\mu_{12}\mu_{13}^* +3 \mu_{12}\mu_{14}^*+2\mu_{12}^*\mu_{13}+|\mu_{13}|^2 +3 \mu_{14}\mu_{12}^*+|\mu_{14}|^2) }{\zeta_1-\zeta_1^{*}},  \\
           m_{12}= \frac{-|\mu_{11}|^2 e^{-2\theta_1-2 \theta_1^*} - e^{ \theta_1 + \theta_1^*}(|\mu_{12}|^2+2\mu_{12}\mu_{13}^* +3 \mu_{12}\mu_{14}^*+2\mu_{12}^*\mu_{13}+|\mu_{13}|^2 +3 \mu_{14}\mu_{12}^*+|\mu_{14}|^2) }{\zeta_2-\zeta_1^{*}}, \\
             m_{21}= \frac{-|\mu_{11}|^2 e^{-2\theta_1-2 \theta_1^*} - e^{ \theta_1 + \theta_1^*}(|\mu_{12}|^2+2\mu_{12}\mu_{13}^* +3 \mu_{12}\mu_{14}^*+2\mu_{12}^*\mu_{13}+|\mu_{13}|^2 +3 \mu_{14}\mu_{12}^*+|\mu_{14}|^2) }{\zeta_1 + \zeta_1^{*}}, \\
              m_{22}= \frac{-|\mu_{11}|^2 e^{-2\theta_1-2 \theta_1^*} + e^{ \theta_1 + \theta_1^*}(|\mu_{12}|^2+2\mu_{12}\mu_{13}^* +3 \mu_{12}\mu_{14}^*+2\mu_{12}^*\mu_{13}+|\mu_{13}|^2 +3 \mu_{14}\mu_{12}^*+|\mu_{14}|^2) }{\zeta_2 + \zeta_1^{*}}. \\
      \end{aligned}
 \right.
\end{equation}

\noindent
{\rotatebox{0}{\includegraphics[width=3.6cm,height=3.0cm,angle=0]{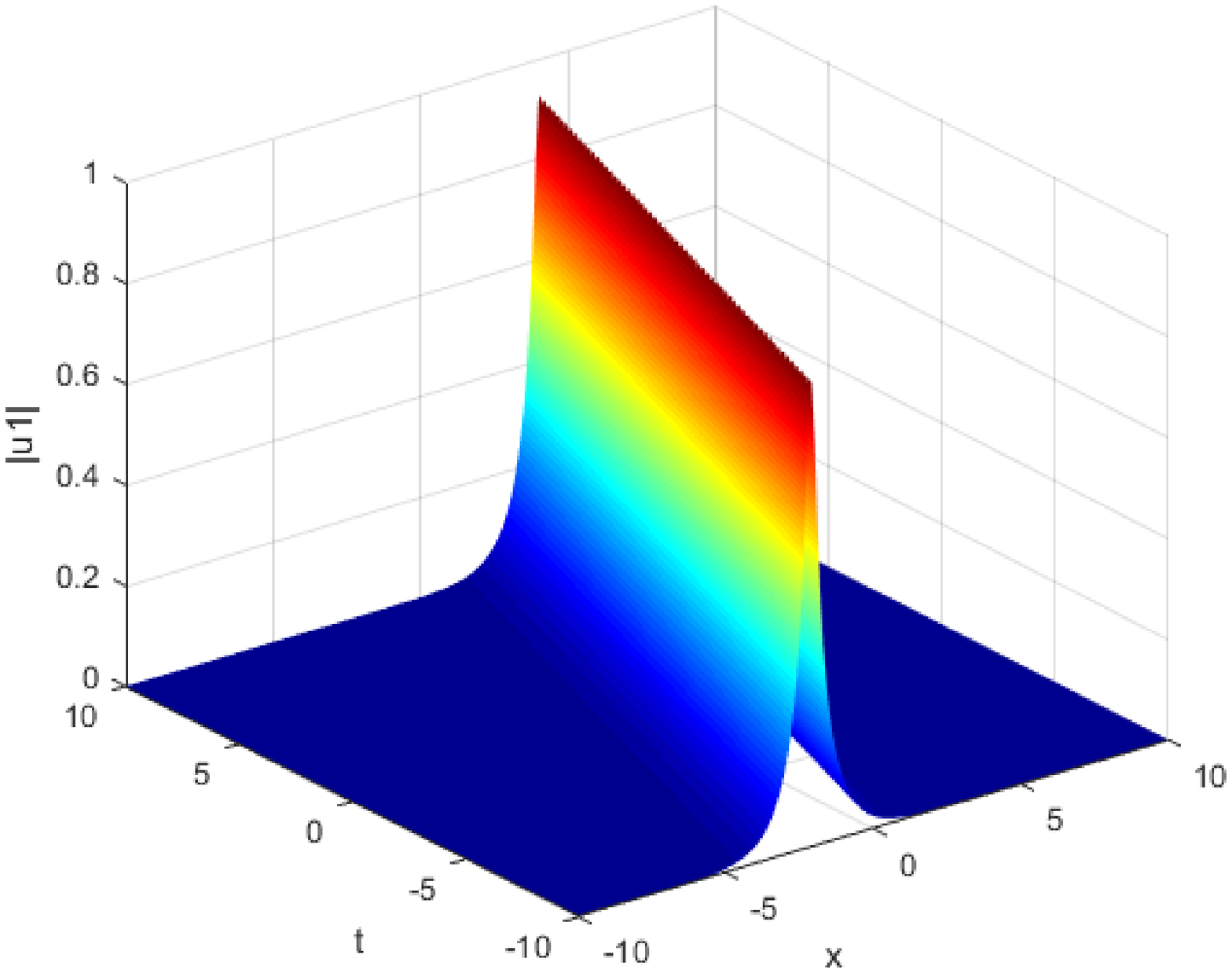}}}
~~~~
{\rotatebox{0}{\includegraphics[width=3.6cm,height=3.0cm,angle=0]{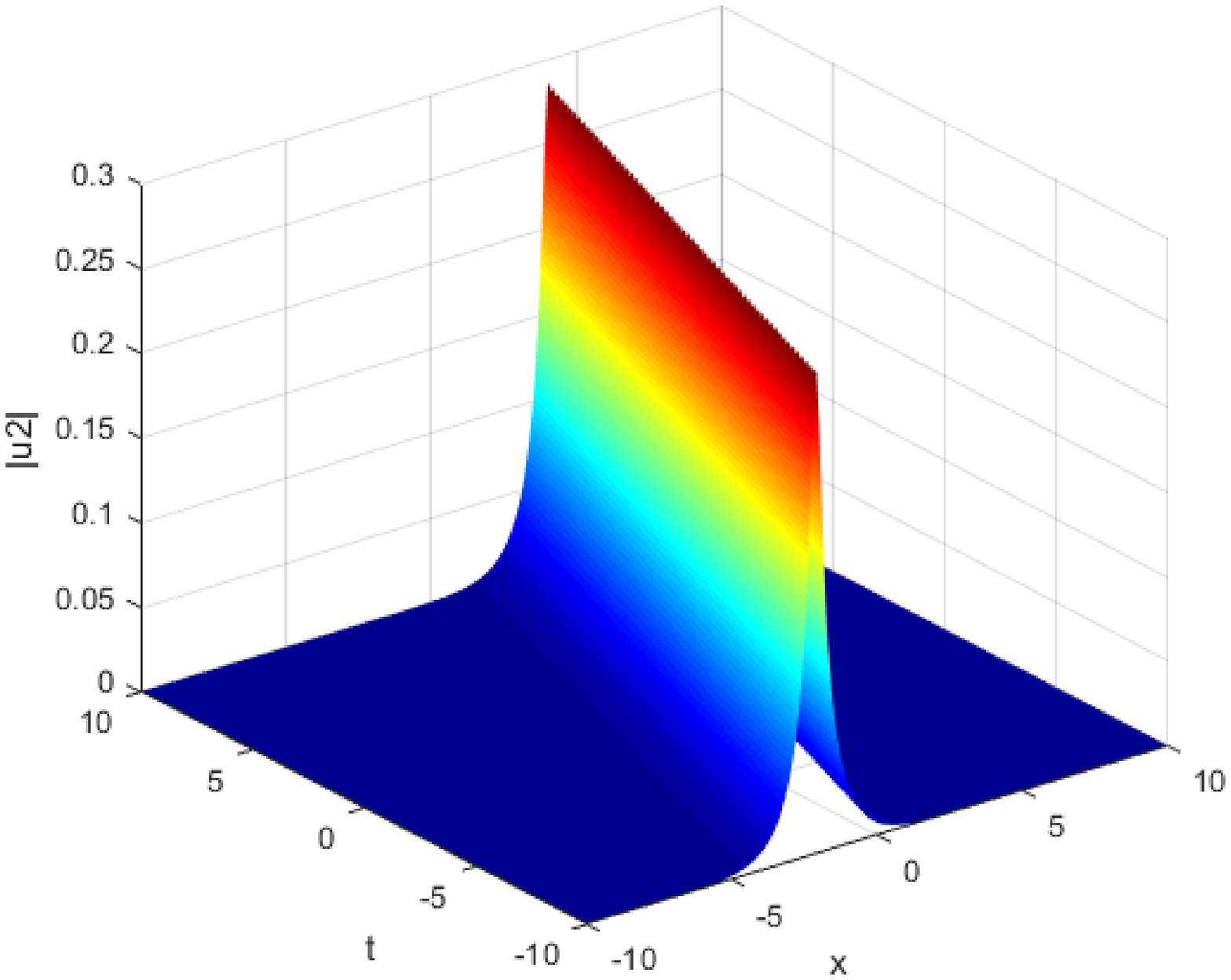}}}
~~~~
{\rotatebox{0}{\includegraphics[width=3.6cm,height=3.0cm,angle=0]{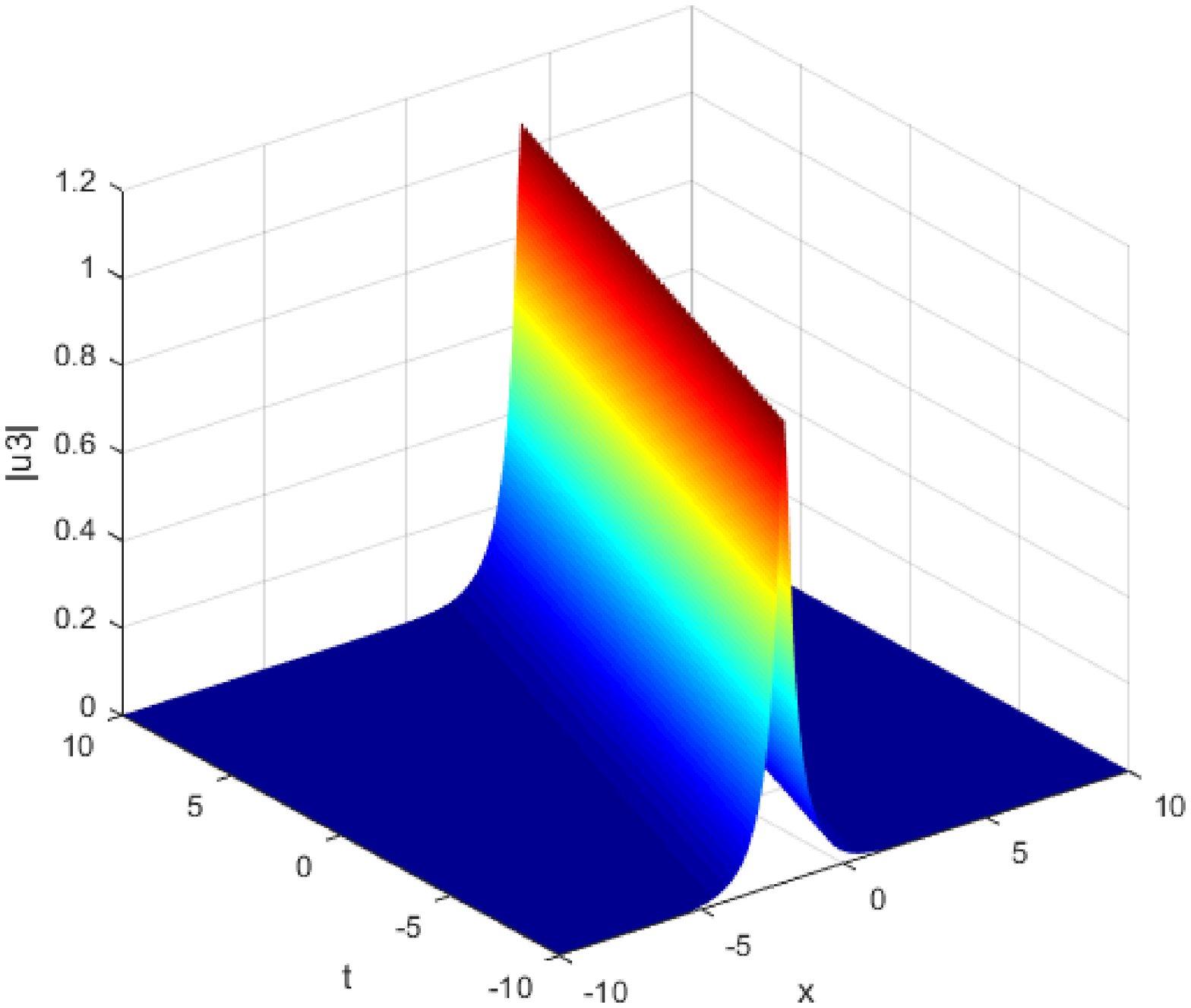}}}

$\ \qquad~~~~~~(\textbf{a})\qquad \ \qquad\qquad\qquad\qquad~(\textbf{b})
\ \qquad\qquad\qquad\qquad\qquad~(\textbf{c})$\\
\noindent
{\rotatebox{0}{\includegraphics[width=3.6cm,height=3.0cm,angle=0]{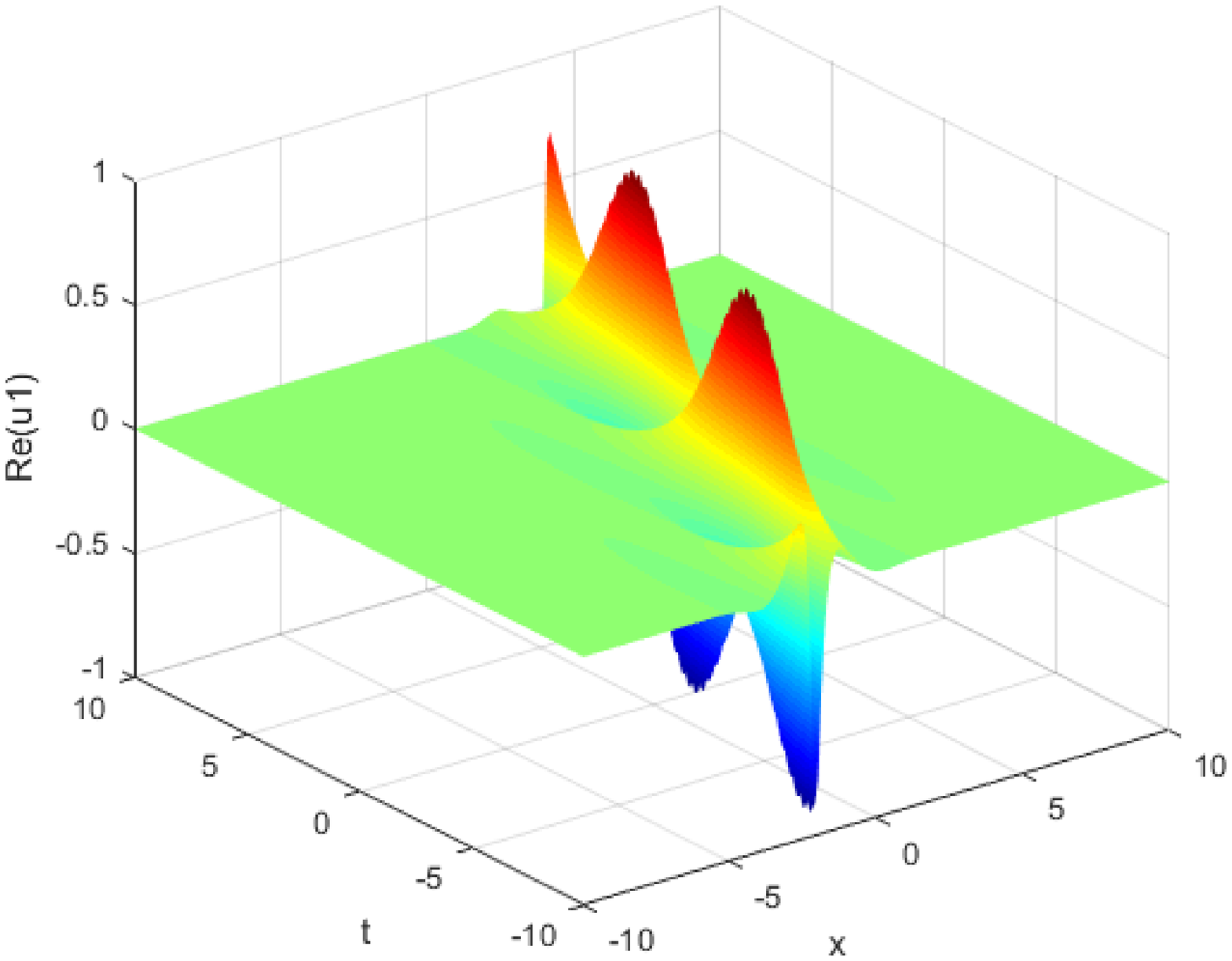}}}
~~~~
{\rotatebox{0}{\includegraphics[width=3.6cm,height=3.0cm,angle=0]{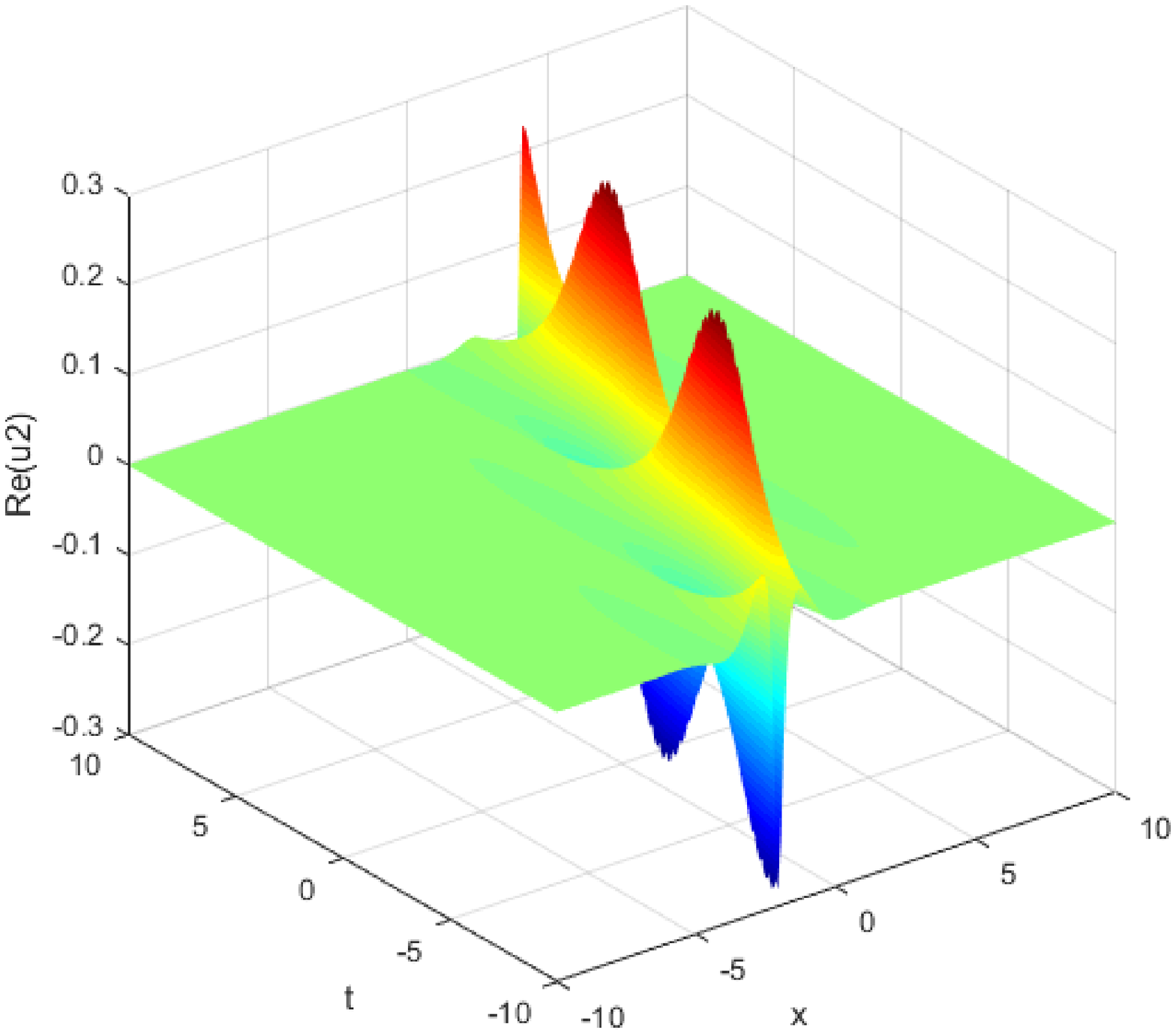}}}
~~~~
{\rotatebox{0}{\includegraphics[width=3.6cm,height=3.0cm,angle=0]{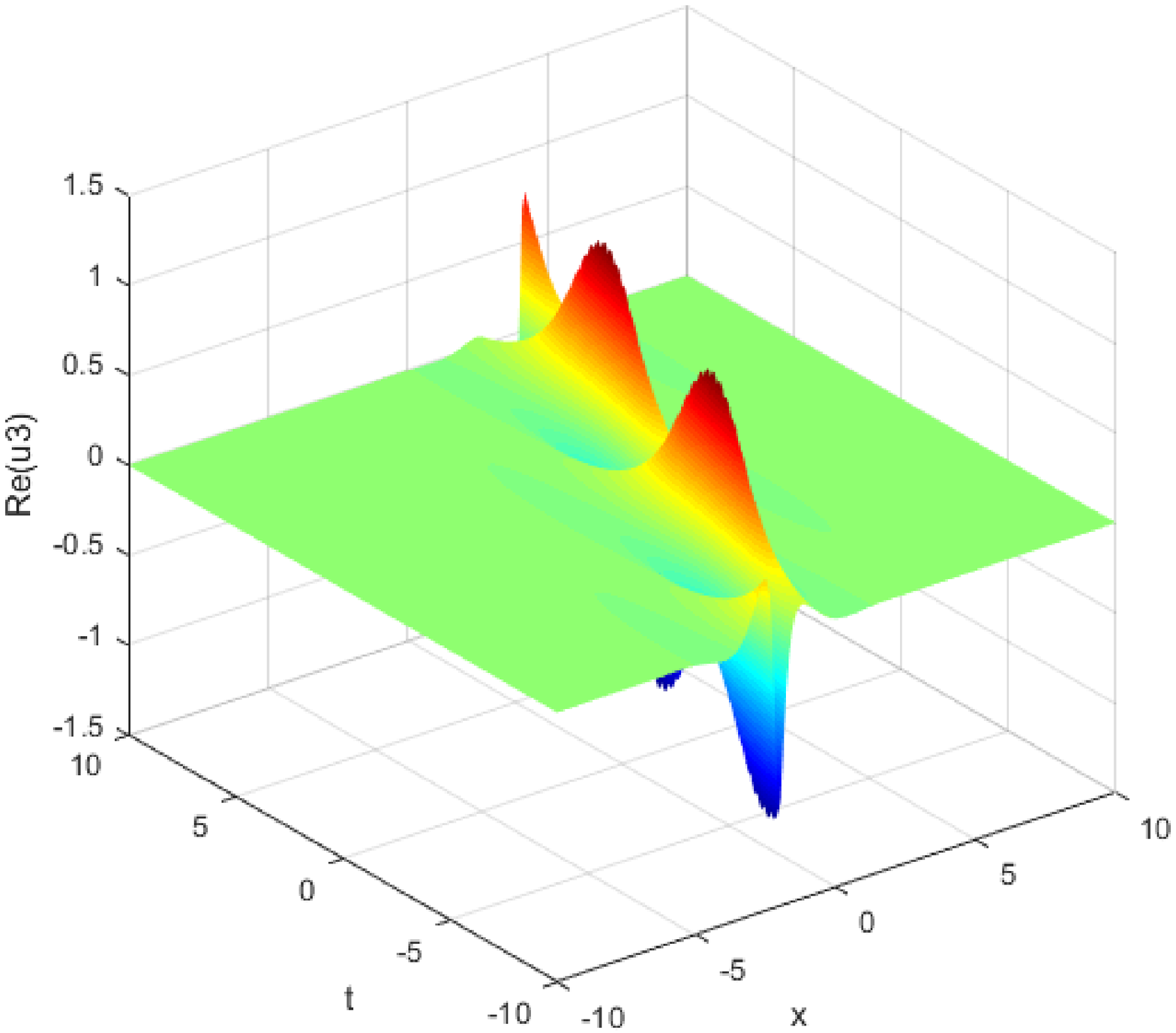}}}

$\ \qquad~~~~~~(\textbf{d})\qquad \ \qquad\qquad\qquad\qquad~(\textbf{e})
\ \qquad\qquad\qquad\qquad\qquad~(\textbf{f})$\\
\noindent
{\rotatebox{0}{\includegraphics[width=3.6cm,height=3.0cm,angle=0]{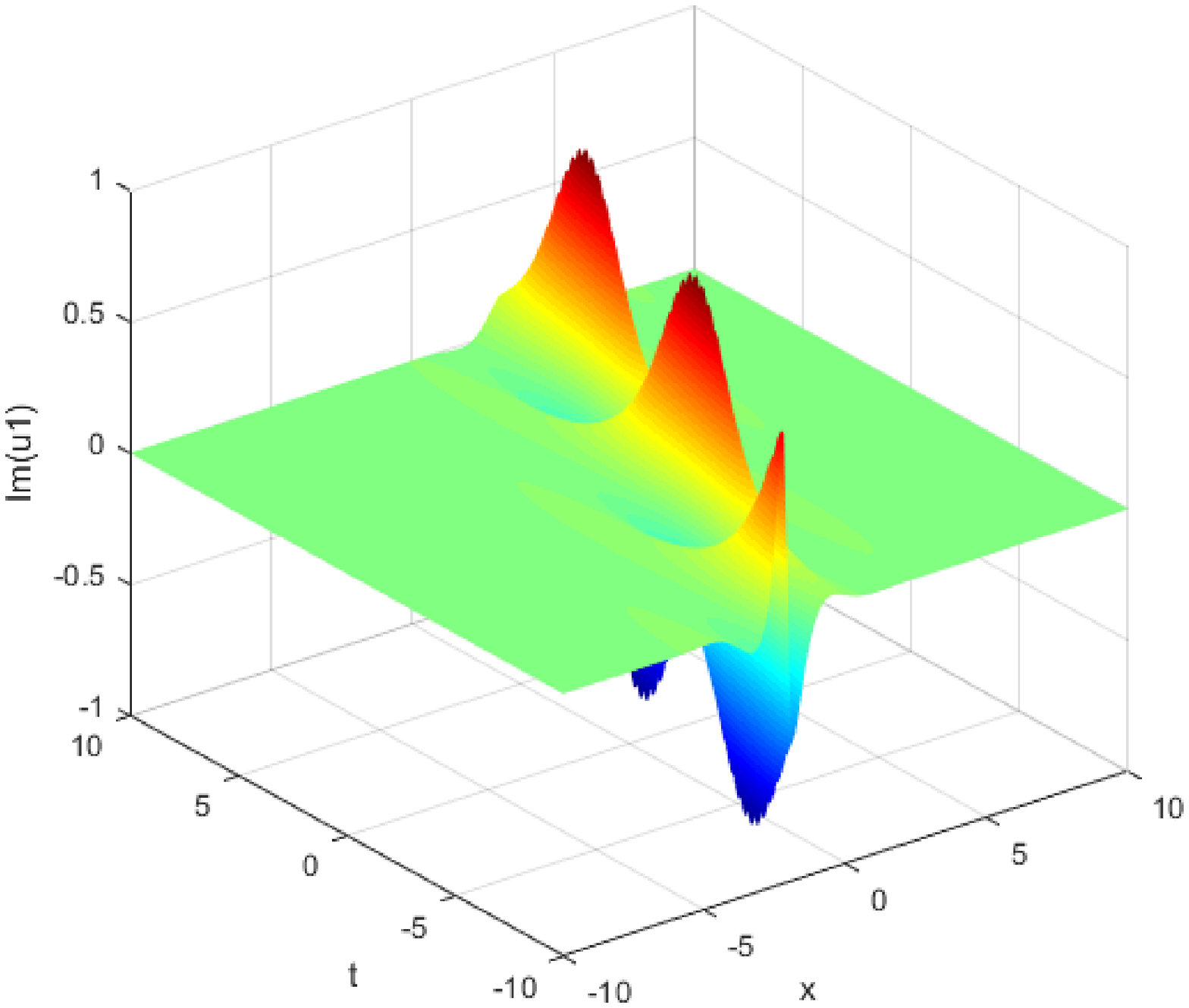}}}
~~~~
{\rotatebox{0}{\includegraphics[width=3.6cm,height=3.0cm,angle=0]{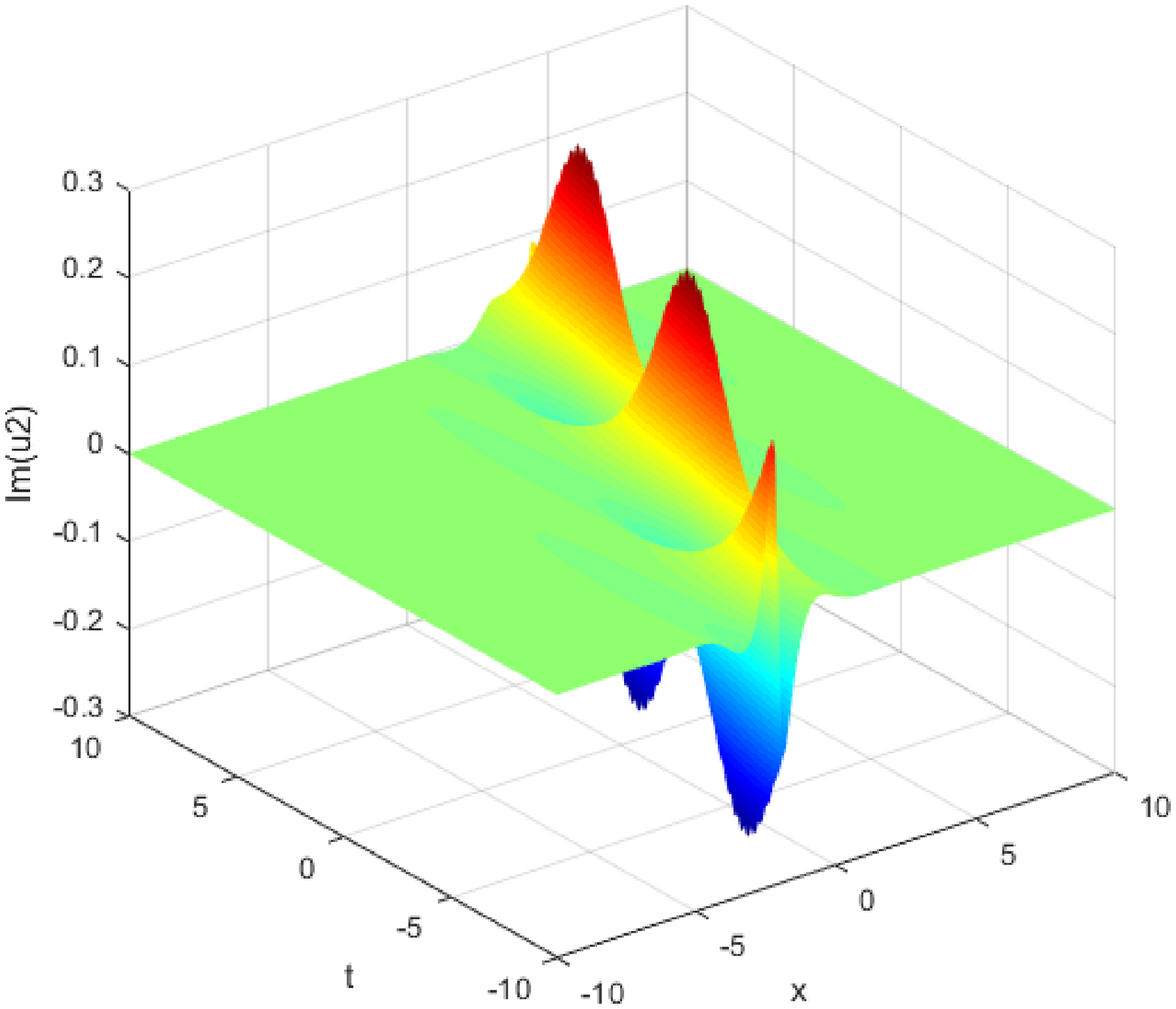}}}
~~~~
{\rotatebox{0}{\includegraphics[width=3.6cm,height=3.0cm,angle=0]{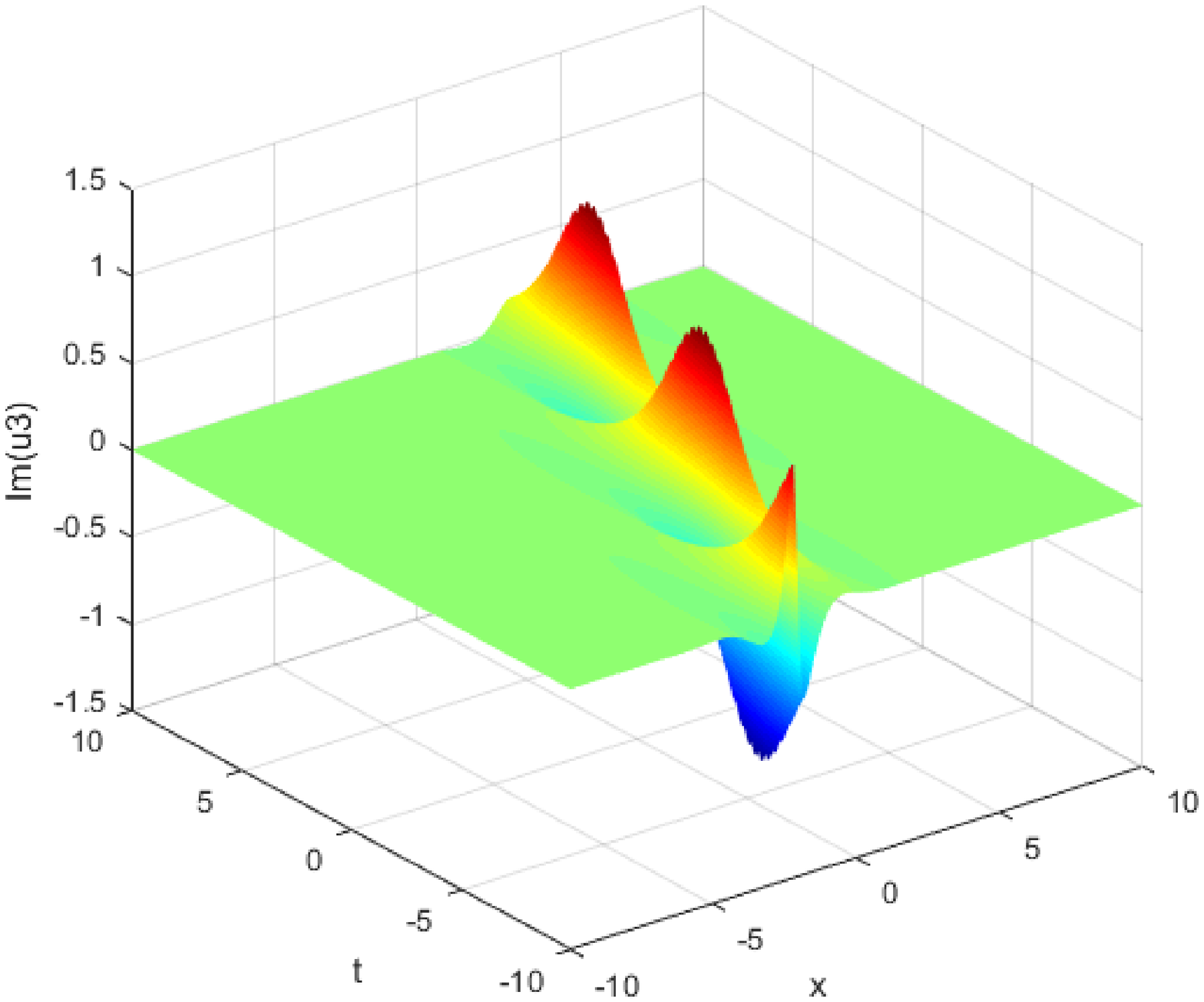}}}

$\ \qquad~~~~~~(\textbf{g})\qquad \ \qquad\qquad\qquad\qquad~(\textbf{h})
\ \qquad\qquad\qquad\qquad\qquad~(\textbf{i})$\\
\noindent { \small \textbf{Figure 5.} One-hump solutions to Eq. \eqref{so44} with parameters $\zeta_1=1.5+0.5i$, $\zeta_2=-\zeta_1$, $\mu_{1,1}=1.1+0.i$, $\mu_{1,2}=0.9+1.5i$,  $\mu_{1,3}=0.2+0.5i$  and $\mu_{1,4}=1.2+ 1.6 i$.
$\textbf{(a)(d)(g)}$: the structures of  $|u_1(x,t)|$, $\mathrm{Re}(u_1)$ and $\mathrm{Im}(u_1)$,
$\textbf{(b)(e)(h)}$:  the structures of  $|u_2(x,t)|$, $\mathrm{Re}(u_2)$ and $\mathrm{Im}(u_2)$,
$\textbf{(c)(f)(i)}$:  the structures of  $|u_3(x,t)|$, $\mathrm{Re}(u_3)$ and $\mathrm{Im}(u_3)$.}  \\

When $N_1=2$,  we can express the solutions to Eq. \eqref{exact}  explicitly
\begin{equation}\label{so444}
  \left\{
   \begin{aligned}
        u_1(x,t) =& i (-\mu_{1,2} \mu_{1,1}^{*} e^{\theta_1 -2\theta_1^*} (M^{-1})_{1,1}
        -\mu_{1,2} \mu_{2,1}^{*} e^{\theta_1 -2\theta_2^*} (M^{-1})_{1,2} \\
         & - \mu_{1,2} \mu_{1,1}^{*} e^{\theta_1 -2\theta_1^*} (M^{-1})_{1,3}
         - \mu_{1,2} \mu_{2,1}^{*} e^{\theta_1 -2\theta_2^*} (M^{-1})_{1,4}           \\
         &  -\mu_{2,2} \mu_{1,1}^{*} e^{\theta_2 -2\theta_1^*} (M^{-1})_{2,1}
        -\mu_{2,2} \mu_{2,1}^{*} e^{\theta_2 -2\theta_2^*} (M^{-1})_{2,2} \\
         & - \mu_{2,2} \mu_{1,1}^{*} e^{\theta_2 -2\theta_1^*} (M^{-1})_{2,3}
         - \mu_{2,2} \mu_{2,1}^{*} e^{\theta_2 -2\theta_2^*} (M^{-1})_{2,4}      \\
         &  +\mu_{1,2} \mu_{1,1}^{*} e^{\theta_1 -2\theta_1^*} (M^{-1})_{3,1}
        +\mu_{1,2} \mu_{2,1}^{*} e^{\theta_1 -2\theta_2^*} (M^{-1})_{3,2} \\
         & +\mu_{1,2} \mu_{1,1}^{*} e^{\theta_1 -2\theta_1^*} (M^{-1})_{3,3}
         +\mu_{1,2} \mu_{2,1}^{*} e^{\theta_1 -2\theta_2^*} (M^{-1})_{3,4}      \\
                  &  +\mu_{2,2} \mu_{1,1}^{*} e^{\theta_2 -2\theta_1^*} (M^{-1})_{4,1}
        +\mu_{2,2} \mu_{2,1}^{*} e^{\theta_2 -2\theta_2^*} (M^{-1})_{4,2} \\
         & +\mu_{2,2} \mu_{1,1}^{*} e^{\theta_2 -2\theta_1^*} (M^{-1})_{4,3}
         +\mu_{2,2} \mu_{2,1}^{*} e^{\theta_2 -2\theta_2^*} (M^{-1})_{4,4} ),     \\
   \end{aligned}
\right.
\end{equation}
\begin{equation}
  \left\{
   \begin{aligned}
     u_2(x,t) =& i (-\mu_{1,3} \mu_{1,1}^{*} e^{\theta_1 -2\theta_1^*} (M^{-1})_{1,1}
        -\mu_{1,3} \mu_{2,1}^{*} e^{\theta_1 -2\theta_2^*} (M^{-1})_{1,2} \\
         & - \mu_{1,3} \mu_{1,1}^{*} e^{\theta_1 -2\theta_1^*} (M^{-1})_{1,3}
         - \mu_{1,3} \mu_{2,1}^{*} e^{\theta_1 -2\theta_2^*} (M^{-1})_{1,4}           \\
         &  -\mu_{2,3} \mu_{1,1}^{*} e^{\theta_2 -2\theta_1^*} (M^{-1})_{2,1}
        -\mu_{2,3} \mu_{2,1}^{*} e^{\theta_2 -2\theta_2^*} (M^{-1})_{2,2} \\
         & - \mu_{2,3} \mu_{1,1}^{*} e^{\theta_2 -2\theta_1^*} (M^{-1})_{2,3}
         - \mu_{2,3} \mu_{2,1}^{*} e^{\theta_2 -2\theta_2^*} (M^{-1})_{2,4}      \\
         &  +\mu_{1,3} \mu_{1,1}^{*} e^{\theta_1 -2\theta_1^*} (M^{-1})_{3,1}
        +\mu_{1,3} \mu_{2,1}^{*} e^{\theta_1 -2\theta_2^*} (M^{-1})_{3,2} \\
         & +\mu_{1,3} \mu_{1,1}^{*} e^{\theta_1 -2\theta_1^*} (M^{-1})_{3,3}
         +\mu_{1,3} \mu_{2,1}^{*} e^{\theta_1 -2\theta_2^*} (M^{-1})_{3,4}      \\
                  &  +\mu_{2,3} \mu_{1,1}^{*} e^{\theta_2 -2\theta_1^*} (M^{-1})_{4,1}
        +\mu_{2,3} \mu_{2,1}^{*} e^{\theta_2 -2\theta_2^*} (M^{-1})_{4,2} \\
         & +\mu_{2,3} \mu_{1,1}^{*} e^{\theta_2 -2\theta_1^*} (M^{-1})_{4,3}
         +\mu_{2,3} \mu_{2,1}^{*} e^{\theta_2 -2\theta_2^*} (M^{-1})_{4,4}  ),    \\
           u_3(x,t) =& i (-\mu_{1,4} \mu_{1,1}^{*} e^{\theta_1 -2\theta_1^*} (M^{-1})_{1,1}
        -\mu_{1,4} \mu_{2,1}^{*} e^{\theta_1 -2\theta_2^*} (M^{-1})_{1,2} \\
         & - \mu_{1,4} \mu_{1,1}^{*} e^{\theta_1 -2\theta_1^*} (M^{-1})_{1,3}
         - \mu_{1,4} \mu_{2,1}^{*} e^{\theta_1 -2\theta_2^*} (M^{-1})_{1,4}    \\
         &  -\mu_{2,4} \mu_{1,1}^{*} e^{\theta_2 -2\theta_1^*} (M^{-1})_{2,1}
        -\mu_{2,4} \mu_{2,1}^{*} e^{\theta_2 -2\theta_2^*} (M^{-1})_{2,2} \\
         & - \mu_{2,4} \mu_{1,1}^{*} e^{\theta_2 -2\theta_1^*} (M^{-1})_{2,3}
         - \mu_{2,4} \mu_{2,1}^{*} e^{\theta_2 -2\theta_2^*} (M^{-1})_{2,4}   \\
     &  +\mu_{1,4} \mu_{1,1}^{*} e^{\theta_1 -2\theta_1^*} (M^{-1})_{3,1}
        +\mu_{1,4} \mu_{2,1}^{*} e^{\theta_1 -2\theta_2^*} (M^{-1})_{3,2} \\
        & +\mu_{1,4} \mu_{1,1}^{*} e^{\theta_1 -2\theta_1^*} (M^{-1})_{3,3}
         +\mu_{1,4} \mu_{2,1}^{*} e^{\theta_1 -2\theta_2^*} (M^{-1})_{3,4}      \\
                  &  +\mu_{2,4} \mu_{1,1}^{*} e^{\theta_2 -2\theta_1^*} (M^{-1})_{4,1}
        +\mu_{2,4} \mu_{2,1}^{*} e^{\theta_2 -2\theta_2^*} (M^{-1})_{4,2} \\
        & +\mu_{2,4} \mu_{1,1}^{*} e^{\theta_2 -2\theta_1^*} (M^{-1})_{4,3}
         +\mu_{2,4} \mu_{2,1}^{*} e^{\theta_2 -2\theta_2^*} (M^{-1})_{4,4}  ),    \\
   \end{aligned}
\right.
\end{equation}

where
\begin{equation}
m_{k,j}=\frac{\hat{\vartheta}_{k}\vartheta_{j}}{\zeta_j-\hat{\zeta_{k}}}, \quad 1 \leq k,j \leq 4,
\end{equation}
with
\begin{equation}
  \zeta_3=-\zeta_1, \quad \zeta_4=-\zeta_2, \quad \hat{\zeta_j}=\zeta_j^{*}, \quad 1 \leq j \leq 4.
\end{equation}

\noindent
{\rotatebox{0}{\includegraphics[width=3.6cm,height=3.5cm,angle=0]{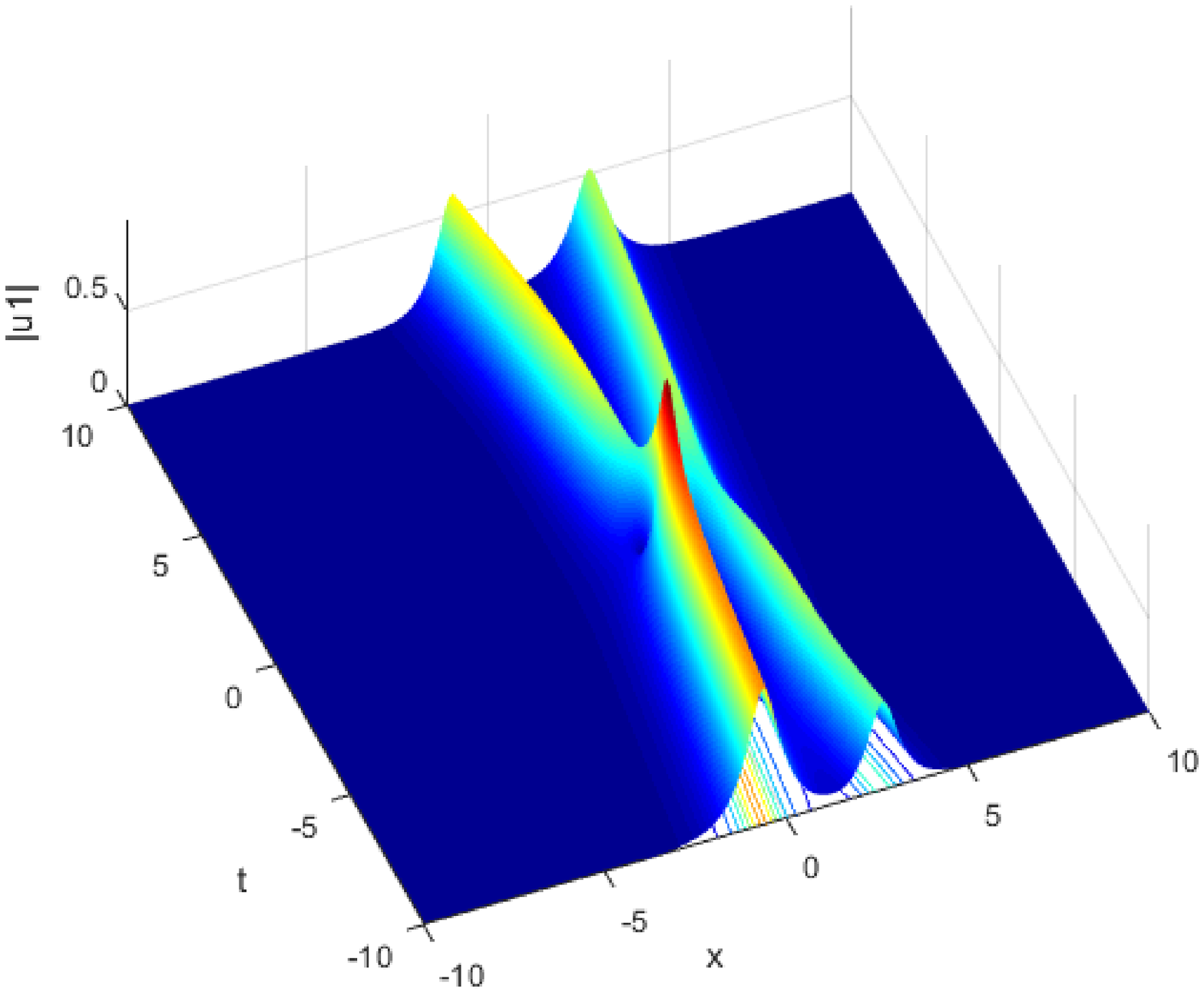}}}
~~~~
{\rotatebox{0}{\includegraphics[width=3.6cm,height=3.5cm,angle=0]{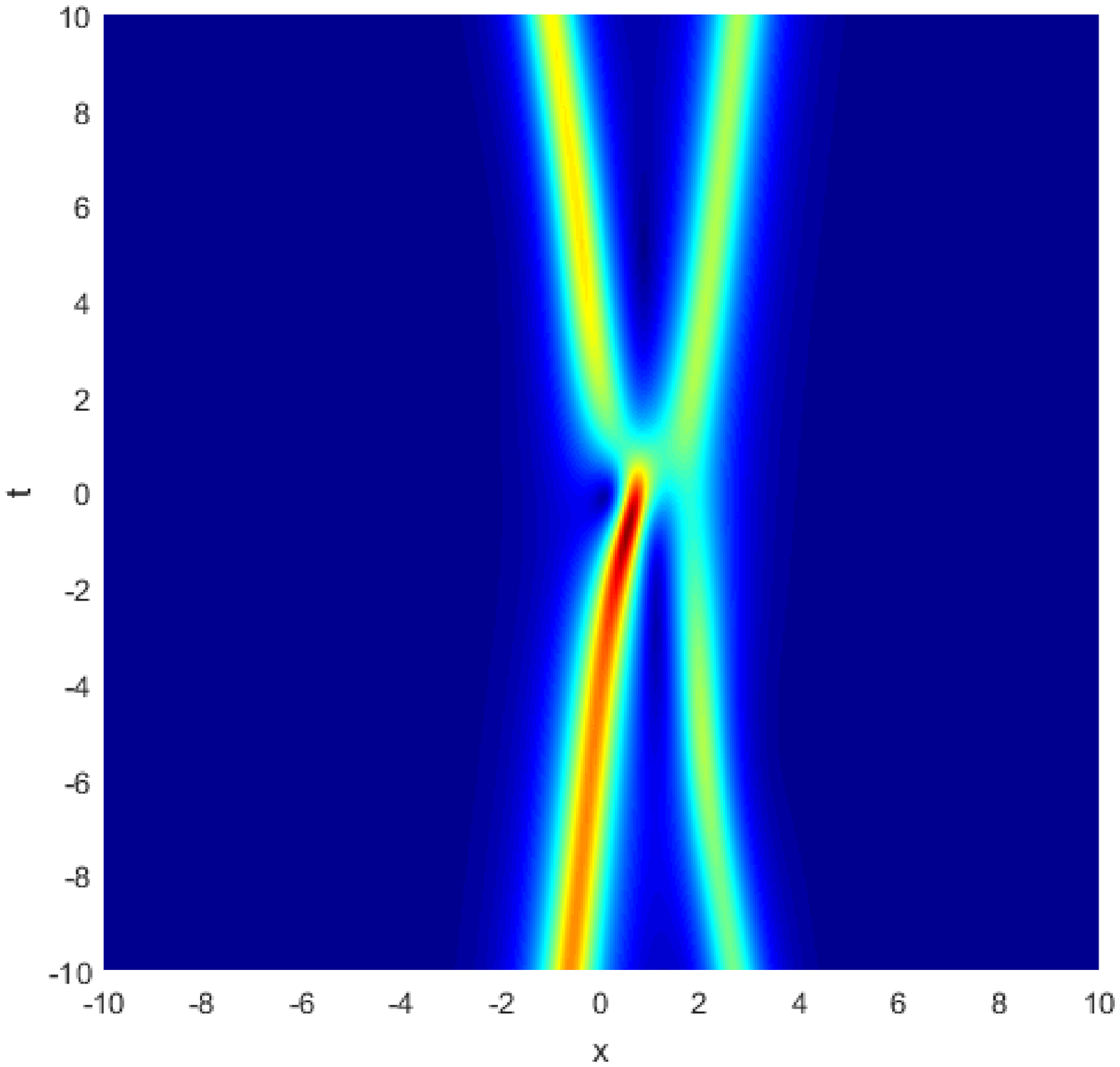}}}
~~~~
{\rotatebox{0}{\includegraphics[width=3.6cm,height=3.5cm,angle=0]{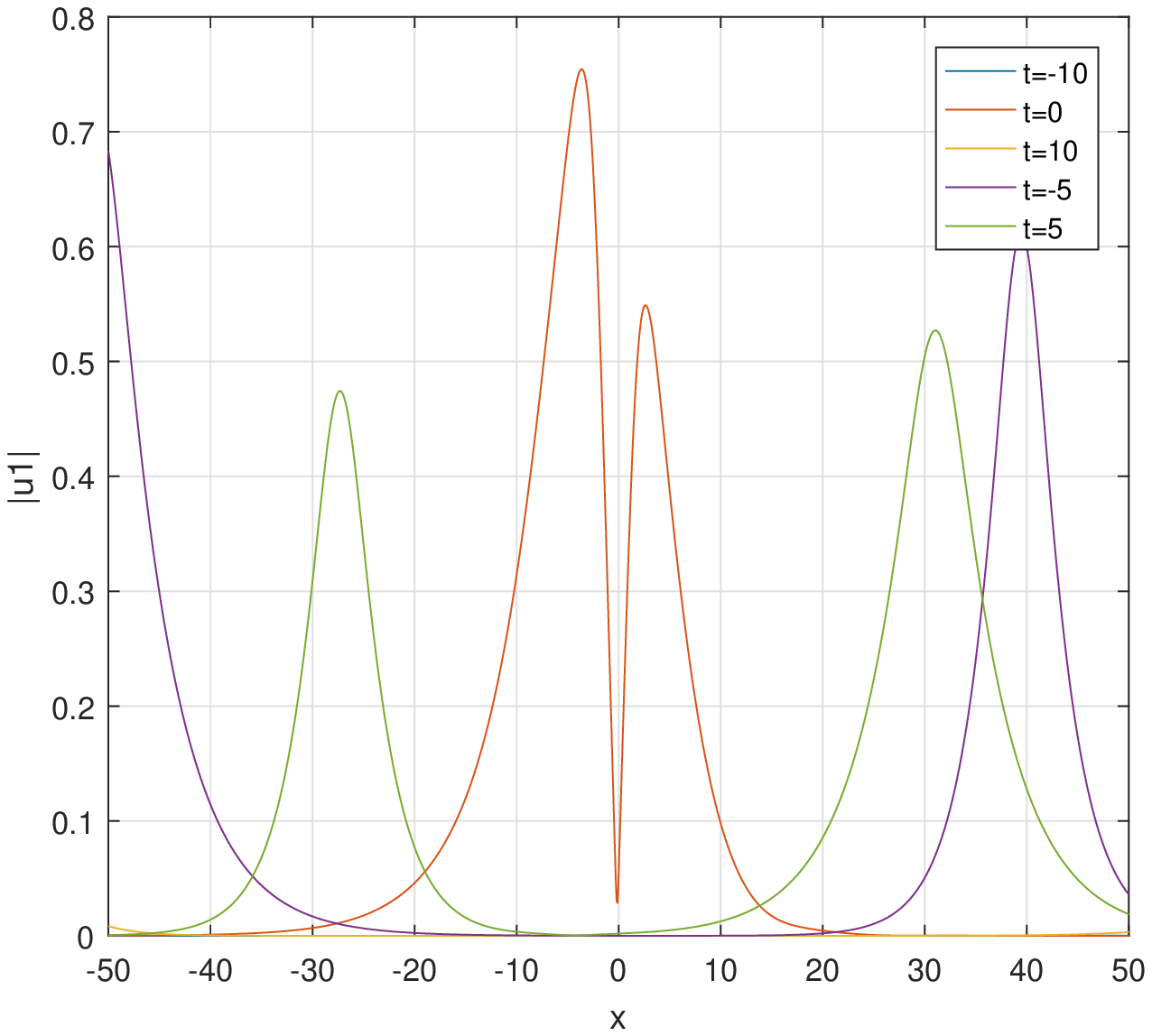}}}

$\ \qquad~~~~~~(\textbf{a})\qquad \ \qquad\qquad\qquad\qquad~(\textbf{b})
\ \qquad\qquad\qquad\qquad\qquad~(\textbf{c})$\\
\noindent
{\rotatebox{0}{\includegraphics[width=3.6cm,height=3.5cm,angle=0]{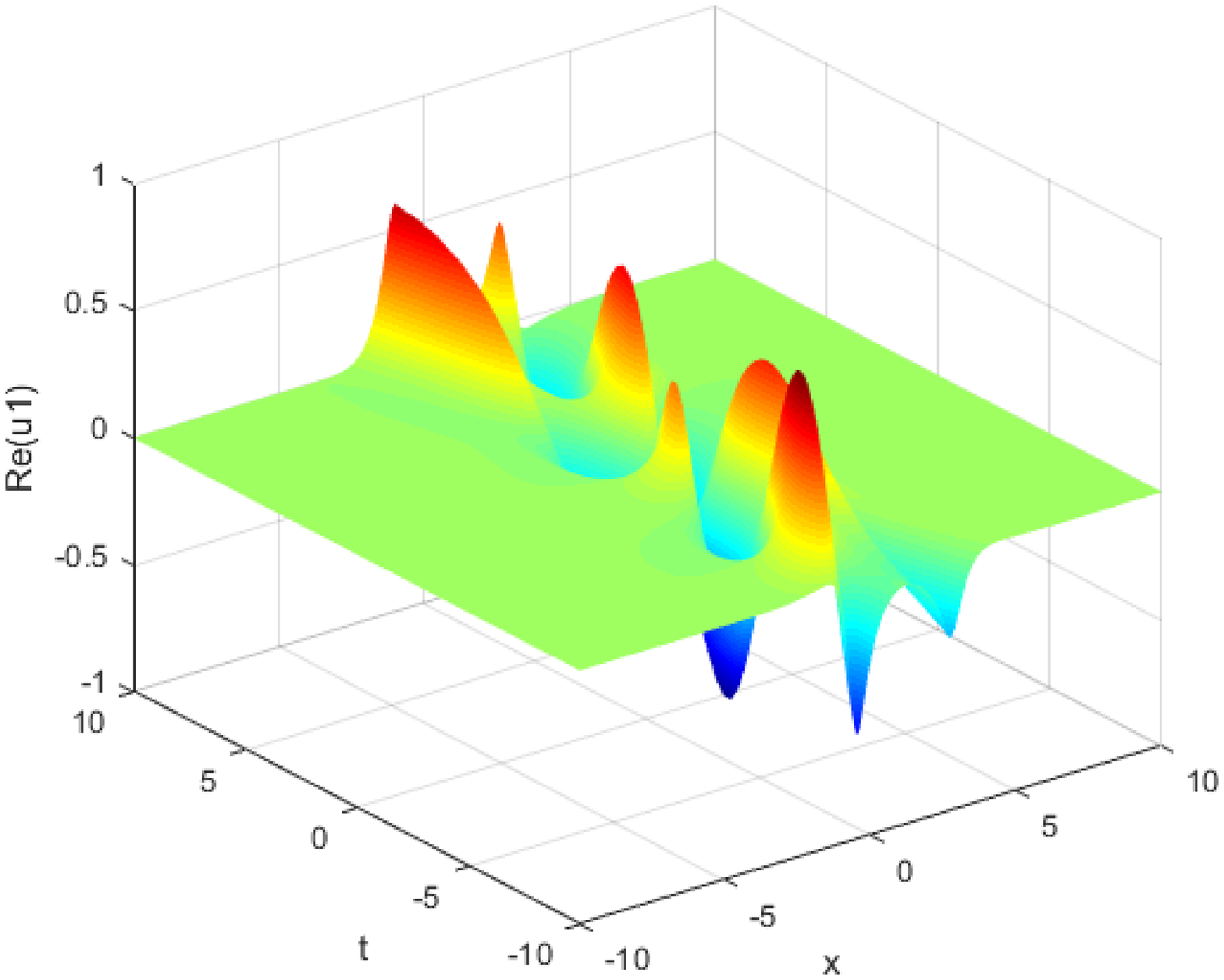}}}
~~~~
{\rotatebox{0}{\includegraphics[width=3.6cm,height=3.5cm,angle=0]{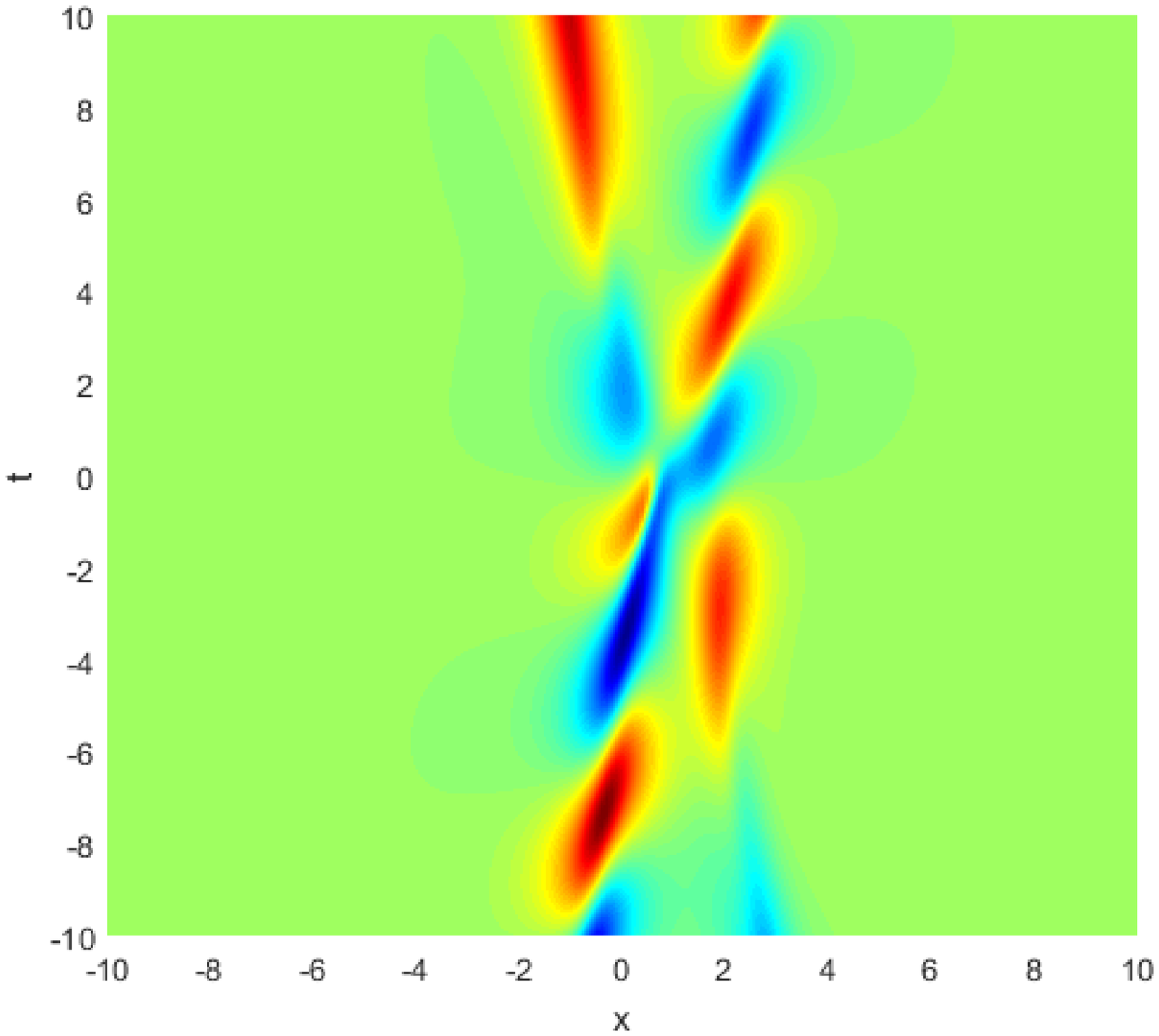}}}
~~~~
{\rotatebox{0}{\includegraphics[width=3.6cm,height=3.5cm,angle=0]{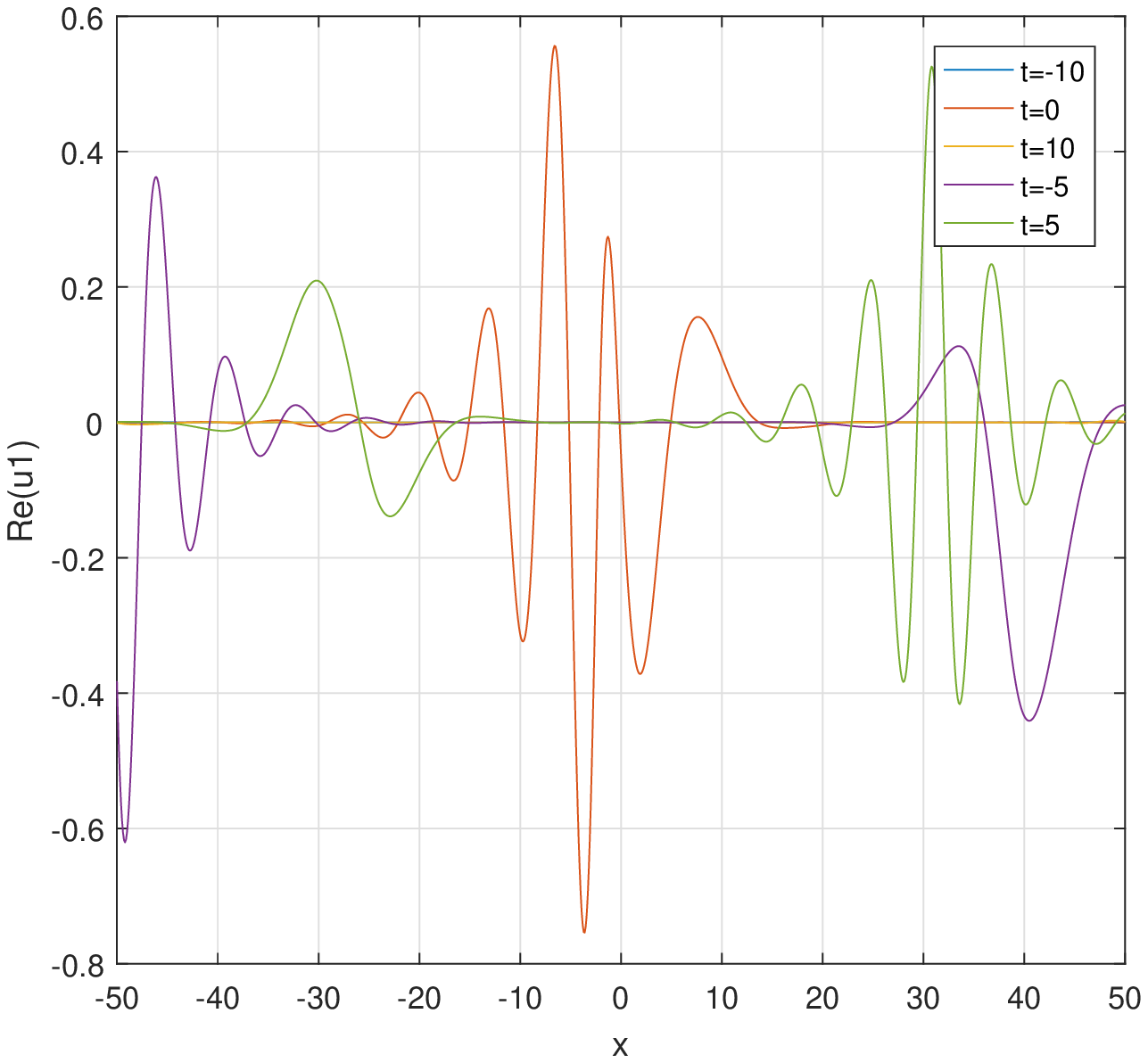}}}

$\ \qquad~~~~~~(\textbf{d})\qquad \ \qquad\qquad\qquad\qquad~(\textbf{e})
\ \qquad\qquad\qquad\qquad\qquad~(\textbf{f})$\\
\noindent
{\rotatebox{0}{\includegraphics[width=3.6cm,height=3.0cm,angle=0]{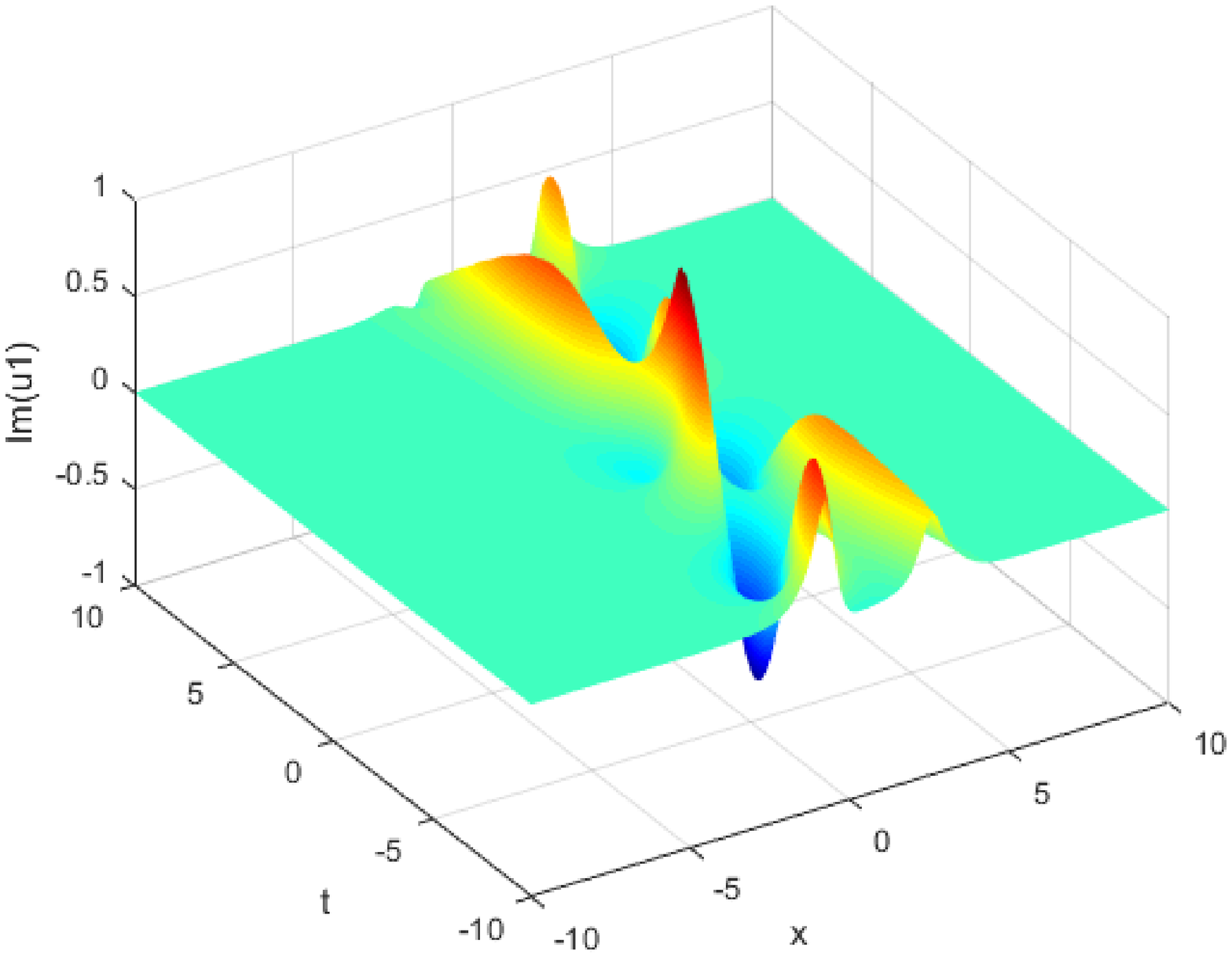}}}
~~~~
{\rotatebox{0}{\includegraphics[width=3.6cm,height=3.0cm,angle=0]{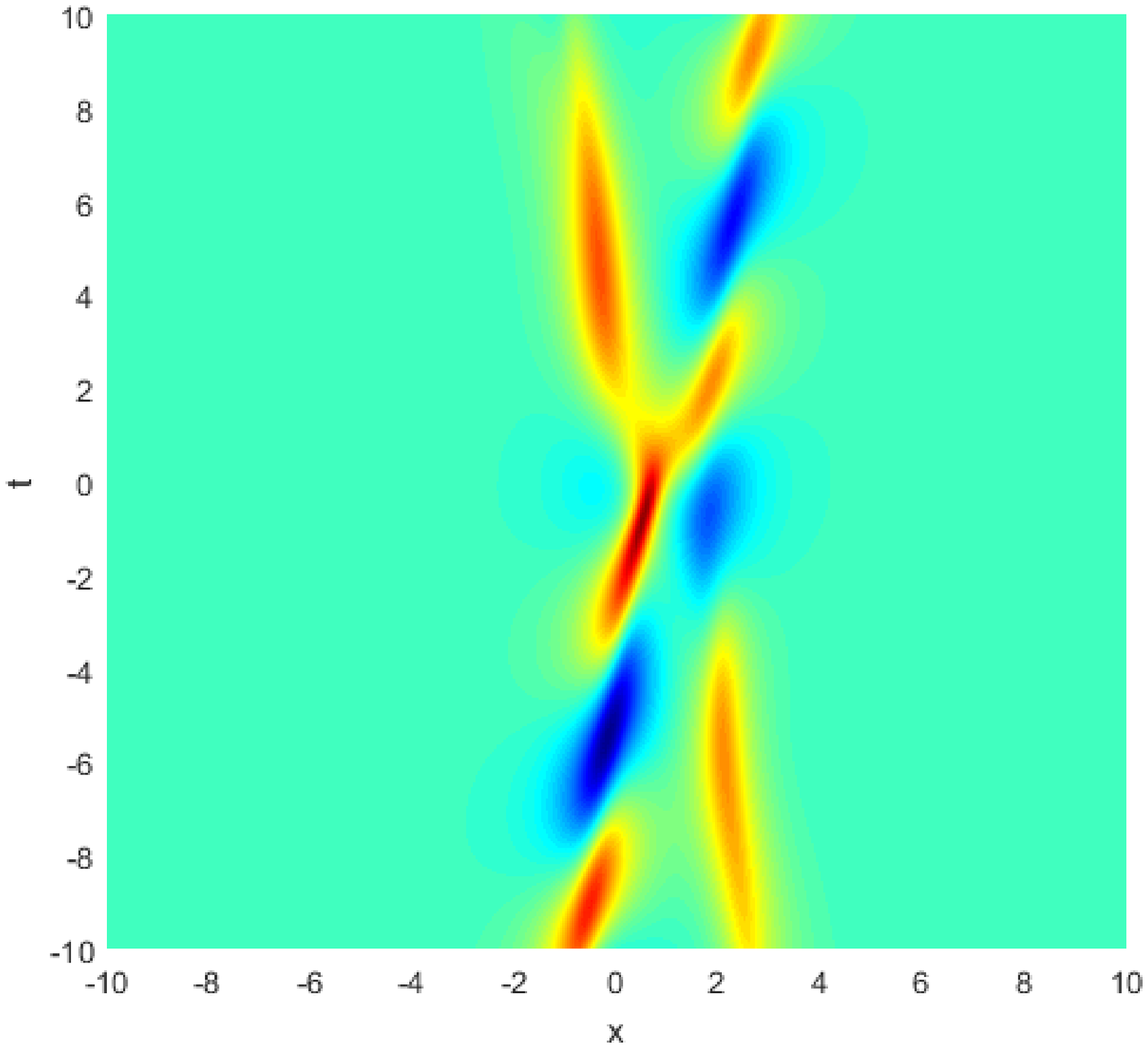}}}
~~~~
{\rotatebox{0}{\includegraphics[width=3.6cm,height=3.0cm,angle=0]{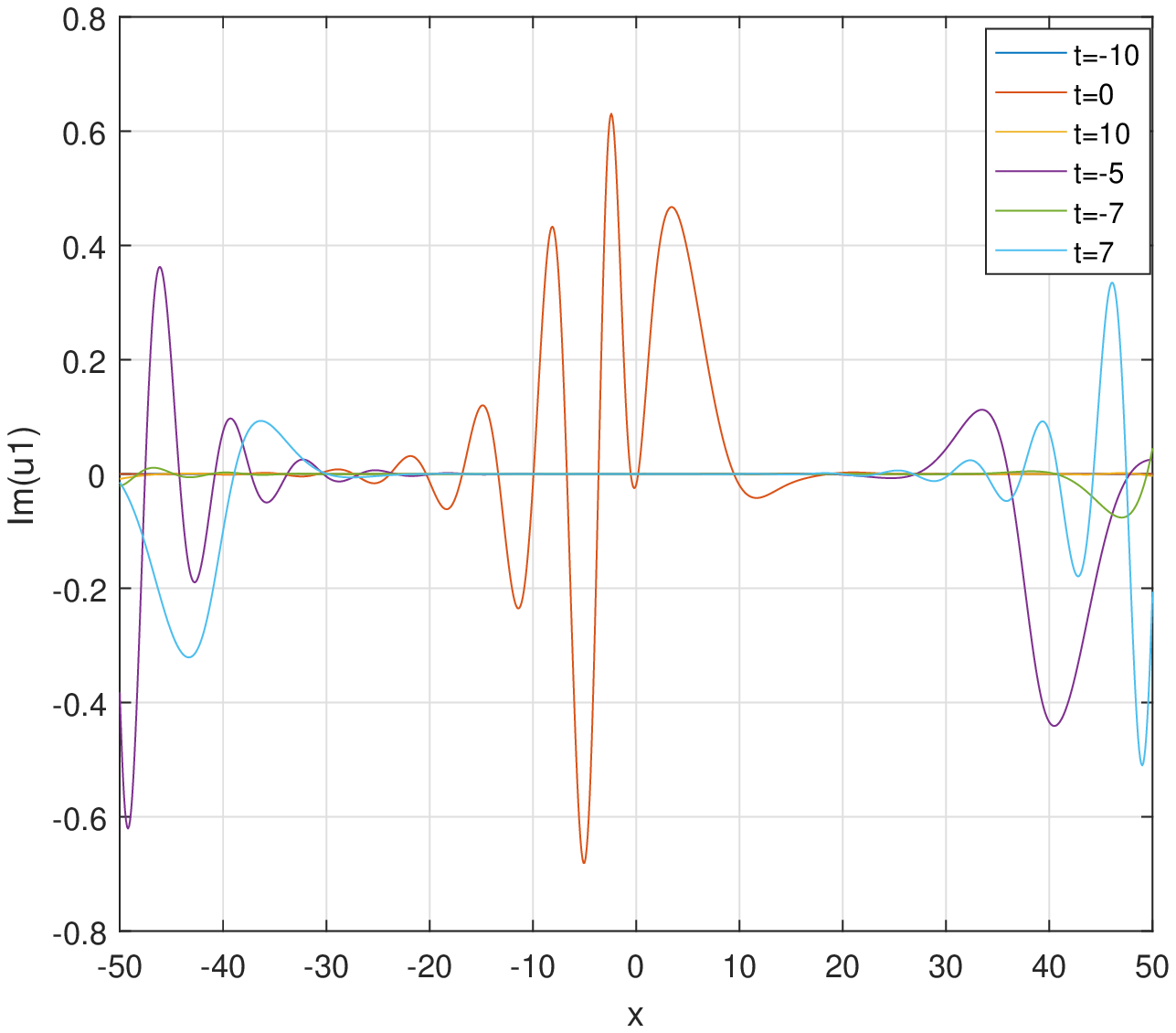}}}

$\ \qquad~~~~~~(\textbf{g})\qquad \ \qquad\qquad\qquad\qquad~(\textbf{h})
\ \qquad\qquad\qquad\qquad\qquad~(\textbf{i})$\\
\noindent { \small \textbf{Figure 6.} Two-soliton  solutions to Eq. \eqref{so444} with parameters $\zeta_1=1+0.9i$, $\zeta_2=0.9+1.3i$, $\mu_{1,1}=1+i$, $\mu_{1,2}=2+i$,  $\mu_{1,3}=3+i$,  $\mu_{1,4}=1+2i$,  $\mu_{2,1}=1-i$, $\mu_{2,2}=2-i$,   $\mu_{2,3}=3-i$ and    $\mu_{2,4}=1+2i$.
$\textbf{(a)(b)(c)}$:  the local structure, density and wave propagation of  $|u_1(x,t)|$,
$\textbf{(d)(e)(f)}$:  the local structure, density and wave propagation of  $\mathrm{Re}(u_1)$,
$\textbf{(g)(h)(i)}$:  the local structure, density and wave propagation of  $\mathrm{Im}(u_1)$.}  \\

The localized structures, density plot and the dynamic propagation behavior of the two soliton solutions are presented in Fig. 6. From Fig. 6, we can  learn the  propagation process and interaction mechanism of the two solitons.  To be more concrete,  the energy of two solitons changes significantly before and after collision, and the direction and the position of them has also changed to some extent.

Next, if we take $N_1=3$,    the solutions to Eq. \eqref{exact}  can be expressed  explicitly by
\begin{equation}
   \begin{aligned}
     u_\nu(x,t)& = i (  -\mu_{1,\nu+1} \mu_{1,1}^{*} e^{\theta_1 - 2 \theta_1^{*}} (M^{-1})_{1,1}
                    -\mu_{1,\nu+1} \mu_{2,1}^{*} e^{\theta_1 - 2 \theta_2^{*}} (M^{-1})_{1,2}
                    -\mu_{1,\nu+1} \mu_{3,1}^{*} e^{\theta_1 - 2 \theta_3^{*}} (M^{-1})_{1,3}  \\
                    & -\mu_{1,\nu+1} \mu_{1,1}^{*} e^{\theta_1 - 2 \theta_1^{*}} (M^{-1})_{1,4}
                    -\mu_{1,\nu+1} \mu_{2,1}^{*} e^{\theta_1 - 2 \theta_2^{*}} (M^{-1})_{1,5}
                     -\mu_{1,\nu+1} \mu_{3,1}^{*} e^{\theta_1 - 2 \theta_3^{*}} (M^{-1})_{1,6}    \\
                   &   -\mu_{2,\nu+1} \mu_{1,1}^{*} e^{\theta_2 - 2 \theta_1^{*}} (M^{-1})_{2,1}
                    -\mu_{2,\nu+1} \mu_{2,1}^{*} e^{\theta_2 - 2 \theta_2^{*}} (M^{-1})_{2,2}
                    -\mu_{2,\nu+1} \mu_{3,1}^{*} e^{\theta_2 - 2 \theta_3^{*}} (M^{-1})_{2,3}  \\
                  &  -\mu_{2,\nu+1} \mu_{1,1}^{*} e^{\theta_2 - 2 \theta_1^{*}} (M^{-1})_{2,4}
                    -\mu_{2,\nu+1} \mu_{2,1}^{*} e^{\theta_2 - 2 \theta_2^{*}} (M^{-1})_{2,5}
                     -\mu_{2,\nu+1} \mu_{3,1}^{*} e^{\theta_2 - 2 \theta_3^{*}} (M^{-1})_{2,6}    \\
                   &      -\mu_{3,\nu+1} \mu_{1,1}^{*} e^{\theta_3 - 2 \theta_1^{*}} (M^{-1})_{3,1}
                    -\mu_{3,\nu+1} \mu_{2,1}^{*} e^{\theta_3 - 2 \theta_2^{*}} (M^{-1})_{3,2}
                    -\mu_{3,\nu+1} \mu_{3,1}^{*} e^{\theta_3 - 2 \theta_3^{*}} (M^{-1})_{3,3}  \\
                   & -\mu_{3,\nu+1} \mu_{1,1}^{*} e^{\theta_3 - 2 \theta_1^{*}} (M^{-1})_{3,4}
                    -\mu_{3,\nu+1} \mu_{2,1}^{*} e^{\theta_3 - 2 \theta_2^{*}} (M^{-1})_{3,5}
                     -\mu_{3,\nu+1} \mu_{3,1}^{*} e^{\theta_3 - 2 \theta_3^{*}} (M^{-1})_{3,6}    \\
                 &    +\mu_{1,\nu+1} \mu_{1,1}^{*} e^{\theta_1 - 2 \theta_1^{*}} (M^{-1})_{4,1}
                    +\mu_{1,\nu+1} \mu_{2,1}^{*} e^{\theta_1 - 2 \theta_2^{*}} (M^{-1})_{4,2}
                    +\mu_{1,\nu+1} \mu_{3,1}^{*} e^{\theta_1 - 2 \theta_3^{*}} (M^{-1})_{4,3}\\
                  &  +\mu_{1,\nu+1} \mu_{1,1}^{*} e^{\theta_1 - 2 \theta_1^{*}} (M^{-1})_{4,4}
                    +\mu_{1,\nu+1} \mu_{2,1}^{*} e^{\theta_1 - 2 \theta_2^{*}} (M^{-1})_{4,5}
                     +\mu_{1,\nu+1} \mu_{3,1}^{*} e^{\theta_1 - 2 \theta_3^{*}} (M^{-1})_{4,6}   \\
                  &                        +\mu_{2,\nu+1} \mu_{1,1}^{*} e^{\theta_2 - 2 \theta_1^{*}} (M^{-1})_{5,1}
                    +\mu_{2,\nu+1} \mu_{2,1}^{*} e^{\theta_2 - 2 \theta_2^{*}} (M^{-1})_{5,2}
                    +\mu_{2,\nu+1} \mu_{3,1}^{*} e^{\theta_2 - 2 \theta_3^{*}} (M^{-1})_{5,3}  \\
                  &  +\mu_{2,\nu+1} \mu_{1,1}^{*} e^{\theta_2 - 2 \theta_1^{*}} (M^{-1})_{5,4}
                    +\mu_{2,\nu+1} \mu_{2,1}^{*} e^{\theta_2 - 2 \theta_2^{*}} (M^{-1})_{5,5}
                     +\mu_{2,\nu+1} \mu_{3,1}^{*} e^{\theta_2 - 2 \theta_3^{*}} (M^{-1})_{5,6}    \\
                   &                       +\mu_{3,\nu+1} \mu_{1,1}^{*} e^{\theta_3 - 2 \theta_1^{*}} (M^{-1})_{6,1}
                    +\mu_{3,\nu+1} \mu_{2,1}^{*} e^{\theta_3 - 2 \theta_2^{*}} (M^{-1})_{6,2}
                    +\mu_{3,\nu+1} \mu_{3,1}^{*} e^{\theta_3 - 2 \theta_3^{*}} (M^{-1})_{6,3}\\
                 &   +\mu_{3,\nu+1} \mu_{1,1}^{*} e^{\theta_3 - 2 \theta_1^{*}} (M^{-1})_{6,4}
                    +\mu_{3,\nu+1} \mu_{2,1}^{*} e^{\theta_3 - 2 \theta_2^{*}} (M^{-1})_{6,5}
                     +\mu_{3,\nu+1} \mu_{3,1}^{*} e^{\theta_3 - 2 \theta_3^{*}} (M^{-1})_{6,6}     ),
   \end{aligned}
\end{equation}
where
\begin{equation}
 \nu=1,2,3, \quad  m_{k,j}=\frac{\hat{\vartheta}_{k}\vartheta_{j}}{\zeta_j-\hat{\zeta_{k}}}, \quad 1 \leq k,j \leq 6,
\end{equation}
with
\begin{equation}
  \zeta_4=-\zeta_1, \quad \zeta_5=-\zeta_2,\quad \zeta_6=-\zeta_3, \quad \hat{\zeta_j}=\zeta_j^{*}, \quad 1 \leq j \leq 6.
\end{equation}

\noindent
{\rotatebox{0}{\includegraphics[width=3.6cm,height=3.5cm,angle=0]{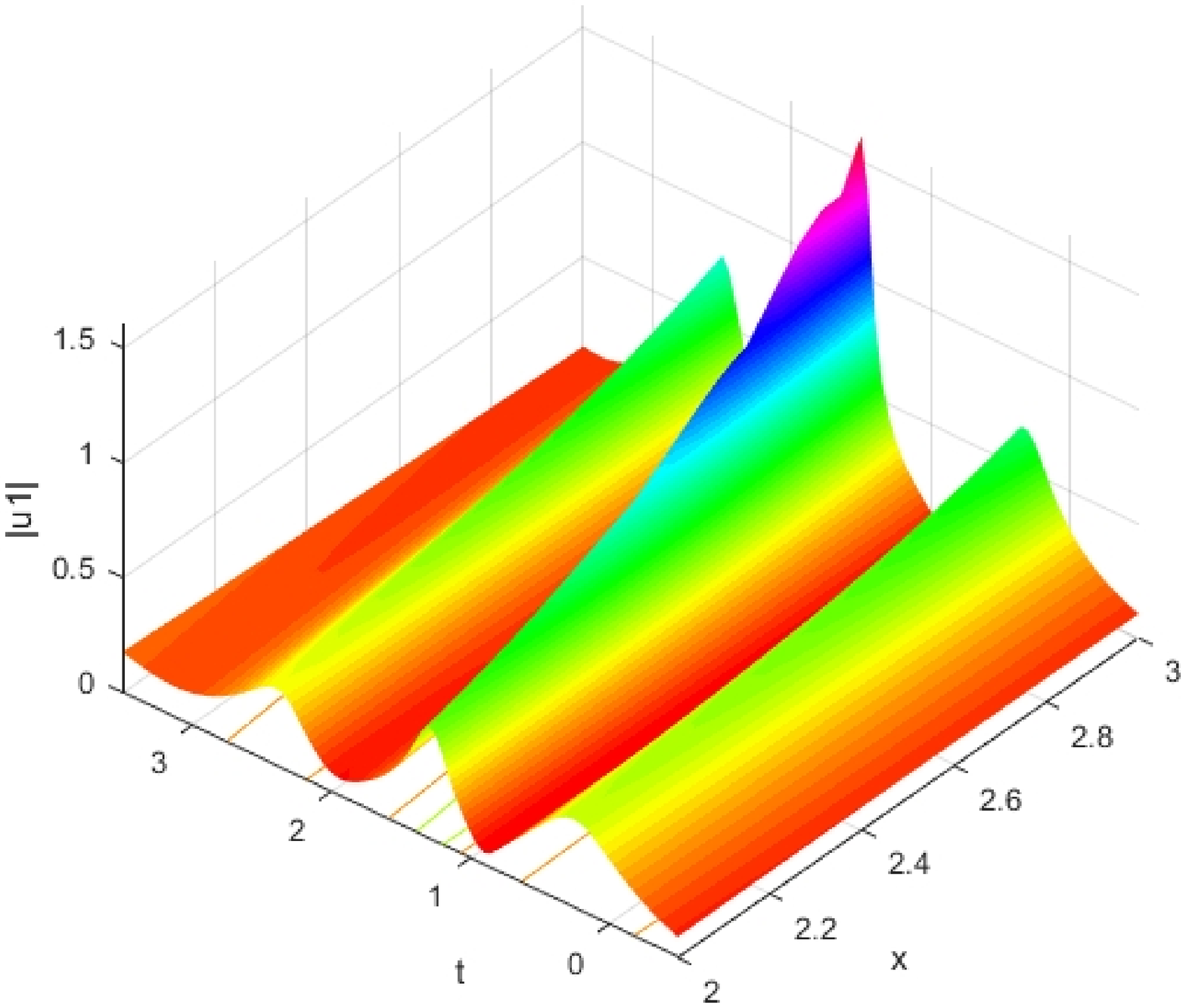}}}
~~~~
{\rotatebox{0}{\includegraphics[width=3.6cm,height=3.5cm,angle=0]{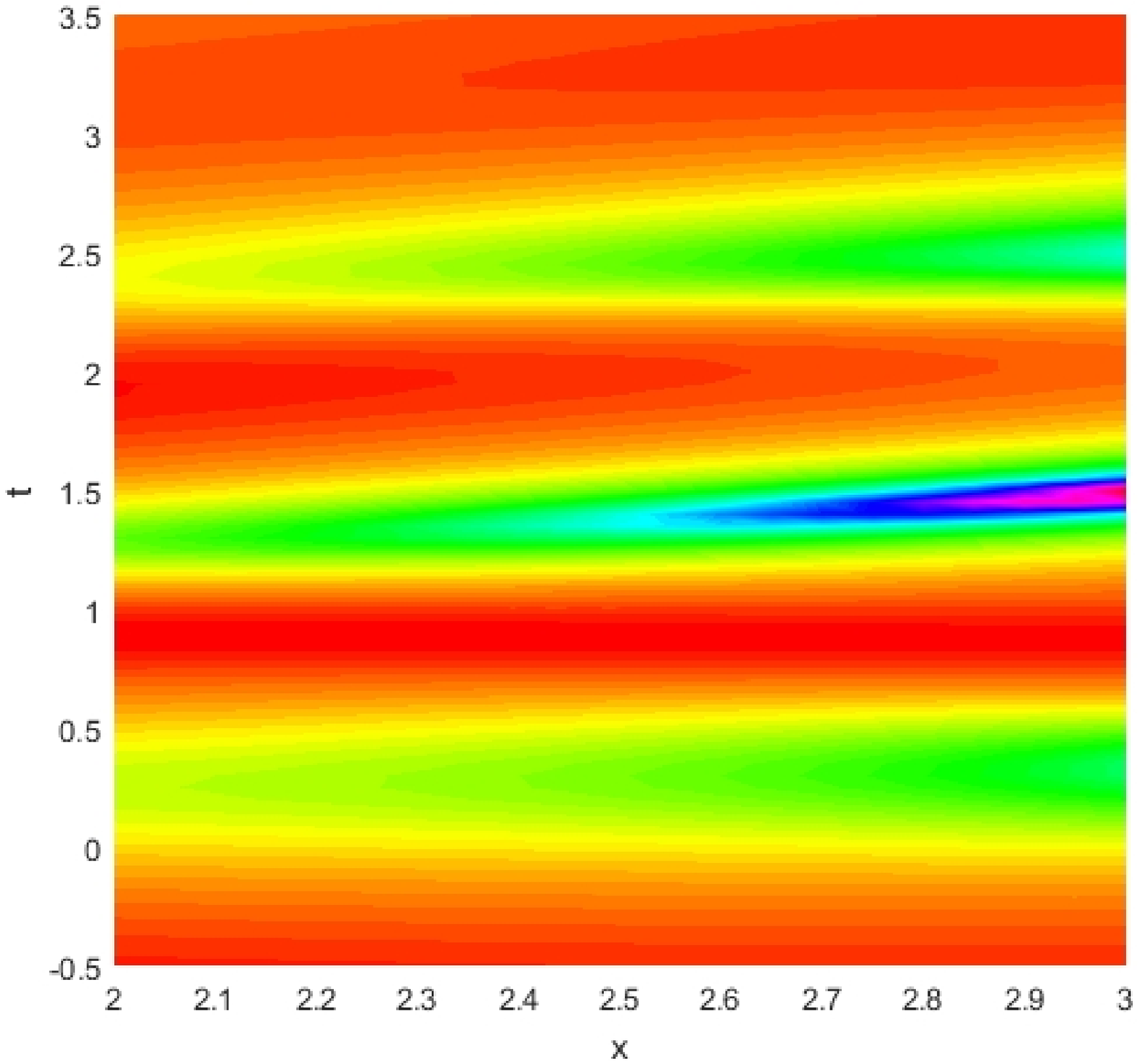}}}
~~~~
{\rotatebox{0}{\includegraphics[width=3.6cm,height=3.5cm,angle=0]{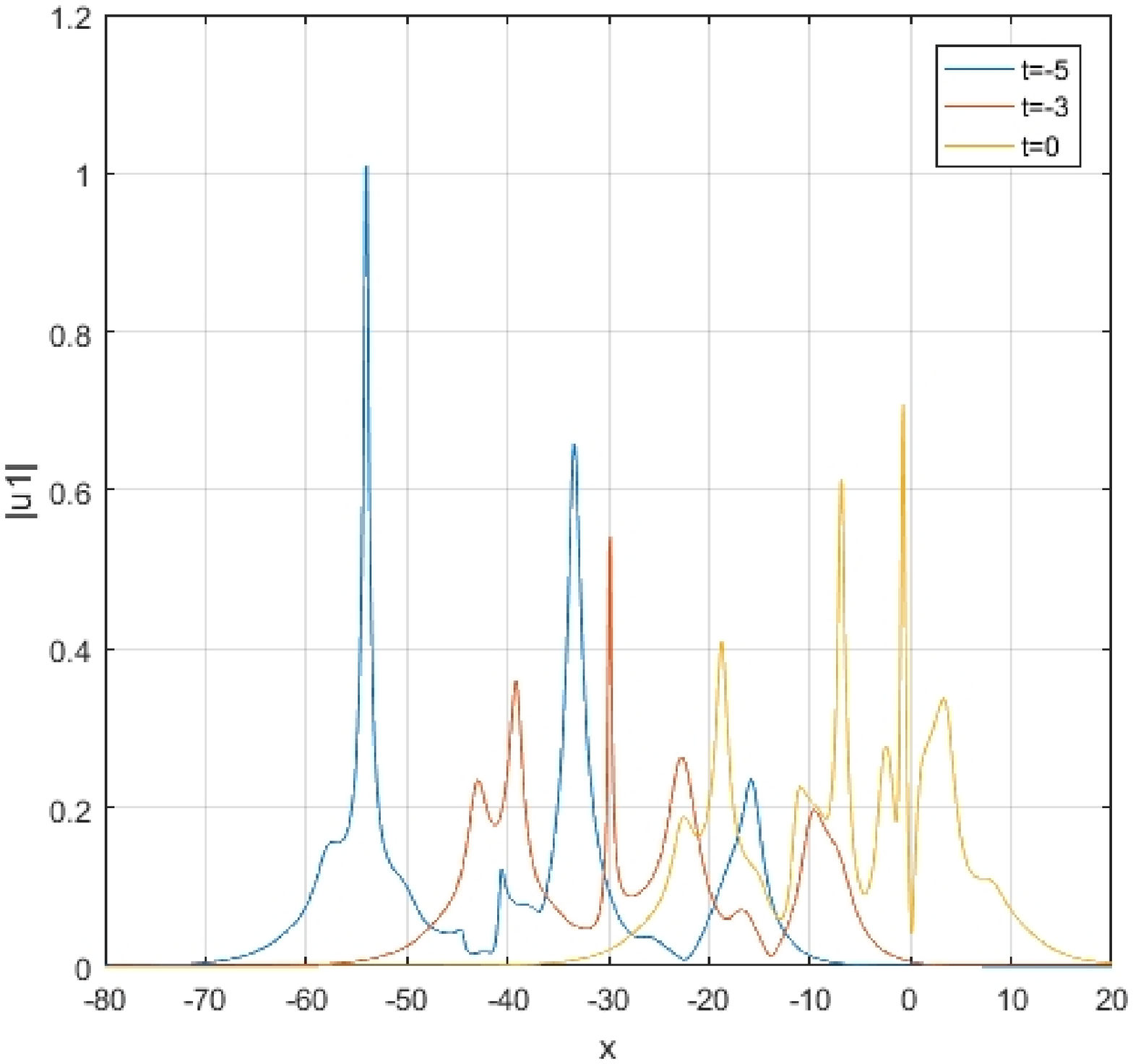}}}

$\ \qquad~~~~~~(\textbf{a})\qquad \ \qquad\qquad\qquad\qquad~(\textbf{b})
\ \qquad\qquad\qquad\qquad\qquad~(\textbf{c})$\\
\noindent { \small \textbf{Figure 7.} Three-soliton  solution to Eq. \eqref{three33} with parameters $\zeta_1=1.2+0.8i$,  $\zeta_2=1.5+0.7i$, $\zeta_3=1.1+0.9i$,  $\mu_{1,1}=0.1$, $\mu_{1,2}=\mu_{2,1}=\mu_{3,4}=0.2$,  $\mu_{1,3}=\mu_{3,1}=0.3$,  $\mu_{1,4}=\mu_{2,2}=\mu_{3,5}=0.4$,  $\mu_{1,5}=\mu_{2,5}=0.5$, $\mu_{2,3}=\mu_{3,2}=0.6$, $\mu_{2,4}=0.8$, $\mu_{3,3}=0.9$ and $\mu_{3,4}=1.2$.
$\textbf{(a)}$: the local structures of the three soliton solutions $u_1(x,t)$,
$\textbf{(b)}$: the density plot of $u_1(x,t)$,
$\textbf{(c)}$: the wave propagation of the  three soliton solutions $u_1(x,t)$.}  \\

\subsection{Case 3:   multi-soliton solutions of  four-component FL equations}

If we take $N=4$ and $v=A^{T}u^{*}$, where
\begin{equation}
    A=
       \begin{pmatrix}
      a_{11}  & a_{21}^{*}  & a_{31}^{*}  & a_{41}^{*}  \\
      a_{21} & a_{22}  & a_{32}^{*}  & a_{42}^{*}   \\
      a_{31}  & a_{32}  & a_{33} & a_{43}^{*}   \\
      a_{41}  & a_{42} & a_{43} & a_{44} \\
    \end{pmatrix}
   =\begin{pmatrix}
      1  & 1-i  & 1-2i & 1-3i  \\
      1+i & 2  & 2-i & 2-2i  \\
      1+2i  & 2+i  & 3 & 3-i  \\
      1+3i  & 2+2i & 3+i & 4  \\
    \end{pmatrix},
\end{equation}
when $N_1=1$,  we can express the solution to Eq. \eqref{exact}  explicitly
\begin{equation}\label{so3}
  \left\{
   \begin{aligned}
        u_1(x,t) =& i (-\mu_{1,2} \mu_{1,1}^{*} e^{\theta_1 -2\theta_1^*} (M^{-1})_{1,1} -\mu_{1,2} \mu_{1,1}^{*} e^{\theta_1 -2\theta_1^*} (M^{-1})_{1,2} \\
         & + \mu_{1,2} \mu_{1,1}^{*} e^{\theta_1 -2\theta_1^*} (M^{-1})_{2,1} + \mu_{1,2} \mu_{1,1}^{*} e^{\theta_1 -2\theta_1^*} (M^{-1})_{2,2}            ),  \\
                 u_2(x,t) =& i (-\mu_{1,3} \mu_{1,1}^{*} e^{\theta_1 -2\theta_1^*} (M^{-1})_{1,1} -\mu_{1,3} \mu_{1,1}^{*} e^{\theta_1 -2\theta_1^*} (M^{-1})_{1,2} \\
         & + \mu_{1,3} \mu_{1,1}^{*} e^{\theta_1 -2\theta_1^*} (M^{-1})_{2,1} + \mu_{1,3} \mu_{1,1}^{*} e^{\theta_1 -2\theta_1^*} (M^{-1})_{2,2}            ),  \\
                          u_3(x,t) =& i (-\mu_{1,4} \mu_{1,1}^{*} e^{\theta_1 -2\theta_1^*} (M^{-1})_{1,1} -\mu_{1,4} \mu_{1,1}^{*} e^{\theta_1 -2\theta_1^*} (M^{-1})_{1,2} \\
         & + \mu_{1,4} \mu_{1,1}^{*} e^{\theta_1 -2\theta_1^*} (M^{-1})_{2,1} + \mu_{1,4} \mu_{1,1}^{*} e^{\theta_1 -2\theta_1^*} (M^{-1})_{2,2}            ),  \\
                          u_4(x,t) =& i (-\mu_{1,5} \mu_{1,1}^{*} e^{\theta_1 -2\theta_1^*} (M^{-1})_{1,1} -\mu_{1,5} \mu_{1,1}^{*} e^{\theta_1 -2\theta_1^*} (M^{-1})_{1,2} \\
         & + \mu_{1,5} \mu_{1,1}^{*} e^{\theta_1 -2\theta_1^*} (M^{-1})_{2,1} + \mu_{1,5} \mu_{1,1}^{*} e^{\theta_1 -2\theta_1^*} (M^{-1})_{2,2}            ),  \\
   \end{aligned}
\right.
\end{equation}
\noindent
{\rotatebox{0}{\includegraphics[width=3.6cm,height=3.5cm,angle=0]{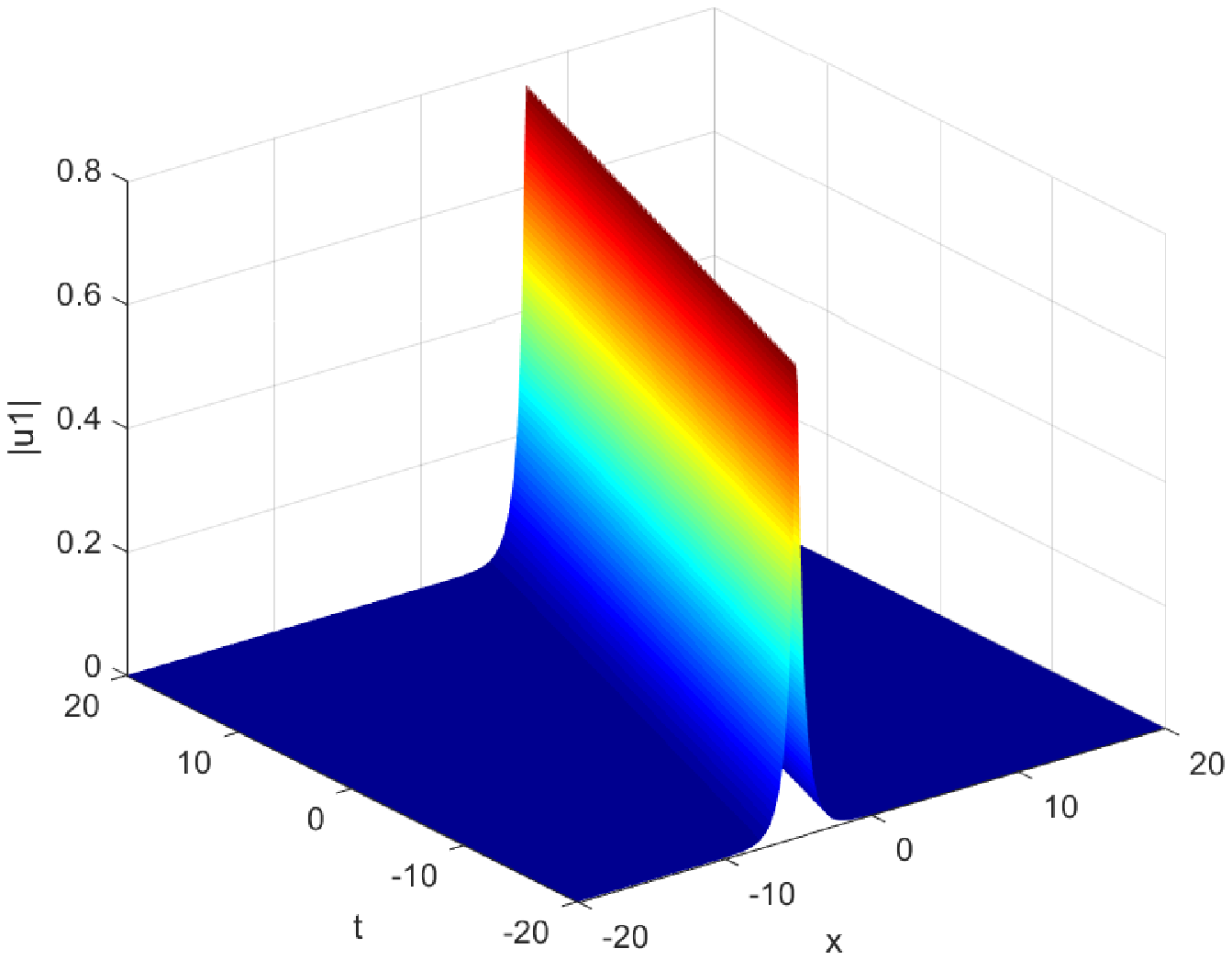}}}
~~~~
{\rotatebox{0}{\includegraphics[width=3.6cm,height=3.5cm,angle=0]{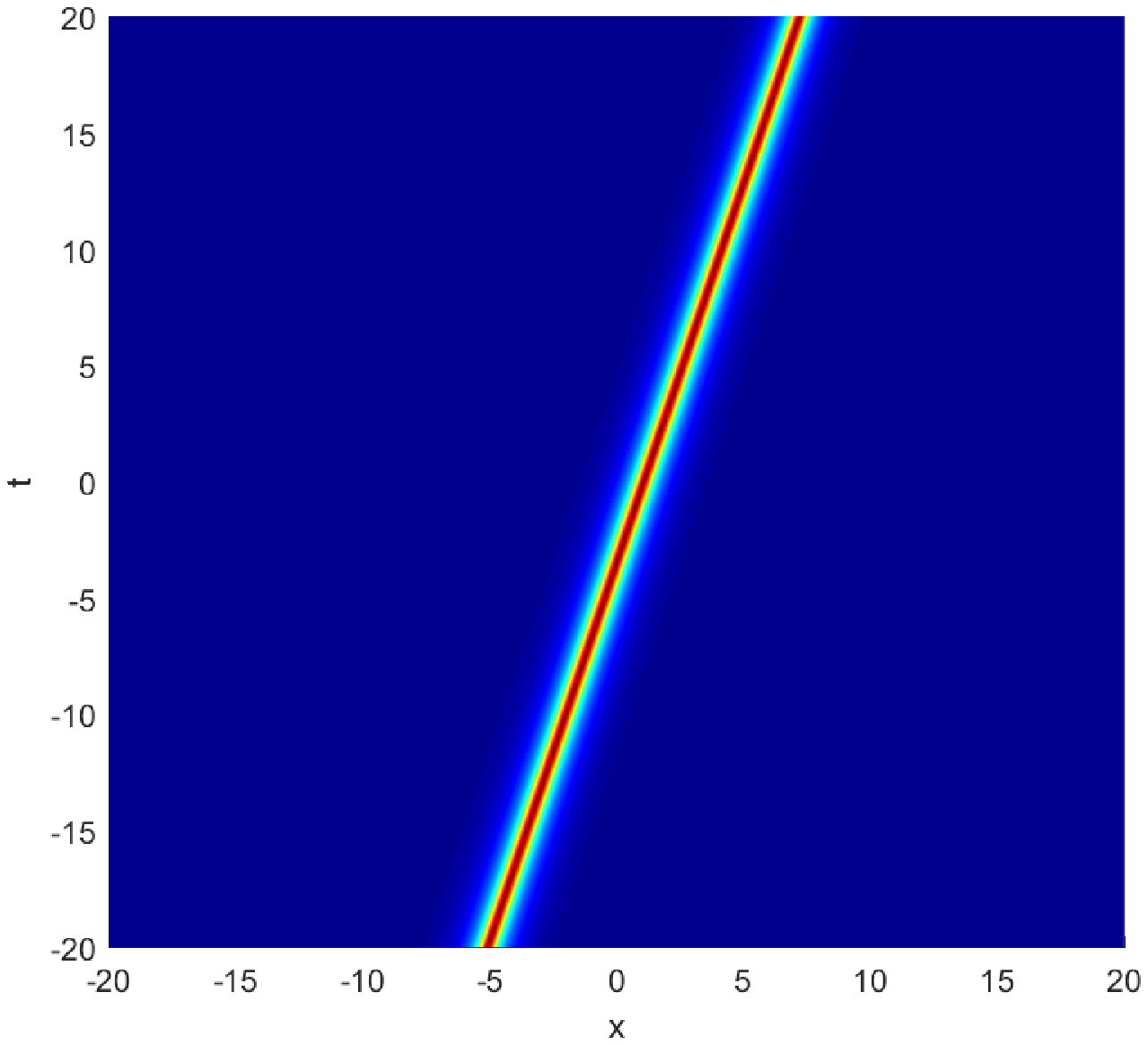}}}
~~~~
{\rotatebox{0}{\includegraphics[width=3.6cm,height=3.5cm,angle=0]{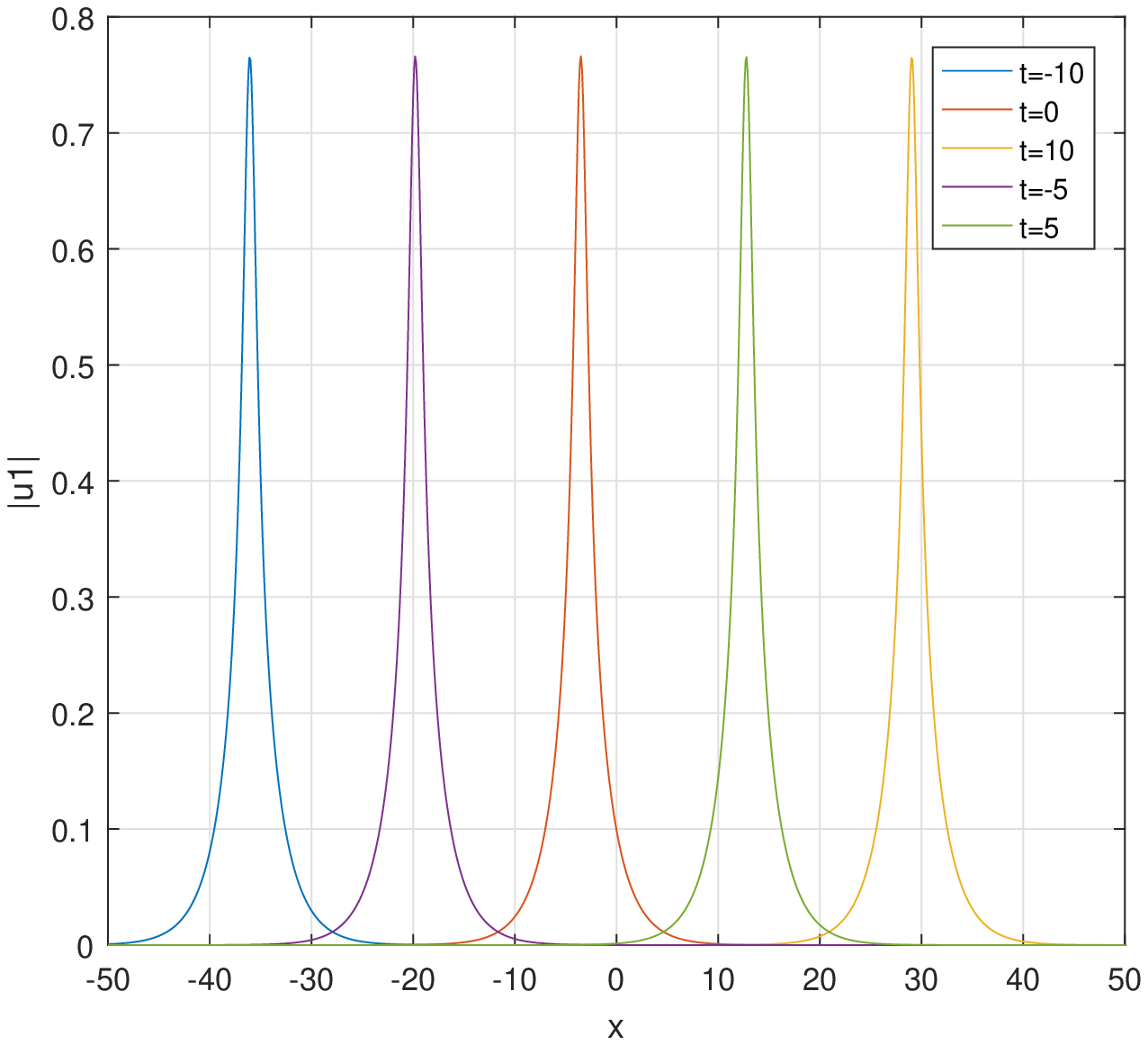}}}

$\ \qquad~~~~~~(\textbf{a})\qquad \ \qquad\qquad\qquad\qquad~(\textbf{b})
\ \qquad\qquad\qquad\qquad\qquad~(\textbf{c})$\\
\noindent
{\rotatebox{0}{\includegraphics[width=3.6cm,height=3.5cm,angle=0]{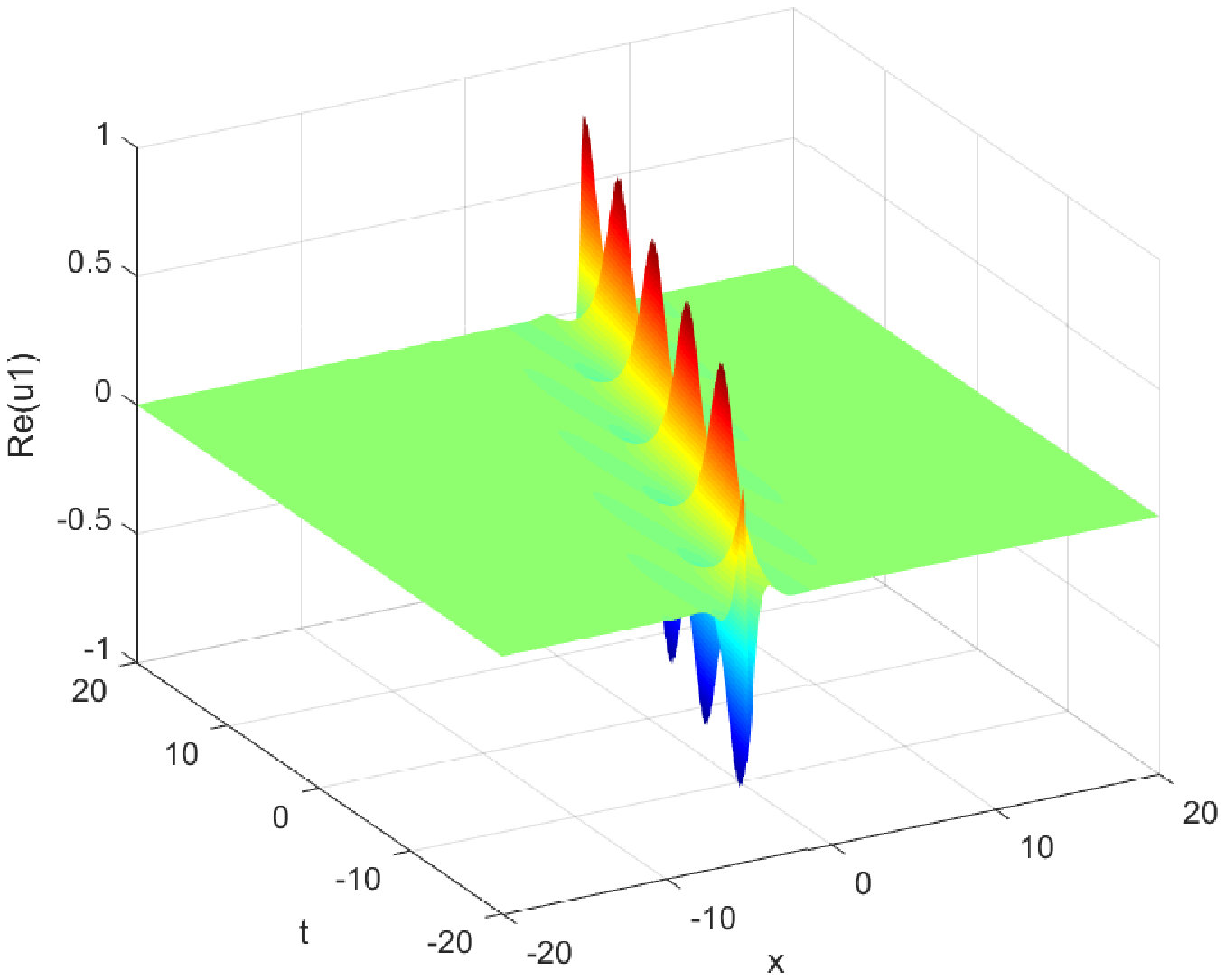}}}
~~~~
{\rotatebox{0}{\includegraphics[width=3.6cm,height=3.5cm,angle=0]{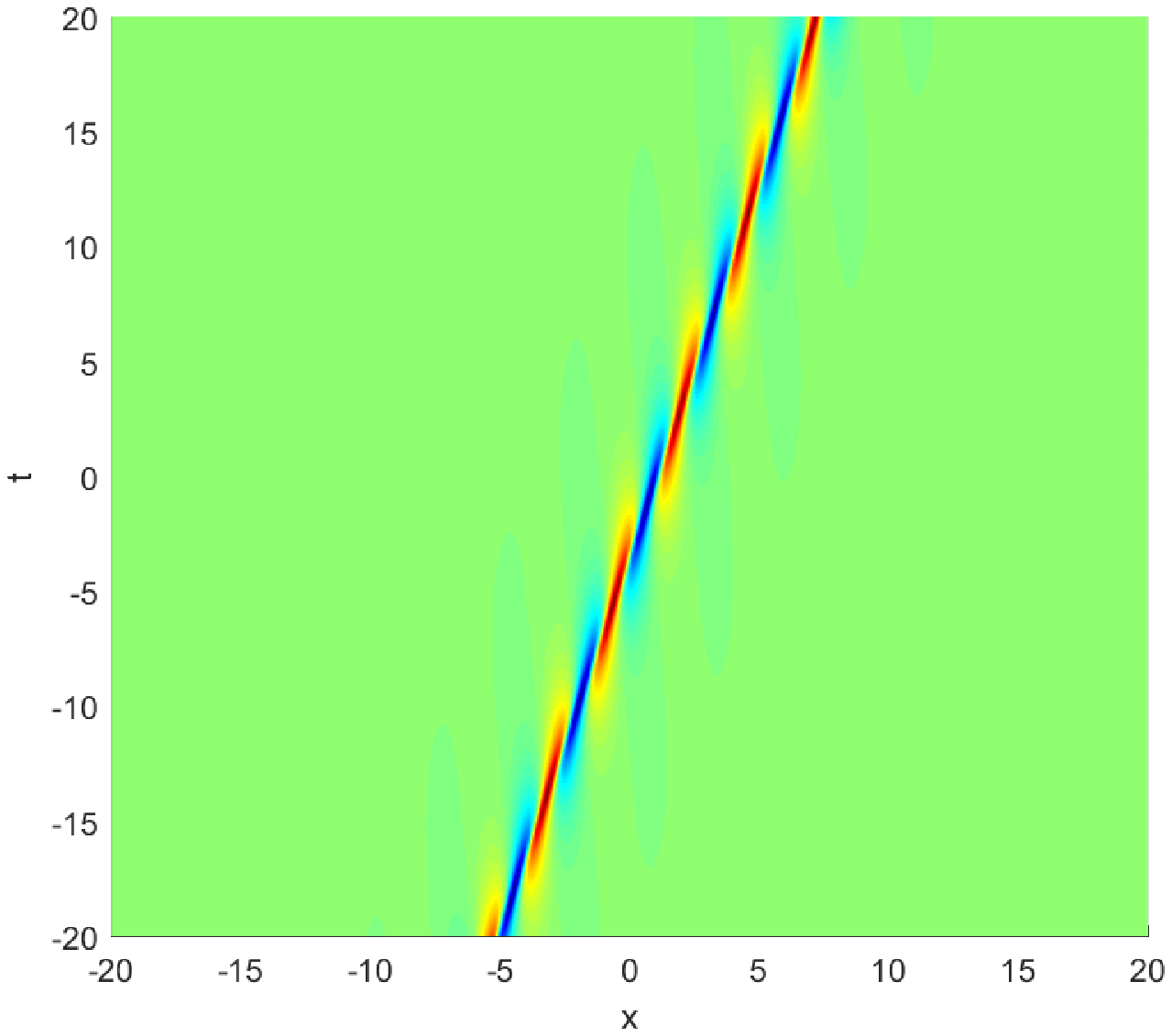}}}
~~~~
{\rotatebox{0}{\includegraphics[width=3.6cm,height=3.5cm,angle=0]{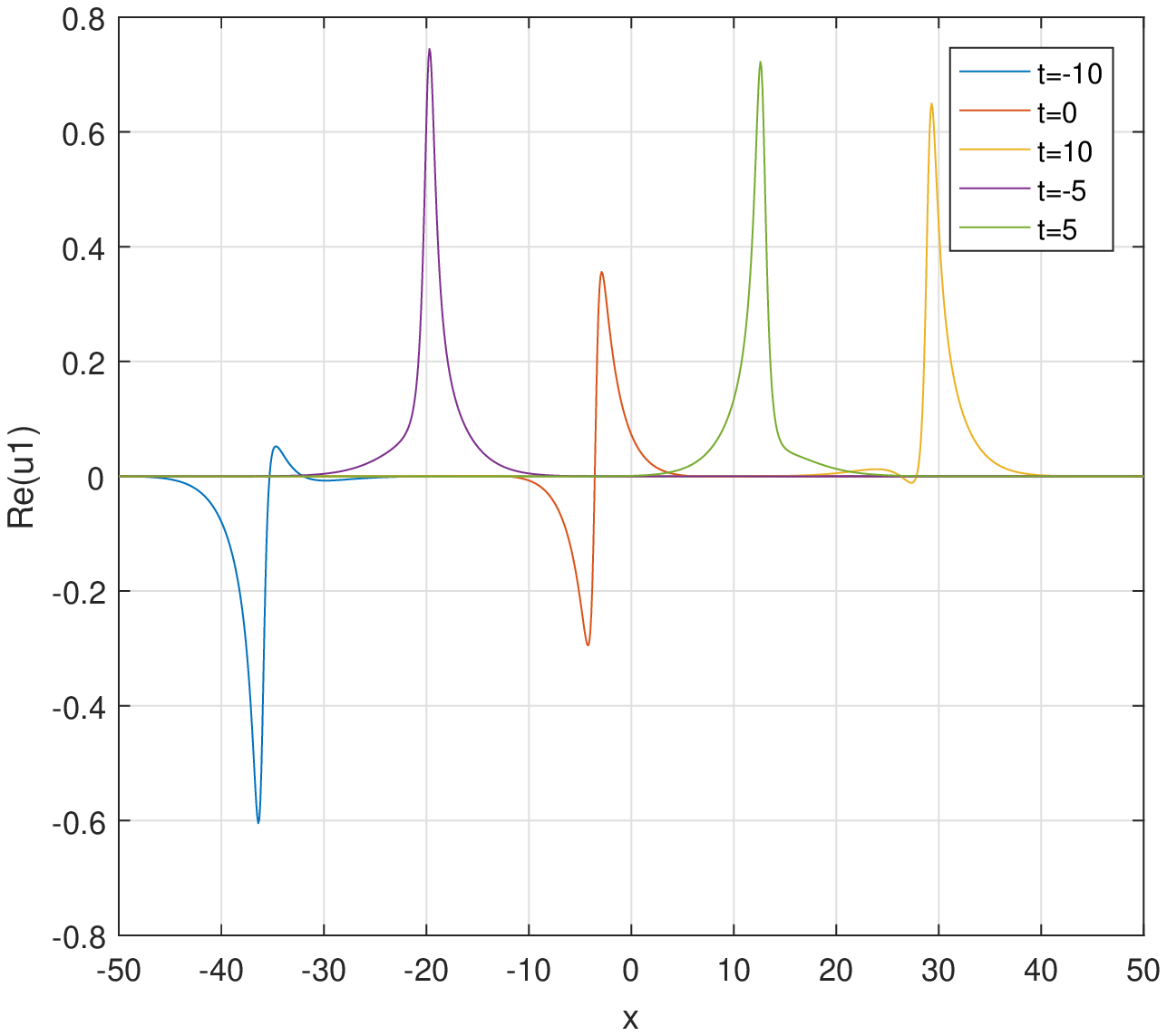}}}

$\ \qquad~~~~~~(\textbf{d})\qquad \ \qquad\qquad\qquad\qquad~(\textbf{e})
\ \qquad\qquad\qquad\qquad\qquad~(\textbf{f})$\\
\noindent
{\rotatebox{0}{\includegraphics[width=3.6cm,height=3.0cm,angle=0]{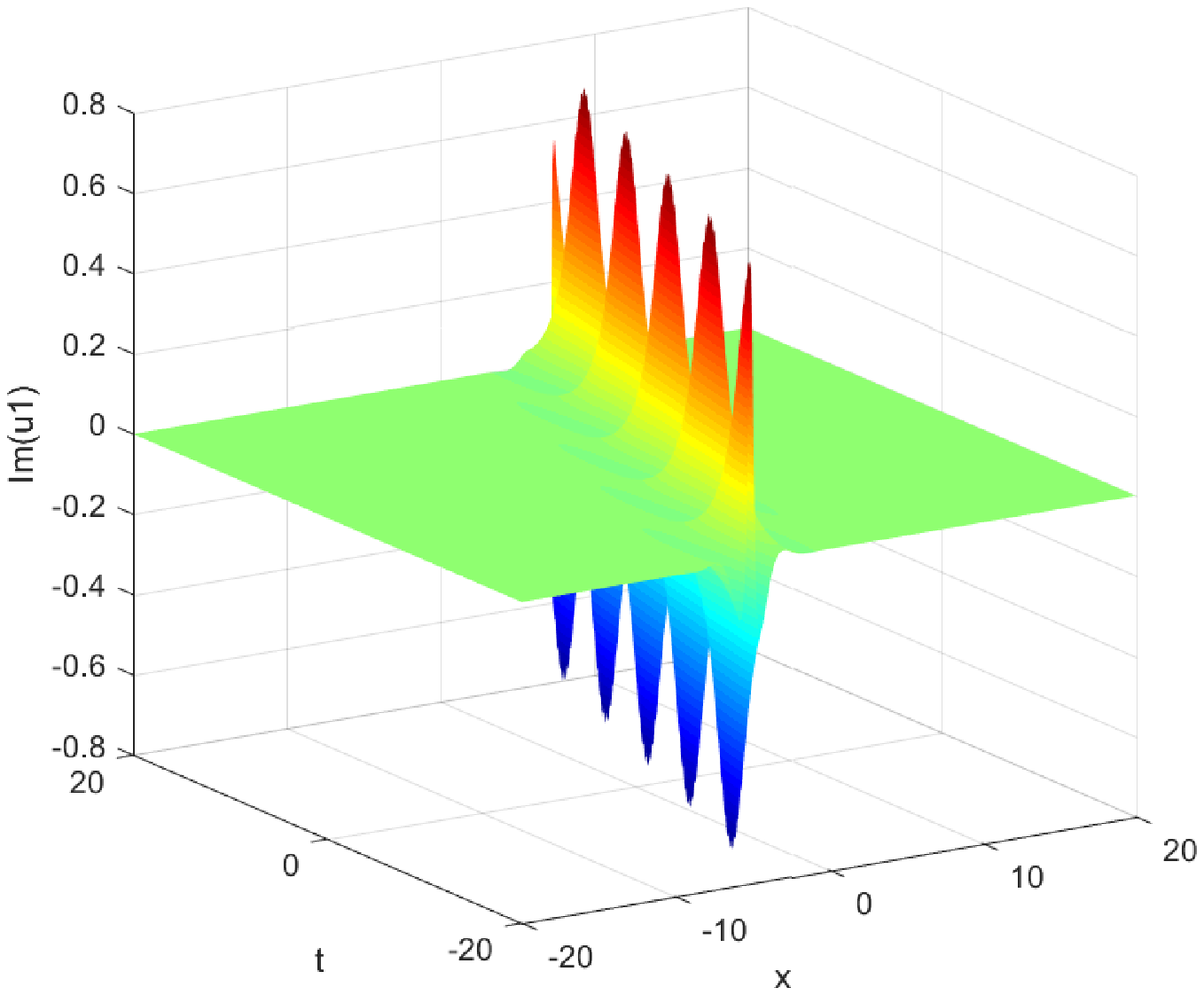}}}
~~~~
{\rotatebox{0}{\includegraphics[width=3.6cm,height=3.0cm,angle=0]{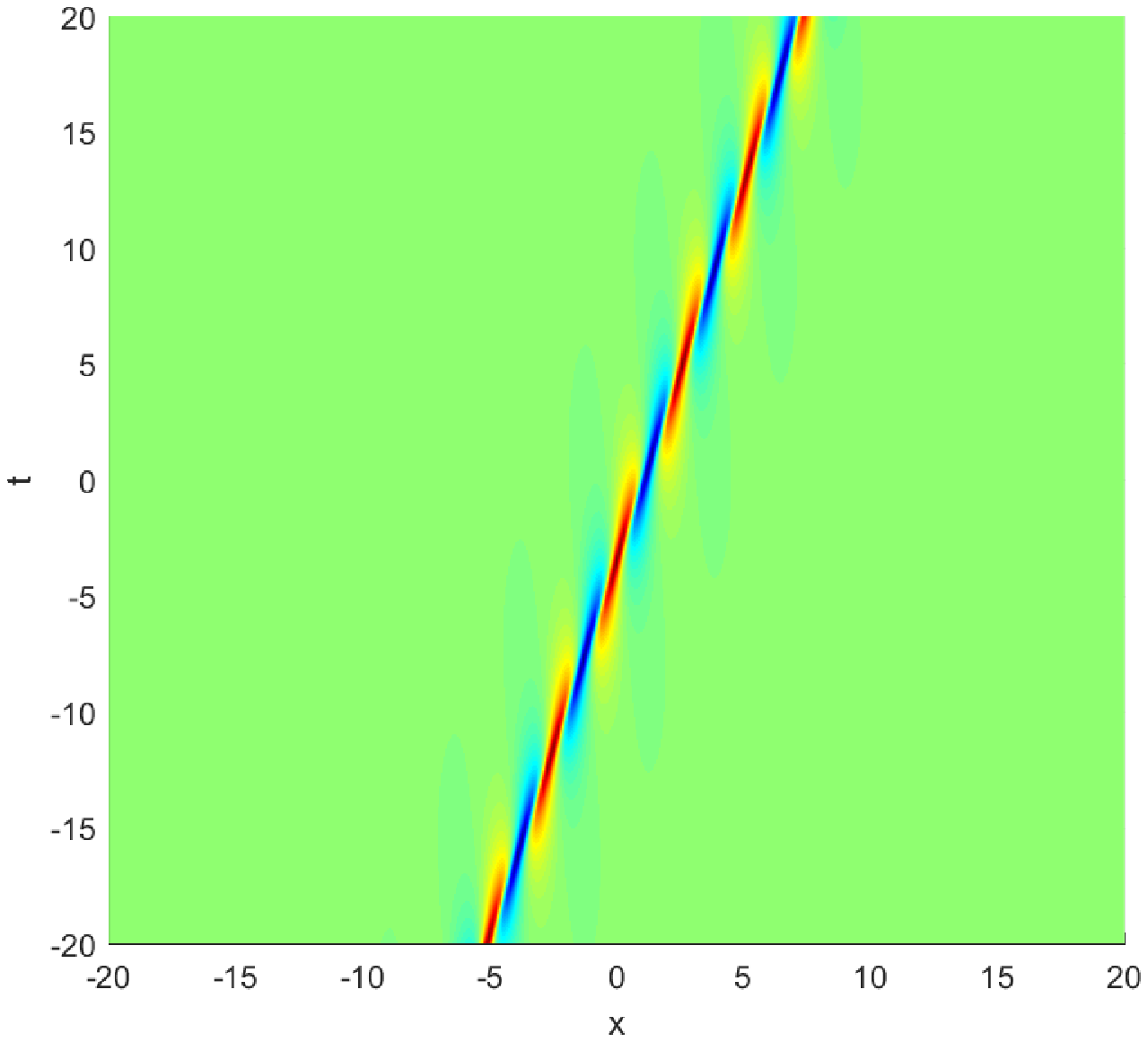}}}
~~~~
{\rotatebox{0}{\includegraphics[width=3.6cm,height=3.0cm,angle=0]{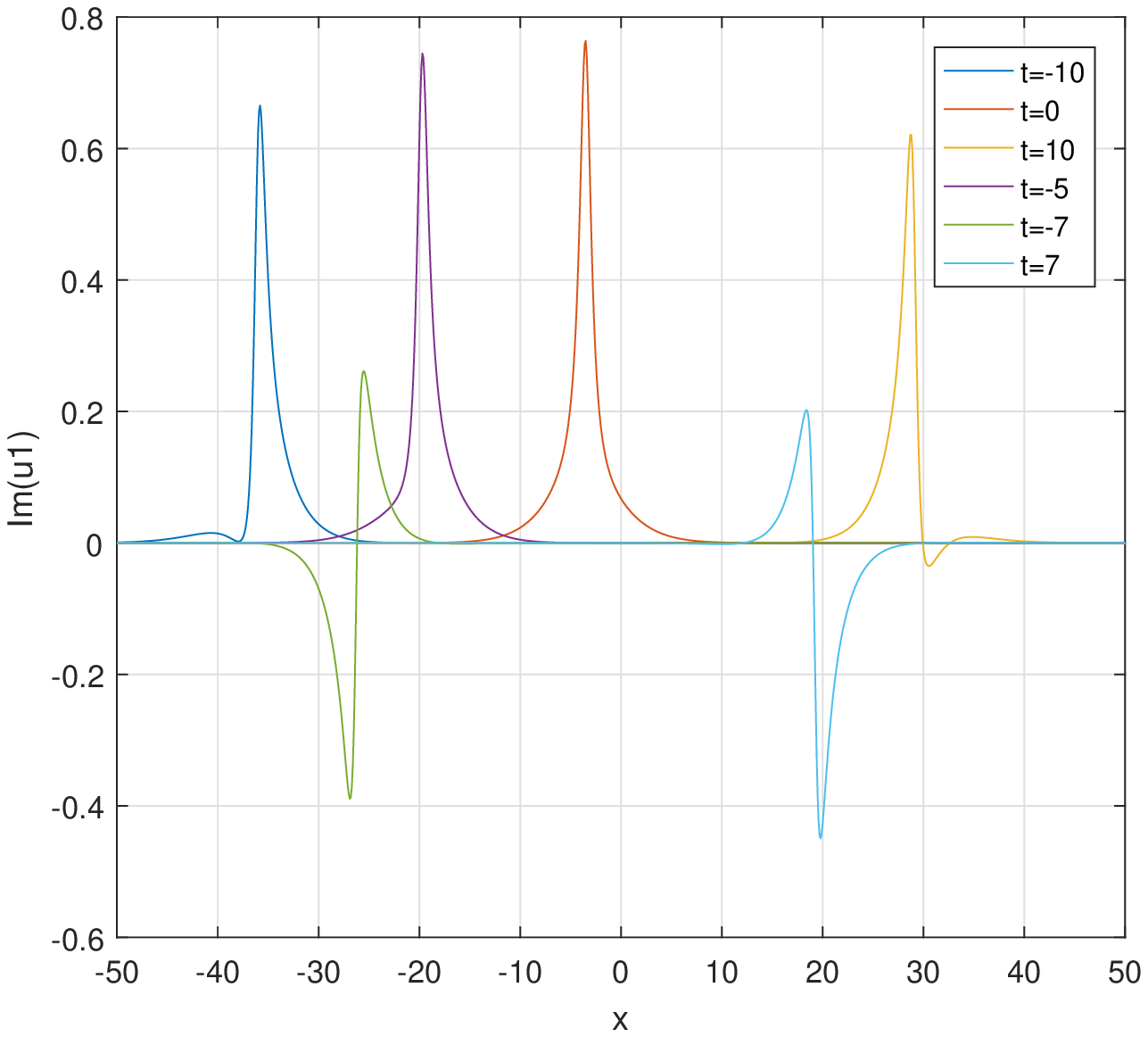}}}

$\ \qquad~~~~~~(\textbf{g})\qquad \ \qquad\qquad\qquad\qquad~(\textbf{h})
\ \qquad\qquad\qquad\qquad\qquad~(\textbf{i})$\\
\noindent { \small \textbf{Figure 8.} One-hump solutions to Eq. \eqref{so3} with parameters $\zeta_1=1.5+0.5i$, $\zeta_2=-\zeta_1$, $\mu_{1,1}=1.1+0.i$, $\mu_{1,2}=0.9+1.5i$,  $\mu_{1,3}=0.2+0.5i$,  $\mu_{1,4}=1.2+1.6i$ and $\mu_{1,5}=0.7+0.8i$.
$\textbf{(a)(b)(c)}$:  the local structure, density and wave propagation of  $|u_1(x,t)|$,
$\textbf{(d)(e)(f)}$:  the local structure, density and wave propagation of  $\mathrm{Re}(u_1)$,
$\textbf{(g)(h)(i)}$:  the local structure, density and wave propagation of  $\mathrm{Im}(u_1)$.}

\noindent
{\rotatebox{0}{\includegraphics[width=3.6cm,height=2.8cm,angle=0]{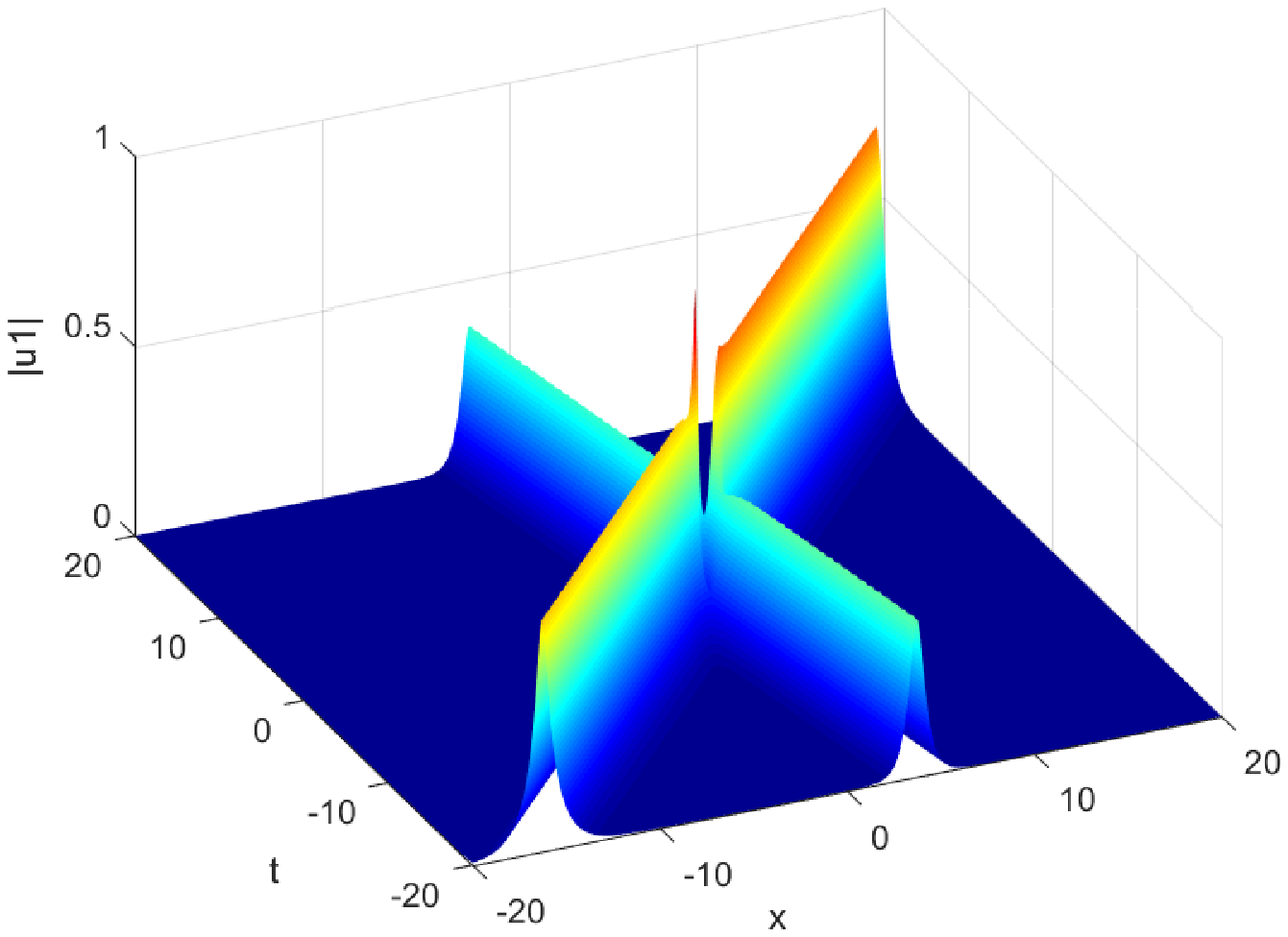}}}
~~~~
{\rotatebox{0}{\includegraphics[width=3.6cm,height=2.8cm,angle=0]{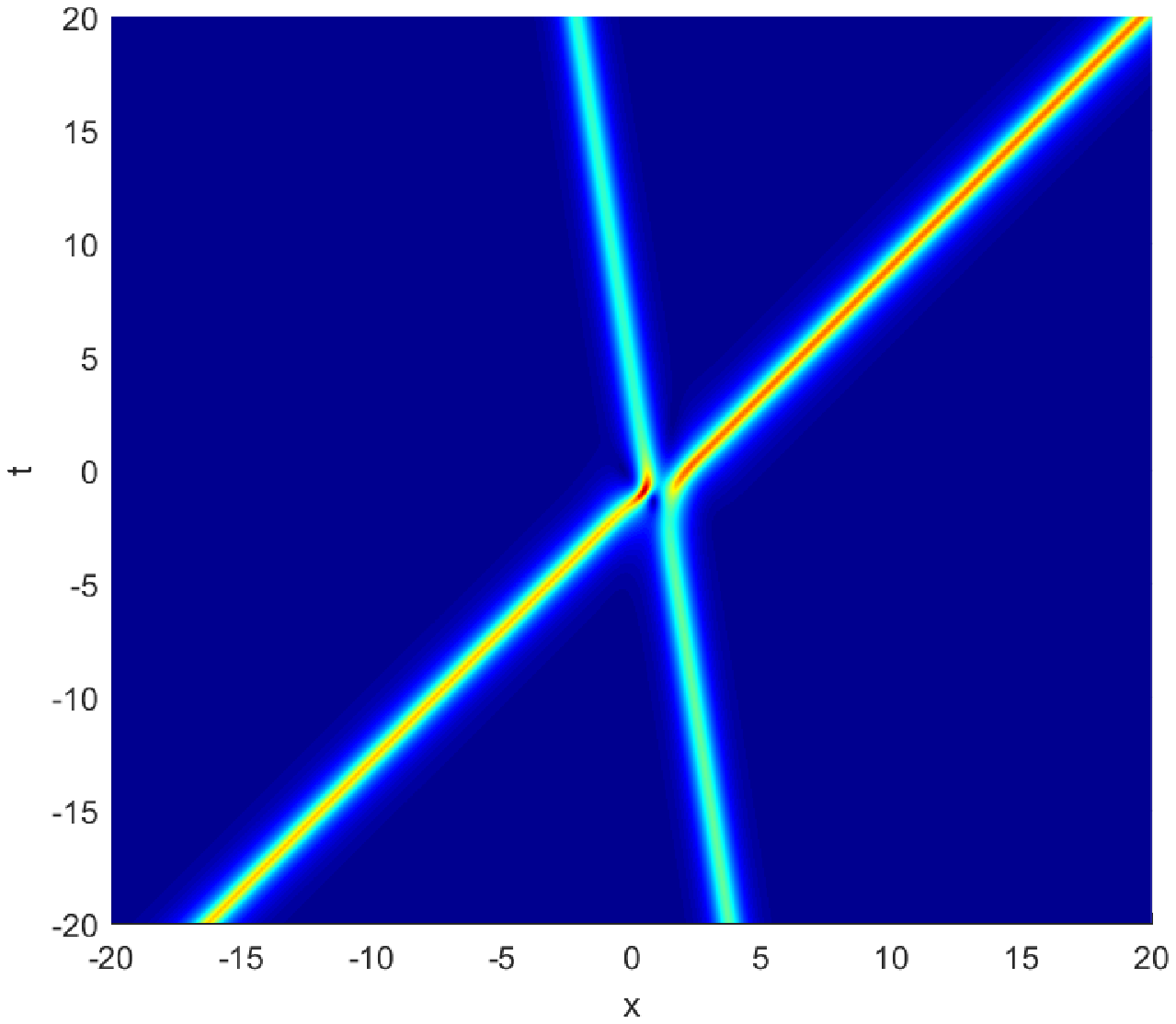}}}
~~~~
{\rotatebox{0}{\includegraphics[width=3.6cm,height=2.8cm,angle=0]{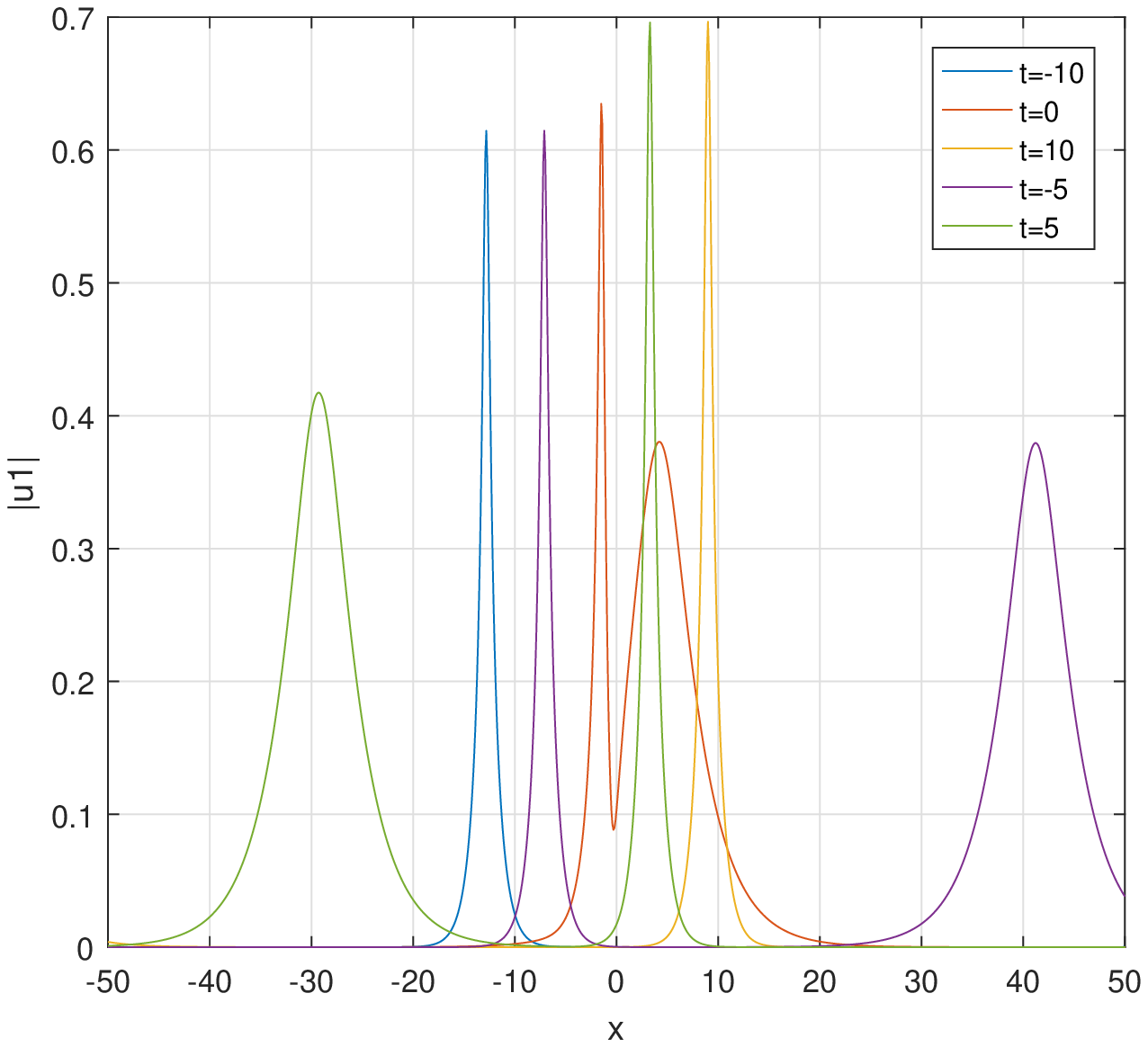}}}

$\ \qquad~~~~~~(\textbf{a})\qquad \ \qquad\qquad\qquad\qquad~(\textbf{b})
\ \qquad\qquad\qquad\qquad\qquad~(\textbf{c})$\\
\noindent
{\rotatebox{0}{\includegraphics[width=3.6cm,height=3.0cm,angle=0]{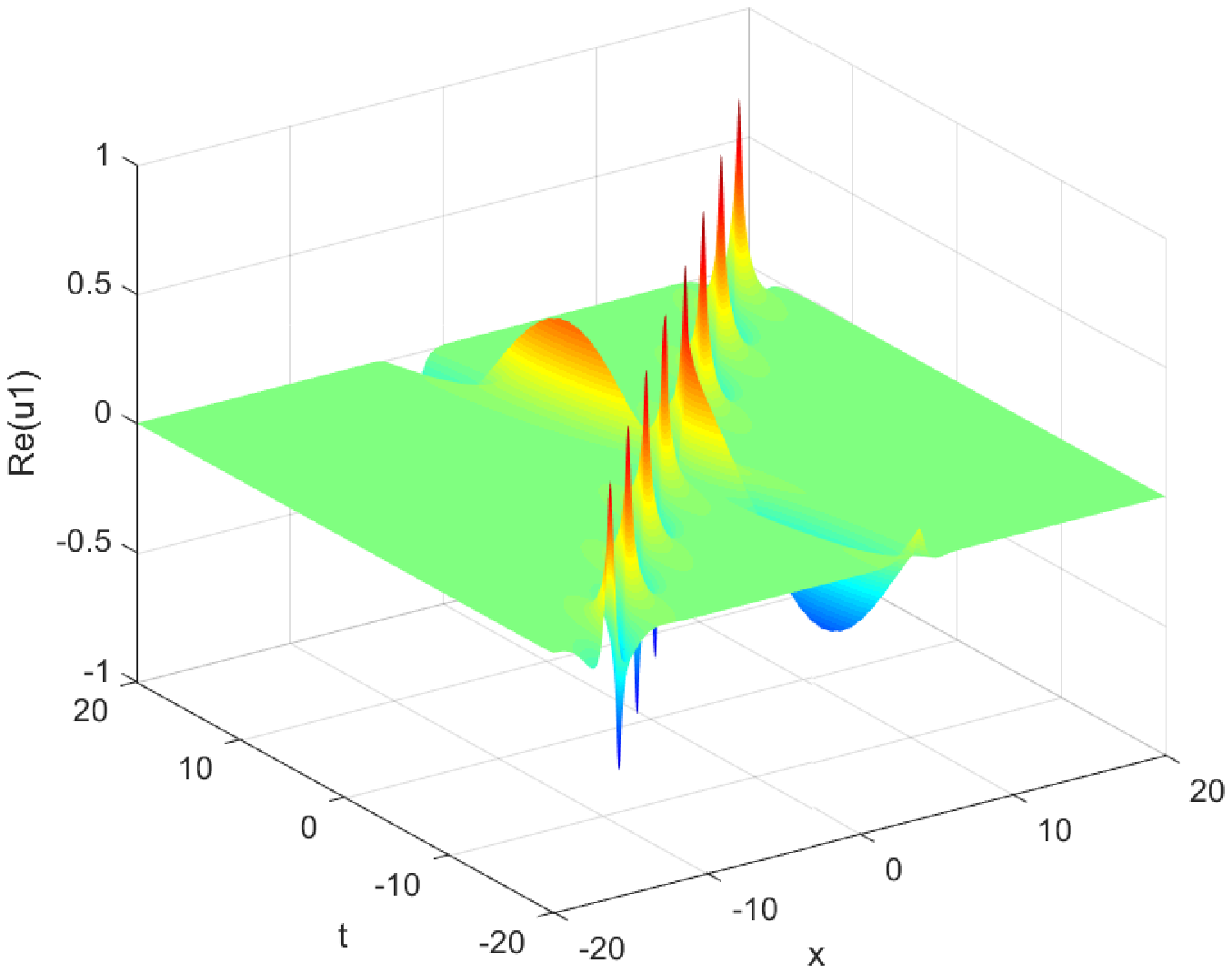}}}
~~~~
{\rotatebox{0}{\includegraphics[width=3.6cm,height=3.0cm,angle=0]{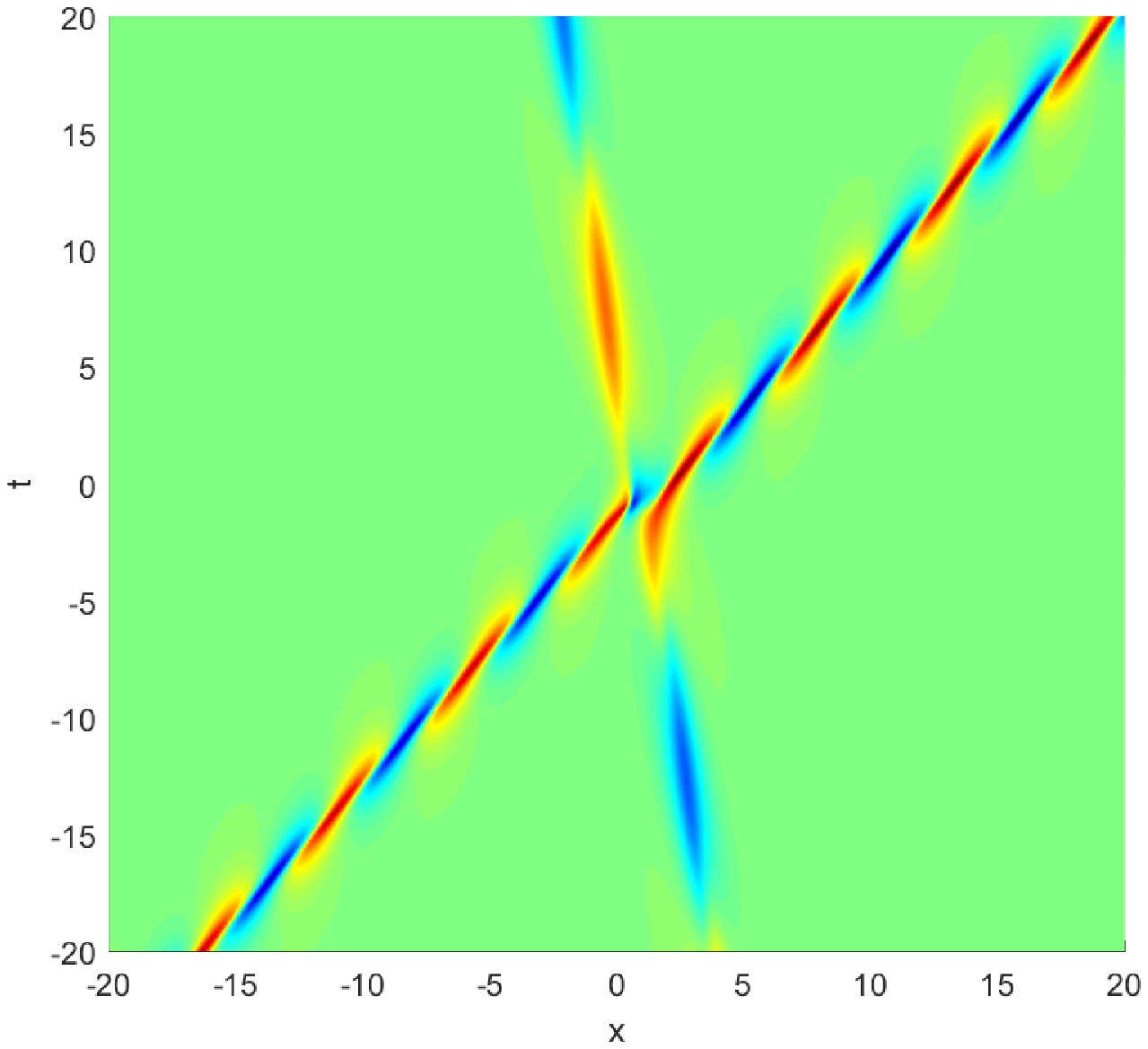}}}
~~~~
{\rotatebox{0}{\includegraphics[width=3.6cm,height=3.0cm,angle=0]{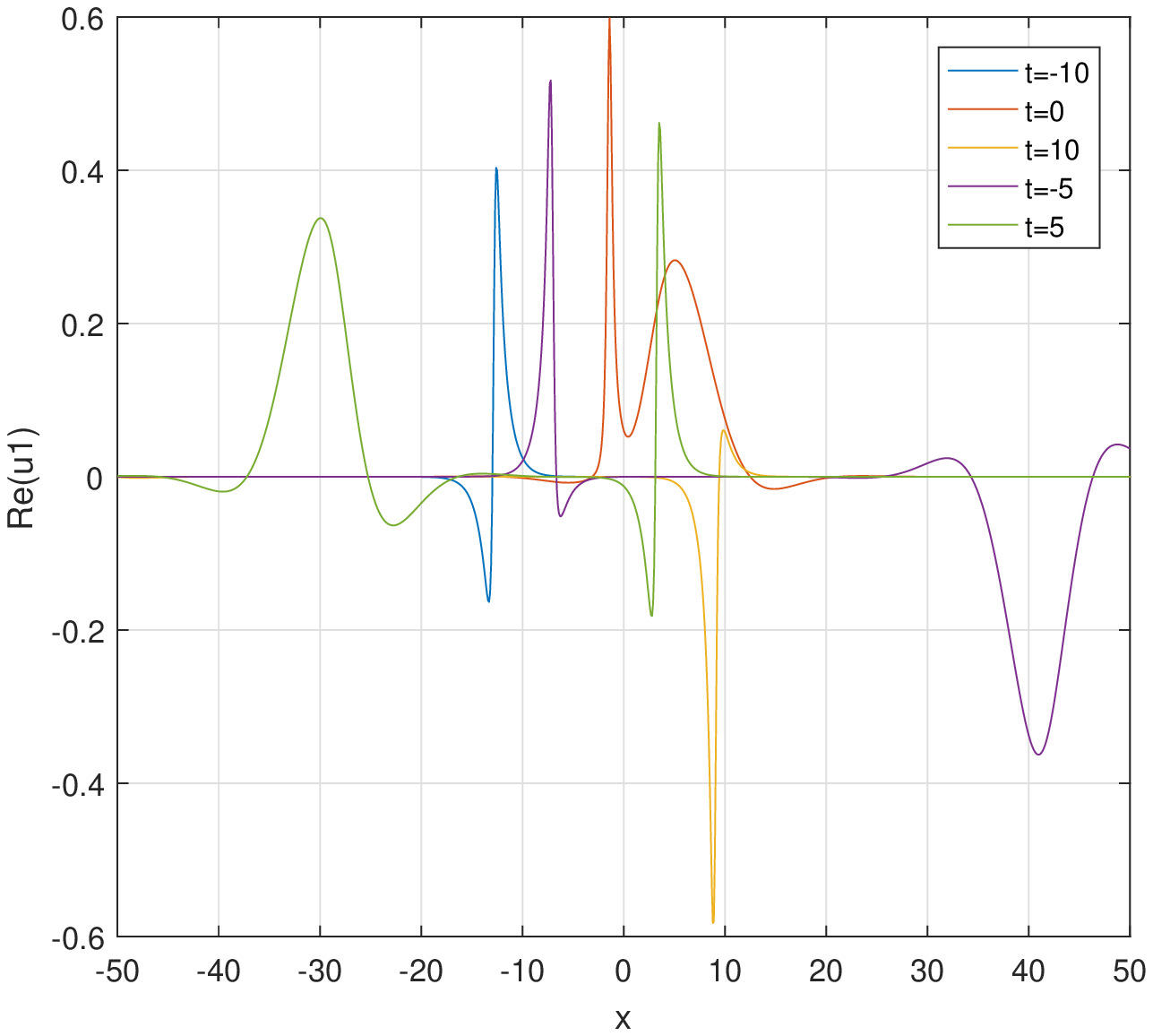}}}

$\ \qquad~~~~~~(\textbf{d})\qquad \ \qquad\qquad\qquad\qquad~(\textbf{e})
\ \qquad\qquad\qquad\qquad\qquad~(\textbf{f})$\\
\noindent
{\rotatebox{0}{\includegraphics[width=3.6cm,height=3.0cm,angle=0]{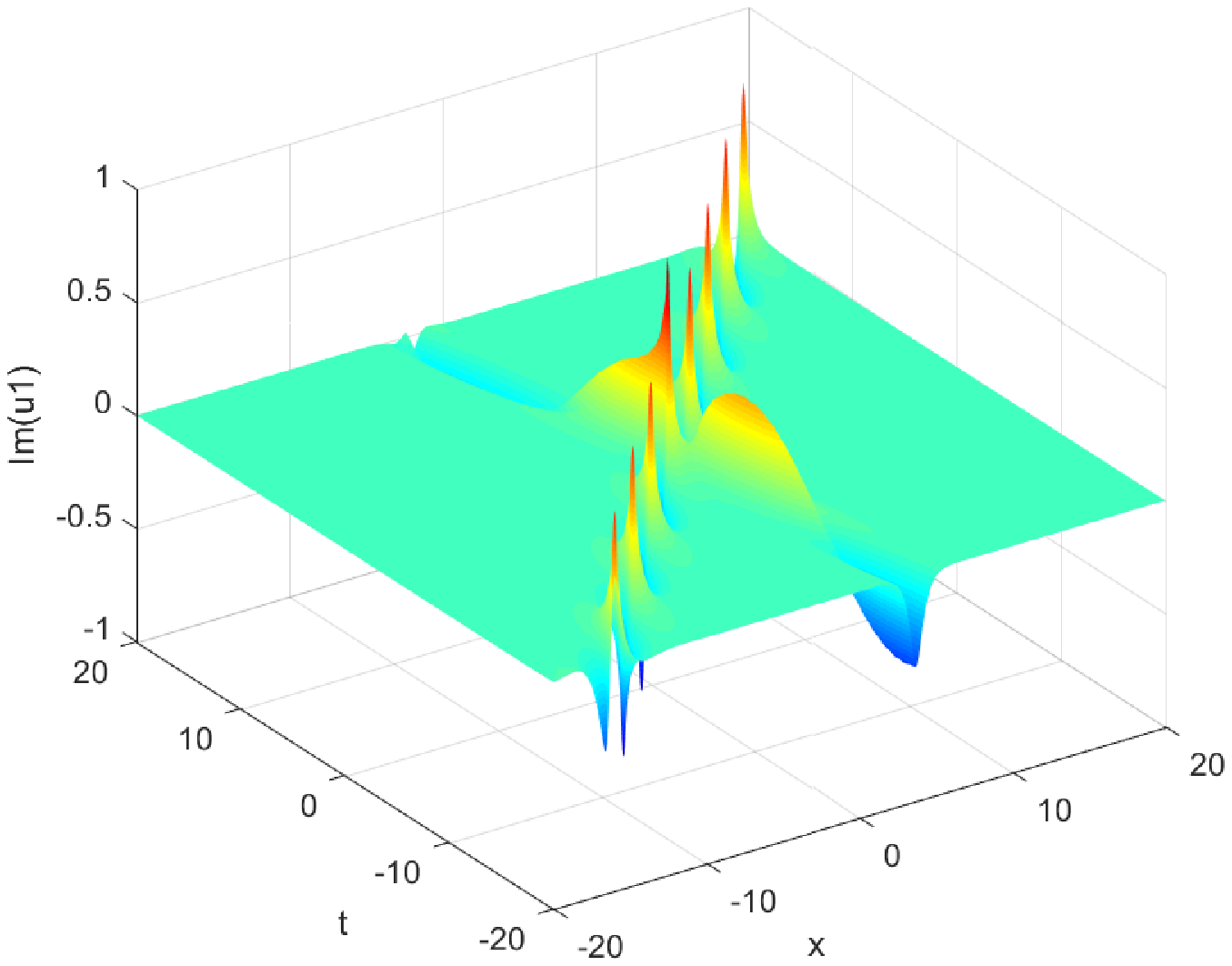}}}
~~~~
{\rotatebox{0}{\includegraphics[width=3.6cm,height=3.0cm,angle=0]{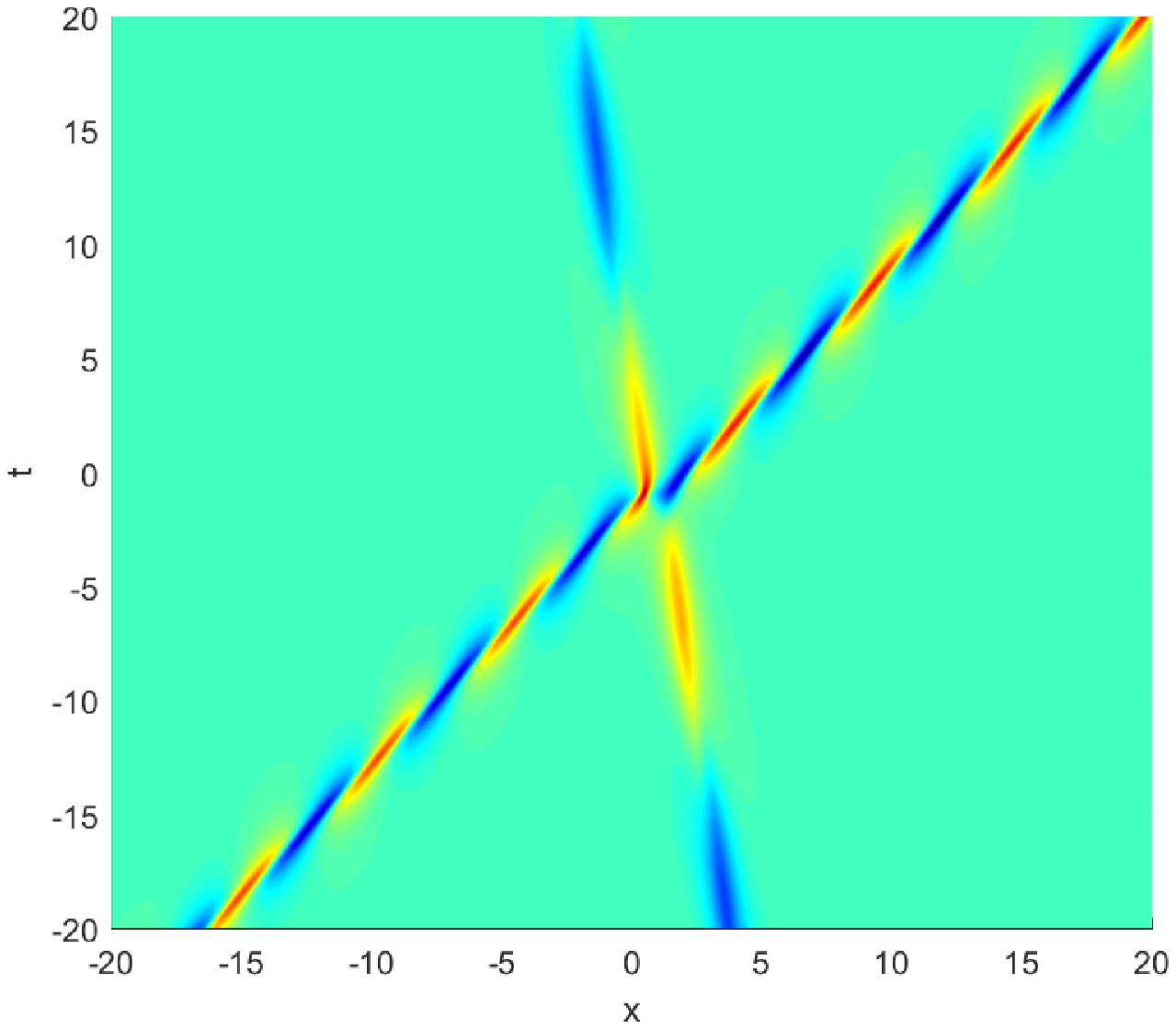}}}
~~~~
{\rotatebox{0}{\includegraphics[width=3.6cm,height=3.0cm,angle=0]{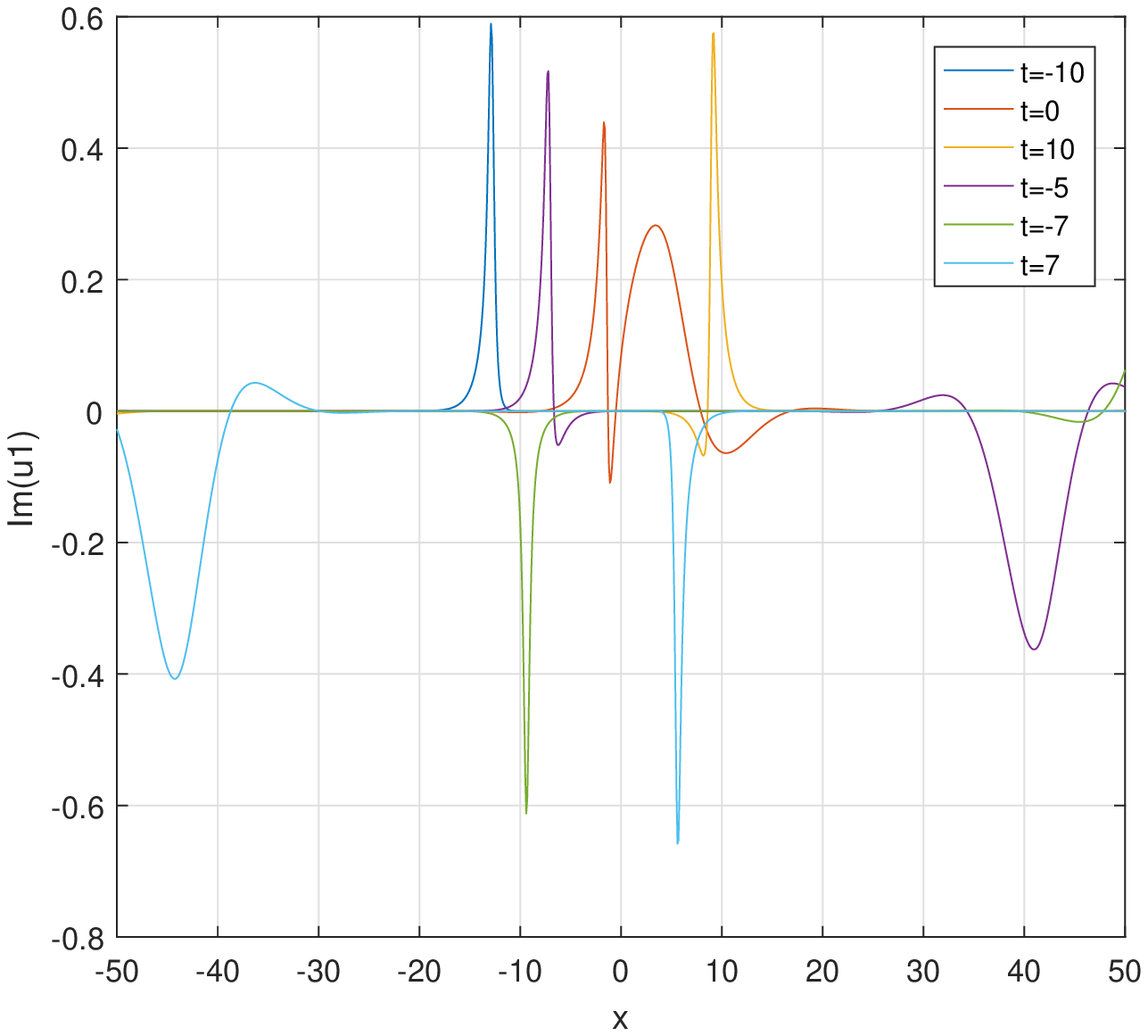}}}

$\ \qquad~~~~~~(\textbf{g})\qquad \ \qquad\qquad\qquad\qquad~(\textbf{h})
\ \qquad\qquad\qquad\qquad\qquad~(\textbf{i})$\\
\noindent { \small \textbf{Figure 9.} Two-soliton  solutions to Eq. \eqref{solu3} with parameters $\zeta_1=1.2+0.5i$, $\zeta_2=0.9+1.3i$, $\mu_{1,1}=0.7$, $\mu_{1,2}=1.5$,  $\mu_{1,3}=0.9$,  $\mu_{1,4}=1.2$, $\mu_{1,5}=0.5$, $\mu_{2,1}=0.8$, $\mu_{2,2}=1.2$,   $\mu_{2,3}=1.4$,    $\mu_{2,4}=0.9$ and    $\mu_{2,5}=0.5$.
$\textbf{(a)(b)(c)}$:  the local structure, density and wave propagation of  $|u_1(x,t)|$,
$\textbf{(d)(e)(f)}$:  the local structure, density and wave propagation of  $\mathrm{Re}(u_1)$,
$\textbf{(g)(h)(i)}$:  the local structure, density and wave propagation of  $\mathrm{Im}(u_1)$.}  \\

where
\begin{equation}
  \left\{
   \begin{aligned}
       m_{1,1}  = &\frac{-|\mu_{1,1}|^2 e^{-2\theta_1-2\theta_1^{*}}}{\zeta_1-\zeta_1^{*}}+\frac{(a_{11}\mu_{1,2}^{*}e^{\theta_1^{*}}+a_{21}\mu_{1,3}^{*}e^{\theta_1^{*}}+a_{31}\mu_{1,4}^{*}e^{\theta_1^{*}}+a_{41}\mu_{1,5}^{*}e^{\theta_1^{*}})\mu_{1,2}e^{\theta_1}}{\zeta_1-\zeta_1^{*}}   \\
& +\frac{(a_{21}^{*}\mu_{1,2}^{*}e^{\theta_1^{*}}+a_{22}\mu_{1,3}^{*}e^{\theta_1^{*}}+a_{32}\mu_{1,4}^{*}e^{\theta_1^{*}}+a_{42}\mu_{1,5}^{*}e^{\theta_1^{*}})\mu_{1,3}e^{\theta_1}}{\zeta_1-\zeta_1^{*}}\\
 &+\frac{(a_{31}^{*}\mu_{1,2}^{*}e^{\theta_1^{*}}+a_{32}^{*}\mu_{1,3}^{*}e^{\theta_1^{*}}+a_{33}\mu_{1,4}^{*}e^{\theta_1^{*}}+a_{43}\mu_{1,5}^{*}e^{\theta_1^{*}})\mu_{1,4}e^{\theta_1}}{\zeta_1-\zeta_1^{*}}\\
&+\frac{(a_{41}^{*}\mu_{1,2}^{*}e^{\theta_1^{*}}+a_{42}^{*}\mu_{1,3}^{*}e^{\theta_1^{*}}+a_{43}^{*}\mu_{1,4}^{*}e^{\theta_1^{*}}+a_{44}\mu_{1,5}^{*}e^{\theta_1^{*}})\mu_{1,5}e^{\theta_1}}{\zeta_1-\zeta_1^{*}},\\
        m_{1,2}  = &\frac{-|\mu_{1,1}|^2 e^{-2\theta_1-2\theta_1^{*}}}{\zeta_2-\zeta_1^{*}}- \frac{(a_{11}\mu_{1,2}^{*}e^{\theta_1^{*}}+a_{21}\mu_{1,3}^{*}e^{\theta_1^{*}}+a_{31}\mu_{1,4}^{*}e^{\theta_1^{*}}+a_{41}\mu_{1,5}^{*}e^{\theta_1^{*}})\mu_{1,2}e^{\theta_1}}{\zeta_2-\zeta_1^{*}}   \\
& -\frac{(a_{21}^{*}\mu_{1,2}^{*}e^{\theta_1^{*}}+a_{22}\mu_{1,3}^{*}e^{\theta_1^{*}}+a_{32}\mu_{1,4}^{*}e^{\theta_1^{*}}+a_{42}\mu_{1,5}^{*}e^{\theta_1^{*}})\mu_{1,3}e^{\theta_1}}{\zeta_2-\zeta_1^{*}}\\
&-\frac{(a_{31}^{*}\mu_{1,2}^{*}e^{\theta_1^{*}}+a_{32}^{*}\mu_{1,3}^{*}e^{\theta_1^{*}}+a_{33}\mu_{1,4}^{*}e^{\theta_1^{*}}+a_{43}\mu_{1,5}^{*}e^{\theta_1^{*}})\mu_{1,4}e^{\theta_1}}{\zeta_2-\zeta_1^{*}}\\
&-\frac{(a_{41}^{*}\mu_{1,2}^{*}e^{\theta_1^{*}}+a_{42}^{*}\mu_{1,3}^{*}e^{\theta_1^{*}}+a_{43}^{*}\mu_{1,4}^{*}e^{\theta_1^{*}}+a_{44}\mu_{1,5}^{*}e^{\theta_1^{*}})\mu_{1,5}e^{\theta_1}}{\zeta_2-\zeta_1^{*}}, \\
       m_{2,1}=  &  \frac{-|\mu_{1,1}|^2 e^{-2\theta_1-2\theta_1^{*}}}{\zeta_1+\zeta_1^{*}}- \frac{(a_{11}\mu_{1,2}^{*}e^{\theta_1^{*}}+a_{21}\mu_{1,3}^{*}e^{\theta_1^{*}}+a_{31}\mu_{1,4}^{*}e^{\theta_1^{*}}+a_{41}\mu_{1,5}^{*}e^{\theta_1^{*}})\mu_{1,2}e^{\theta_1}}{\zeta_1+\zeta_1^{*}} \\
&-\frac{(a_{21}^{*}\mu_{1,2}^{*}e^{\theta_1^{*}}+a_{22}\mu_{1,3}^{*}e^{\theta_1^{*}}+a_{32}\mu_{1,4}^{*}e^{\theta_1^{*}}+a_{42}\mu_{1,5}^{*}e^{\theta_1^{*}})\mu_{1,3}e^{\theta_1}}{\zeta_1+\zeta_1^{*}}\\
&-\frac{(a_{31}^{*}\mu_{1,2}^{*}e^{\theta_1^{*}}+a_{32}^{*}\mu_{1,3}^{*}e^{\theta_1^{*}}+a_{33}\mu_{1,4}^{*}e^{\theta_1^{*}}+a_{43}\mu_{1,5}^{*}e^{\theta_1^{*}})\mu_{1,4}e^{\theta_1}}{\zeta_1+\zeta_1^{*}}\\
&-\frac{(a_{41}^{*}\mu_{1,2}^{*}e^{\theta_1^{*}}+a_{42}^{*}\mu_{1,3}^{*}e^{\theta_1^{*}}+a_{43}^{*}\mu_{1,4}^{*}e^{\theta_1^{*}}+a_{44}\mu_{1,5}^{*}e^{\theta_1^{*}})\mu_{1,5}e^{\theta_1}}{\zeta_1+\zeta_1^{*}}, \\
m_{2,2}=  &  \frac{-|\mu_{1,1}|^2 e^{-2\theta_1-2\theta_1^{*}}}{\zeta_2+\zeta_1^{*}}   + \frac{(a_{11}\mu_{1,2}^{*}e^{\theta_1^{*}}+a_{21}\mu_{1,3}^{*}e^{\theta_1^{*}}+a_{31}\mu_{1,4}^{*}e^{\theta_1^{*}}+a_{41}\mu_{1,5}^{*}e^{\theta_1^{*}})\mu_{1,2}e^{\theta_1}}{\zeta_2+\zeta_1^{*}}   \\
& + \frac{(a_{21}^{*}\mu_{1,2}^{*}e^{\theta_1^{*}}+a_{22}\mu_{1,3}^{*}e^{\theta_1^{*}}+a_{32}\mu_{1,4}^{*}e^{\theta_1^{*}}+a_{42}\mu_{1,5}^{*}e^{\theta_1^{*}})\mu_{1,3}e^{\theta_1}}{\zeta_2+\zeta_1^{*}}\\
 &+\frac{(a_{31}^{*}\mu_{1,2}^{*}e^{\theta_1^{*}}+a_{32}^{*}\mu_{1,3}^{*}e^{\theta_1^{*}}+a_{33}\mu_{1,4}^{*}e^{\theta_1^{*}}+a_{43}\mu_{1,5}^{*}e^{\theta_1^{*}})\mu_{1,4}e^{\theta_1}}{\zeta_2+\zeta_1^{*}} \\
&+ \frac{(a_{41}^{*}\mu_{1,2}^{*}e^{\theta_1^{*}}+a_{42}^{*}\mu_{1,3}^{*}e^{\theta_1^{*}}+a_{43}^{*}\mu_{1,4}^{*}e^{\theta_1^{*}}+a_{44}\mu_{1,5}^{*}e^{\theta_1^{*}})\mu_{1,5}e^{\theta_1}}{\zeta_2+\zeta_1^{*}} . \\
   \end{aligned}
\right.
\end{equation}

When $N_1=2$,  we can express the solutions to Eq. \eqref{exact}  explicitly
\begin{equation}\label{solu3}
  \left\{
   \begin{aligned}
        u_1(x,t) =& i (-\mu_{1,2} \mu_{1,1}^{*} e^{\theta_1 -2\theta_1^*} (M^{-1})_{1,1}
        -\mu_{1,2} \mu_{2,1}^{*} e^{\theta_1 -2\theta_2^*} (M^{-1})_{1,2} \\
         & - \mu_{1,2} \mu_{1,1}^{*} e^{\theta_1 -2\theta_1^*} (M^{-1})_{1,3}
         - \mu_{1,2} \mu_{2,1}^{*} e^{\theta_1 -2\theta_2^*} (M^{-1})_{1,4}           \\
         &  -\mu_{2,2} \mu_{1,1}^{*} e^{\theta_2 -2\theta_1^*} (M^{-1})_{2,1}
        -\mu_{2,2} \mu_{2,1}^{*} e^{\theta_2 -2\theta_2^*} (M^{-1})_{2,2} \\
         & - \mu_{2,2} \mu_{1,1}^{*} e^{\theta_2 -2\theta_1^*} (M^{-1})_{2,3}
         - \mu_{2,2} \mu_{2,1}^{*} e^{\theta_2 -2\theta_2^*} (M^{-1})_{2,4}      \\
         &  +\mu_{1,2} \mu_{1,1}^{*} e^{\theta_1 -2\theta_1^*} (M^{-1})_{3,1}
        +\mu_{1,2} \mu_{2,1}^{*} e^{\theta_1 -2\theta_2^*} (M^{-1})_{3,2} \\
         & +\mu_{1,2} \mu_{1,1}^{*} e^{\theta_1 -2\theta_1^*} (M^{-1})_{3,3}
         +\mu_{1,2} \mu_{2,1}^{*} e^{\theta_1 -2\theta_2^*} (M^{-1})_{3,4}      \\
                  &  +\mu_{2,2} \mu_{1,1}^{*} e^{\theta_2 -2\theta_1^*} (M^{-1})_{4,1}
        +\mu_{2,2} \mu_{2,1}^{*} e^{\theta_2 -2\theta_2^*} (M^{-1})_{4,2} \\
         & +\mu_{2,2} \mu_{1,1}^{*} e^{\theta_2 -2\theta_1^*} (M^{-1})_{4,3}
         +\mu_{2,2} \mu_{2,1}^{*} e^{\theta_2 -2\theta_2^*} (M^{-1})_{4,4} ),     \\
                 u_2(x,t) =& i (-\mu_{1,3} \mu_{1,1}^{*} e^{\theta_1 -2\theta_1^*} (M^{-1})_{1,1}
        -\mu_{1,3} \mu_{2,1}^{*} e^{\theta_1 -2\theta_2^*} (M^{-1})_{1,2} \\
         & - \mu_{1,3} \mu_{1,1}^{*} e^{\theta_1 -2\theta_1^*} (M^{-1})_{1,3}
         - \mu_{1,3} \mu_{2,1}^{*} e^{\theta_1 -2\theta_2^*} (M^{-1})_{1,4}           \\
         &  -\mu_{2,3} \mu_{1,1}^{*} e^{\theta_2 -2\theta_1^*} (M^{-1})_{2,1}
        -\mu_{2,3} \mu_{2,1}^{*} e^{\theta_2 -2\theta_2^*} (M^{-1})_{2,2} \\
         & - \mu_{2,3} \mu_{1,1}^{*} e^{\theta_2 -2\theta_1^*} (M^{-1})_{2,3}
         - \mu_{2,3} \mu_{2,1}^{*} e^{\theta_2 -2\theta_2^*} (M^{-1})_{2,4}      \\
         &  +\mu_{1,3} \mu_{1,1}^{*} e^{\theta_1 -2\theta_1^*} (M^{-1})_{3,1}
        +\mu_{1,3} \mu_{2,1}^{*} e^{\theta_1 -2\theta_2^*} (M^{-1})_{3,2} \\
         & +\mu_{1,3} \mu_{1,1}^{*} e^{\theta_1 -2\theta_1^*} (M^{-1})_{3,3}
         +\mu_{1,3} \mu_{2,1}^{*} e^{\theta_1 -2\theta_2^*} (M^{-1})_{3,4}      \\
                  &  +\mu_{2,3} \mu_{1,1}^{*} e^{\theta_2 -2\theta_1^*} (M^{-1})_{4,1}
        +\mu_{2,3} \mu_{2,1}^{*} e^{\theta_2 -2\theta_2^*} (M^{-1})_{4,2} \\
         & +\mu_{2,3} \mu_{1,1}^{*} e^{\theta_2 -2\theta_1^*} (M^{-1})_{4,3}
         +\mu_{2,3} \mu_{2,1}^{*} e^{\theta_2 -2\theta_2^*} (M^{-1})_{4,4}  ),    \\
         u_3(x,t) =& i (-\mu_{1,4} \mu_{1,1}^{*} e^{\theta_1 -2\theta_1^*} (M^{-1})_{1,1}
        -\mu_{1,4} \mu_{2,1}^{*} e^{\theta_1 -2\theta_2^*} (M^{-1})_{1,2} \\
         & - \mu_{1,4} \mu_{1,1}^{*} e^{\theta_1 -2\theta_1^*} (M^{-1})_{1,3}
         - \mu_{1,4} \mu_{2,1}^{*} e^{\theta_1 -2\theta_2^*} (M^{-1})_{1,4}           \\
         &  -\mu_{2,4} \mu_{1,1}^{*} e^{\theta_2 -2\theta_1^*} (M^{-1})_{2,1}
        -\mu_{2,4} \mu_{2,1}^{*} e^{\theta_2 -2\theta_2^*} (M^{-1})_{2,2} \\
         & - \mu_{2,4} \mu_{1,1}^{*} e^{\theta_2 -2\theta_1^*} (M^{-1})_{2,3}
         - \mu_{2,4} \mu_{2,1}^{*} e^{\theta_2 -2\theta_2^*} (M^{-1})_{2,4}      \\
         &  +\mu_{1,4} \mu_{1,1}^{*} e^{\theta_1 -2\theta_1^*} (M^{-1})_{3,1}
        +\mu_{1,4} \mu_{2,1}^{*} e^{\theta_1 -2\theta_2^*} (M^{-1})_{3,2} \\
         & +\mu_{1,4} \mu_{1,1}^{*} e^{\theta_1 -2\theta_1^*} (M^{-1})_{3,3}
         +\mu_{1,4} \mu_{2,1}^{*} e^{\theta_1 -2\theta_2^*} (M^{-1})_{3,4}      \\
                  &  +\mu_{2,4} \mu_{1,1}^{*} e^{\theta_2 -2\theta_1^*} (M^{-1})_{4,1}
        +\mu_{2,4} \mu_{2,1}^{*} e^{\theta_2 -2\theta_2^*} (M^{-1})_{4,2} \\
         & +\mu_{2,4} \mu_{1,1}^{*} e^{\theta_2 -2\theta_1^*} (M^{-1})_{4,3}
         +\mu_{2,4} \mu_{2,1}^{*} e^{\theta_2 -2\theta_2^*} (M^{-1})_{4,4}  ),    \\
         u_4(x,t) =& i (-\mu_{1,5} \mu_{1,1}^{*} e^{\theta_1 -2\theta_1^*} (M^{-1})_{1,1}
        -\mu_{1,5} \mu_{2,1}^{*} e^{\theta_1 -2\theta_2^*} (M^{-1})_{1,2} \\
         & - \mu_{1,5} \mu_{1,1}^{*} e^{\theta_1 -2\theta_1^*} (M^{-1})_{1,3}
         - \mu_{1,5} \mu_{2,1}^{*} e^{\theta_1 -2\theta_2^*} (M^{-1})_{1,4}           \\
         &  -\mu_{2,5} \mu_{1,1}^{*} e^{\theta_2 -2\theta_1^*} (M^{-1})_{2,1}
        -\mu_{2,5} \mu_{2,1}^{*} e^{\theta_2 -2\theta_2^*} (M^{-1})_{2,2} \\
         & - \mu_{2,5} \mu_{1,1}^{*} e^{\theta_2 -2\theta_1^*} (M^{-1})_{2,3}
         - \mu_{2,5} \mu_{2,1}^{*} e^{\theta_2 -2\theta_2^*} (M^{-1})_{2,4}      \\
         &  +\mu_{1,5} \mu_{1,1}^{*} e^{\theta_1 -2\theta_1^*} (M^{-1})_{3,1}
        +\mu_{1,5} \mu_{2,1}^{*} e^{\theta_1 -2\theta_2^*} (M^{-1})_{3,2} \\
         & +\mu_{1,5} \mu_{1,1}^{*} e^{\theta_1 -2\theta_1^*} (M^{-1})_{3,3}
         +\mu_{1,5} \mu_{2,1}^{*} e^{\theta_1 -2\theta_2^*} (M^{-1})_{3,4}      \\
                  &  +\mu_{2,5} \mu_{1,1}^{*} e^{\theta_2 -2\theta_1^*} (M^{-1})_{4,1}
        +\mu_{2,5} \mu_{2,1}^{*} e^{\theta_2 -2\theta_2^*} (M^{-1})_{4,2} \\
         & +\mu_{2,5} \mu_{1,1}^{*} e^{\theta_2 -2\theta_1^*} (M^{-1})_{4,3}
         +\mu_{2,5} \mu_{2,1}^{*} e^{\theta_2 -2\theta_2^*} (M^{-1})_{4,4}  ),    \\
   \end{aligned}
\right.
\end{equation}
where
\begin{equation}
m_{k,j}=\frac{\hat{\vartheta}_{k}\vartheta_{j}}{\zeta_j-\hat{\zeta_{k}}}, \quad 1 \leq k,j \leq 4,
\end{equation}
with
\begin{equation}
  \zeta_3=-\zeta_1, \quad \zeta_4=-\zeta_2, \quad \hat{\zeta_j}=\zeta_j^{*}, \quad 1 \leq j \leq 4.
\end{equation}

In Figs. 8 and 9, we display the localized structures, density plot and the wave propagation of the one-soliton and two-soliton solutions. It is interesting that whatever the solutions are single-soliton  or two-soliton solutions,  their real part and  image part are all breather-like solitons.

Simliar to Case 1 and Case 2,   the three soliton solutions  are given by
\begin{equation}\label{three33}
   \begin{aligned}
     u_\nu(x,t)& = i (  -\mu_{1,\nu +1} \mu_{1,1}^{*} e^{\theta_1 - 2 \theta_1^{*}} (M^{-1})_{1,1}
                    -\mu_{1,\nu +1} \mu_{2,1}^{*} e^{\theta_1 - 2 \theta_2^{*}} (M^{-1})_{1,2}
                    -\mu_{1,\nu +1} \mu_{3,1}^{*} e^{\theta_1 - 2 \theta_3^{*}} (M^{-1})_{1,3}  \\
                    & -\mu_{1,\nu +1} \mu_{1,1}^{*} e^{\theta_1 - 2 \theta_1^{*}} (M^{-1})_{1,4}
                    -\mu_{1,\nu +1} \mu_{2,1}^{*} e^{\theta_1 - 2 \theta_2^{*}} (M^{-1})_{1,5}
                     -\mu_{1,\nu +1} \mu_{3,1}^{*} e^{\theta_1 - 2 \theta_3^{*}} (M^{-1})_{1,6}    \\
                   &   -\mu_{2,\nu +1} \mu_{1,1}^{*} e^{\theta_2 - 2 \theta_1^{*}} (M^{-1})_{2,1}
                    -\mu_{2,\nu +1} \mu_{2,1}^{*} e^{\theta_2 - 2 \theta_2^{*}} (M^{-1})_{2,2}
                    -\mu_{2,\nu +1} \mu_{3,1}^{*} e^{\theta_2 - 2 \theta_3^{*}} (M^{-1})_{2,3}  \\
                  &  -\mu_{2,\nu +1} \mu_{1,1}^{*} e^{\theta_2 - 2 \theta_1^{*}} (M^{-1})_{2,4}
                    -\mu_{2,\nu +1} \mu_{2,1}^{*} e^{\theta_2 - 2 \theta_2^{*}} (M^{-1})_{2,5}
                     -\mu_{2,\nu +1} \mu_{3,1}^{*} e^{\theta_2 - 2 \theta_3^{*}} (M^{-1})_{2,6}    \\
                   &      -\mu_{3,\nu +1} \mu_{1,1}^{*} e^{\theta_3 - 2 \theta_1^{*}} (M^{-1})_{3,1}
                    -\mu_{3,\nu +1} \mu_{2,1}^{*} e^{\theta_3 - 2 \theta_2^{*}} (M^{-1})_{3,2}
                    -\mu_{3,\nu +1} \mu_{3,1}^{*} e^{\theta_3 - 2 \theta_3^{*}} (M^{-1})_{3,3}  \\
                   & -\mu_{3,\nu +1} \mu_{1,1}^{*} e^{\theta_3 - 2 \theta_1^{*}} (M^{-1})_{3,4}
                    -\mu_{3,\nu +1} \mu_{2,1}^{*} e^{\theta_3 - 2 \theta_2^{*}} (M^{-1})_{3,5}
                     -\mu_{3,\nu +1} \mu_{3,1}^{*} e^{\theta_3 - 2 \theta_3^{*}} (M^{-1})_{3,6}    \\
                 &    +\mu_{1,\nu +1} \mu_{1,1}^{*} e^{\theta_1 - 2 \theta_1^{*}} (M^{-1})_{4,1}
                    +\mu_{1,\nu +1} \mu_{2,1}^{*} e^{\theta_1 - 2 \theta_2^{*}} (M^{-1})_{4,2}
                    +\mu_{1,\nu +1} \mu_{3,1}^{*} e^{\theta_1 - 2 \theta_3^{*}} (M^{-1})_{4,3}\\
                  &  +\mu_{1,\nu +1} \mu_{1,1}^{*} e^{\theta_1 - 2 \theta_1^{*}} (M^{-1})_{4,4}
                    +\mu_{1,\nu +1} \mu_{2,1}^{*} e^{\theta_1 - 2 \theta_2^{*}} (M^{-1})_{4,5}
                     +\mu_{1,\nu +1} \mu_{3,1}^{*} e^{\theta_1 - 2 \theta_3^{*}} (M^{-1})_{4,6}   \\
                  &                        +\mu_{2,\nu +1} \mu_{1,1}^{*} e^{\theta_2 - 2 \theta_1^{*}} (M^{-1})_{5,1}
                    +\mu_{2,\nu +1} \mu_{2,1}^{*} e^{\theta_2 - 2 \theta_2^{*}} (M^{-1})_{5,2}
                    +\mu_{2,\nu +1} \mu_{3,1}^{*} e^{\theta_2 - 2 \theta_3^{*}} (M^{-1})_{5,3}  \\
                  &  +\mu_{2,\nu +1} \mu_{1,1}^{*} e^{\theta_2 - 2 \theta_1^{*}} (M^{-1})_{5,4}
                    +\mu_{2,\nu +1} \mu_{2,1}^{*} e^{\theta_2 - 2 \theta_2^{*}} (M^{-1})_{5,5}
                     +\mu_{2,\nu +1} \mu_{3,1}^{*} e^{\theta_2 - 2 \theta_3^{*}} (M^{-1})_{5,6}    \\
                   &                       +\mu_{3,\nu +1} \mu_{1,1}^{*} e^{\theta_3 - 2 \theta_1^{*}} (M^{-1})_{6,1}
                    +\mu_{3,\nu +1} \mu_{2,1}^{*} e^{\theta_3 - 2 \theta_2^{*}} (M^{-1})_{6,2}
                    +\mu_{3,\nu +1} \mu_{3,1}^{*} e^{\theta_3 - 2 \theta_3^{*}} (M^{-1})_{6,3}\\
                 &   +\mu_{3,\nu +1} \mu_{1,1}^{*} e^{\theta_3 - 2 \theta_1^{*}} (M^{-1})_{6,4}
                    +\mu_{3,\nu +1} \mu_{2,1}^{*} e^{\theta_3 - 2 \theta_2^{*}} (M^{-1})_{6,5}
                     +\mu_{3,\nu +1} \mu_{3,1}^{*} e^{\theta_3 - 2 \theta_3^{*}} (M^{-1})_{6,6}     ),
   \end{aligned}
\end{equation}
where
\begin{equation}
 \nu=1,2,3,4, \quad  m_{k,j}=\frac{\hat{\vartheta}_{k}\vartheta_{j}}{\zeta_j-\hat{\zeta_{k}}}, \quad 1 \leq k,j \leq 6,
\end{equation}
with
\begin{equation}
  \zeta_4=-\zeta_1, \quad \zeta_5=-\zeta_2,\quad \zeta_6=-\zeta_3, \quad \hat{\zeta_j}=\zeta_j^{*}, \quad 1 \leq j \leq 6.
\end{equation}

\newpage
\noindent
{\rotatebox{0}{\includegraphics[width=3.6cm,height=3.6cm,angle=0]{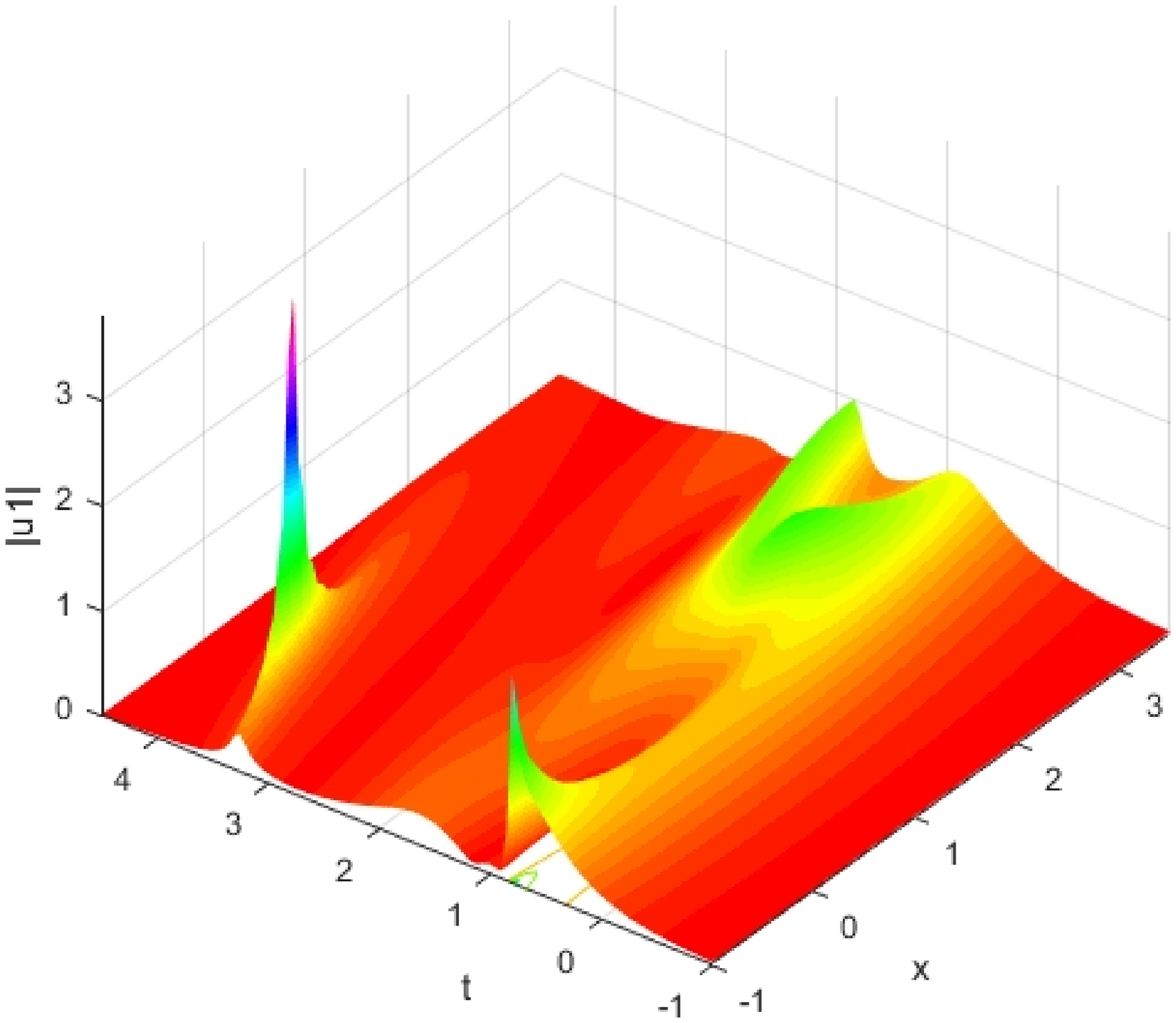}}}
~~~~
{\rotatebox{0}{\includegraphics[width=3.6cm,height=3.6cm,angle=0]{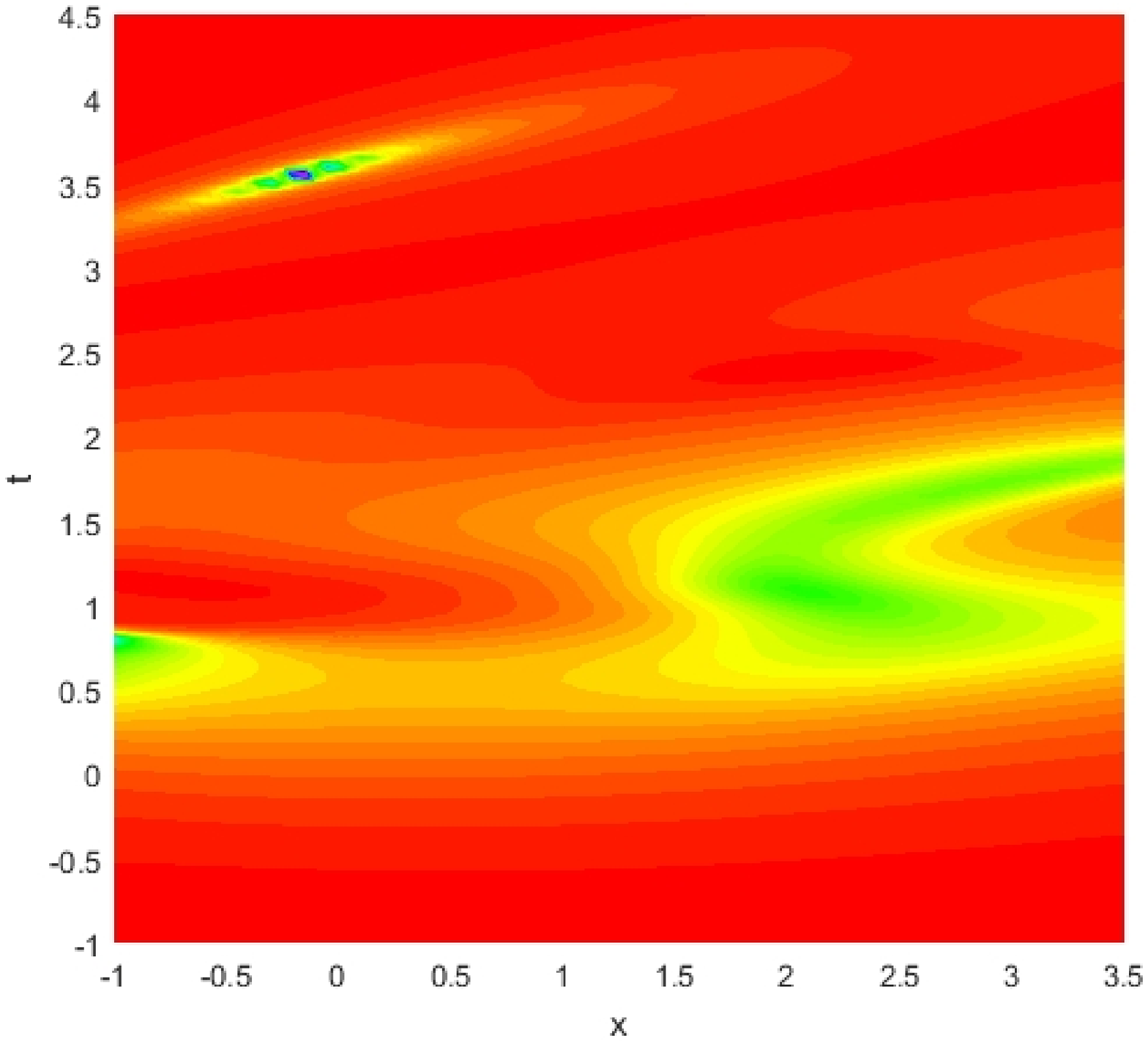}}}
~~~~
{\rotatebox{0}{\includegraphics[width=3.6cm,height=3.6cm,angle=0]{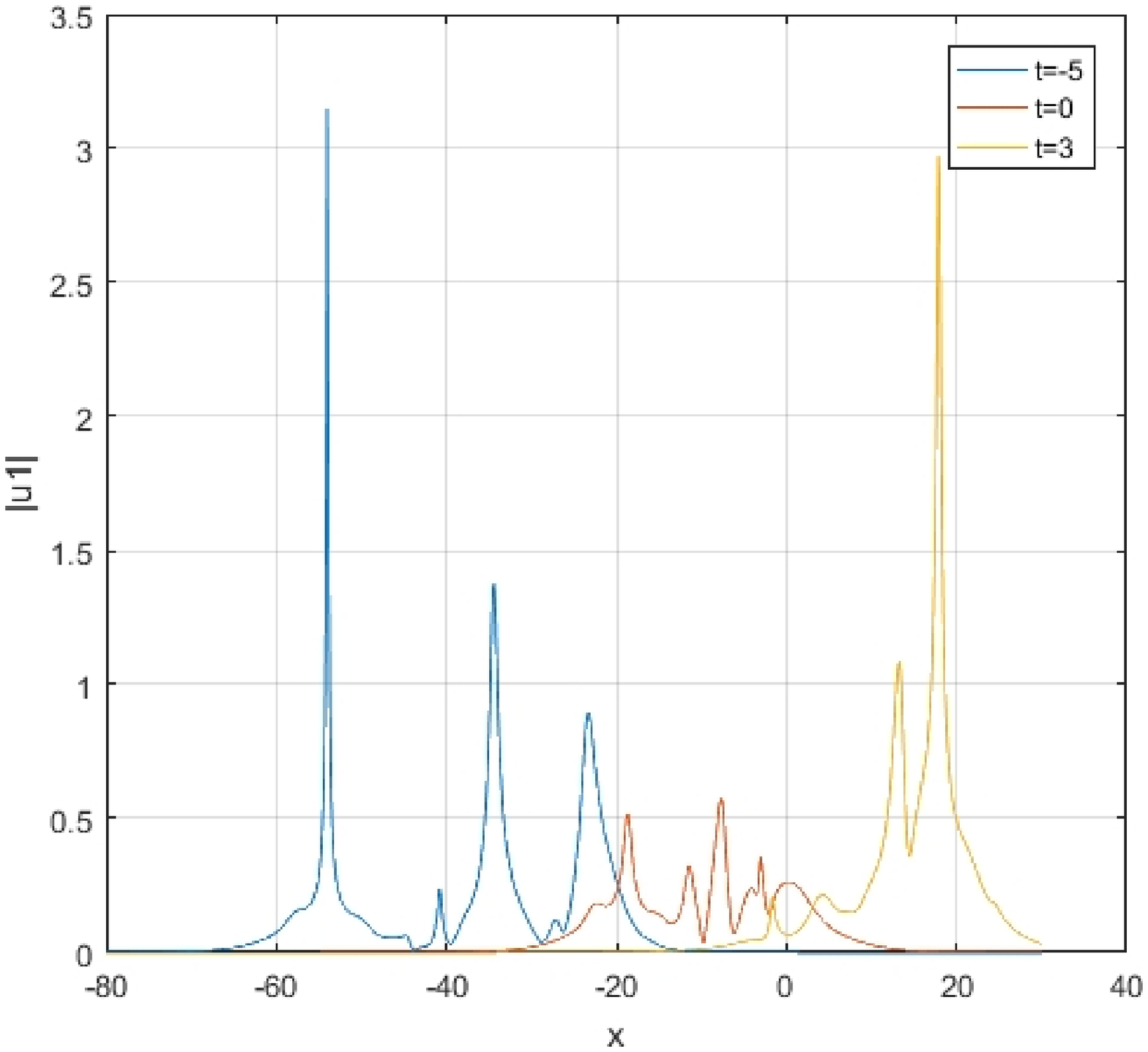}}}

$\ \qquad~~~~~~(\textbf{a})\qquad \ \qquad\qquad\qquad\qquad~(\textbf{b})
\ \qquad\qquad\qquad\qquad\qquad~(\textbf{c})$\\
\noindent { \small \textbf{Figure 10.} Three-soliton  solution to Eq. \eqref{three33} with parameters $\zeta_1=1.2+0.8i$,  $\zeta_2=1.5+0.7i$, $\zeta_3=1.1+0.9i$,  $\mu_{1,1}=0.1$, $\mu_{1,2}=\mu_{2,1}=\mu_{3,4}=0.2$,  $\mu_{1,3}=\mu_{3,1}=0.3$,  $\mu_{1,4}=\mu_{2,2}=\mu_{3,5}=0.4$,  $\mu_{1,5}=\mu_{2,5}=0.5$, $\mu_{2,3}=\mu_{3,2}=0.6$, $\mu_{2,4}=0.8$  and $\mu_{3,3}=0.9$.
$\textbf{(a)}$: the local structures of the three soliton solutions $u_1(x,t)$,
$\textbf{(b)}$: the density plot of $u_1(x,t)$,
$\textbf{(c)}$: the wave propagation of the  three soliton solutions $u_1(x,t)$.}  \\

\section{Conclusions}
Based on the previous work \cite{guo2012riemann},  the main purpose of our work is  to investigate a generalized $N$-component  FL equations  via the Riemann-Hilbert approach,  practically speaking,  which is to greatly promote the results of previous work.  In this work, the spectral analysis of the associated Lax pair is first carried out and a Riemann-Hilbert problem is established. After that, via solving the presented Riemann-Hilbert problem with reflectionless case, the $N$-soliton solution to the $N$-component  FL equations are obtained  at last. Furthermore, by selecting specific values of the involved parameters, a few plots of one-, two- and three- soliton solutions  are made to display the localized structures and dynamic  propagation  behaviors.

\section*{Acknowledgements}

This work was supported by  the Natural Science Foundation of Jiangsu Province under Grant No. BK20181351, the National Natural Science Foundation of China under Grant No. 11975306, the Six Talent Peaks Project in Jiangsu Province under Grant No. JY-059, the Qinglan Project of Jiangsu Province of China,  and the Fundamental Research Fund for the Central Universities under the Grant Nos. 2019ZDPY07 and 2019QNA35.

\end{document}